\pdfoutput=1
\documentclass[UKenglish,PAPER,cernpreprint,paper=a4,maketitle=true,texlive=2011]{atlasdoc}
\usepackage{textcomp} 
\usepackage[block=none]{atlaspackage}
\usepackage{atlascontribute}
\usepackage{atlasphysics}
\graphicspath{{logos/}{figures/}{.}}
\usepackage{pmssmpaper-defs}

\usepackage{etoolbox}
\newtoggle{texfig}
\toggletrue{texfig}
\hypersetup{pdftitle={ATLAS paper},pdfauthor={The ATLAS Collaboration}}

\AtlasTitle{Summary of the ATLAS experiment's sensitivity to supersymmetry after LHC \runone{} --- interpreted in the phenomenological MSSM }

\author{The ATLAS Collaboration}

\AtlasRefCode{SUSY-2014-08}

\PreprintIdNumber{CERN-PH-EP-2015-214}

 \AtlasJournal{JHEP}

\AtlasAbstract{%
A summary of the constraints from the ATLAS experiment on 
$R$-parity-conserving supersymmetry is presented. Results from \nanalyses{} separate ATLAS 
searches are considered, each based on analysis of up to \maxlumi{} of 
proton--proton collision data at centre-of-mass energies of $\sqrt{s}$ = 7 and 
\SI{8}{\TeV} at the Large Hadron Collider. 
The results are interpreted in the context of the 
19-parameter phenomenological minimal supersymmetric standard model, 
in which the lightest supersymmetric particle is a neutralino,
taking into account constraints from previous precision electroweak and flavour 
measurements as well as from dark matter related measurements.
The results are presented in terms of constraints on supersymmetric particle 
masses and are compared to limits from simplified models. The impact of ATLAS 
searches on parameters such as the dark matter relic density, 
the couplings of the observed Higgs boson,
and the degree of electroweak fine-tuning is also shown. 
Spectra for surviving supersymmetry model points with low fine-tunings are presented.
}

\AtlasCoverSupportingNote{
\AtlasTitleText --- supporting documentation}{https://cds.cern.ch/record/1747267}

\AtlasCoverCommentsDeadline{Thursday, 13 August 2015}

\AtlasCoverAnalysisTeam{
Y.~Abulaiti, A.~Barr, P.~Bechtle, I.~Besana, C.~Bock, R.~Caminal~Armadans, A.~Cortes-Gonzalez, D.~Cote, \O.~Dale, C.~David, D.~Delgove, C.~Deluca, P.~Dondero, L.~Duflot, L.~Escobar, W.~Fawcett, E.~Feng, S.~Gadatsch, T.~Gillam, B.~Gjelsten, Z.~Grout, L.~Heinrich, S.~Henrot-Versille, T.~Javurek, A.~Kastanas, S.~Kazama, D.~Krauss, T.~Lari, F.~Legger, W.~Liebig, A.~Lipniacka, J.~Lorenz, C.~Macdonald, A.~Mann, M.~Marjanovic, B.~Martin dit Latour, M.~Martinez-Perez, A.~Marzin, J.~Maurer, F.~Meloni, P.~Pani, B.~Petersen, T.~Potter, X.~Poveda, O.~Ricken, T.~Rizzo, N.~Rompotis, F.~Salvatore, H.~Sandaker, A.~De Santo, N.~Santoyo, S.~Schaepe, M.~Schultens, L.~Smestad, V.~Tudorache, R.~Ueno, P.~Urrejola, C.~Wanotayaroj, D.~Xu, S.~Yamamoto, T.~Yamanaka, X.~Zhuang
}

\AtlasCoverEdBoardMember{Sascha~Caron~(chair)}
\AtlasCoverEdBoardMember{Ian~Hinchliffe}
\AtlasCoverEdBoardMember{Simone~Amoroso}
\AtlasCoverEdBoardMember{Stephane~Willocq}

\AtlasCoverEgroupEditors{\url{mailto:atlas-SUSY-2014-08-editors@cern.ch}}

\AtlasCoverEgroupEdBoard{\url{mailto:atlas-SUSY-2014-08-editorial-board@cern.ch}}

\begin{document}

\tableofcontents

\section{Introduction}
\label{sec:intro}
During the first run of the Large Hadron Collider (LHC), 
the ATLAS~\cite{PERF-2007-01} and CMS~\cite{CMS-TDR-08-001} 
Collaborations performed a wide range of searches for supersymmetry
(SUSY)~\cite{Miyazawa:1966,Ramond:1971gb,Golfand:1971iw,Neveu:1971rx,Neveu:1971iv,Gervais:1971ji,Volkov:1973ix,Wess:1973kz,Wess:1974tw}, 
using proton-proton ($pp$) collision data at centre-of-mass energies 
$(\sqrt{s})$ of 7 and \SI{8}{\TeV}. SUSY, a theoretically favoured framework for 
extending the Standard Model (SM), is able to address some of its unanswered 
questions, particularly the hierarchy 
problem~\cite{Weinberg:1975gm,Gildener:1976ai,Weinberg:1979bn,Susskind:1978ms}, 
which is related to the fine-tuning needed to obtain the correct mass
for the observed Higgs boson. SUSY can also provide credible dark matter 
candidates~\cite{Goldberg:1983nd,Ellis:1983ew} 
and can improve the unification of the electroweak and strong 
interactions~\cite{Dimopoulos:1981yj,Sakai:1981gr,Ibanez:1981yh,Einhorn:1981sx,Marciano:1981un,Giunti:1991ta,Ellis:1990wk,Amaldi:1991cn,Langacker:1991an}.

The minimal supersymmetric extension of the Standard Model 
(MSSM)~\cite{Fayet:1976et,Fayet:1977yc,Farrar:1978xj,Fayet:1979sa,Dimopoulos:1981zb}
predicts partners for each of the SM states. 
It predicts a pair of scalar partners -- one for each fermion chirality --
for each of the SM quarks and leptons. 
These spin-zero partner particles are known as squarks ($\tilde{q}$) and
sleptons ($\tilde{\ell}$) respectively.
In the first two generations the pair of chiral partners is largely unmixed,
so the mass states can be labelled $\tilde{e}_{\rm L}$ and $\tilde{e}_{\rm R}$, 
where the ${\rm L}$ and ${\rm R}$ subscripts 
denote the scalar partners of the left- and right-handed Standard Model fermion states respectively.
In the third generation of quarks and leptons the mixing between the scalars is larger, 
and the mixed states are labelled by their mass indices 
e.g.~$\tilde{t}_1$ and $\tilde{t}_2$, where  $\tilde{t}_1$ is lighter by
construction. Each state in the SM gluon colour octet has a spin-half partner
known as a gluino $\tilde{g}$. 
There are a total of eight spin-half partners of the electroweak gauge and Higgs
bosons: the neutral bino (superpartner of the U(1) gauge field); the winos,
which are a charged pair and a neutral particle (superpartners of the $W$ bosons
of the SU(2)$_{\rm L}$ gauge fields); and the Higgsinos, which are two neutral particles
and a charged pair (superpartners of the Higgs field's degrees of freedom). 
The bino, winos and Higgsinos
mix to form four charged states called charginos 
$\tilde{\chi}_{1,2}^\pm$, and four
neutral states known as neutralinos $\tilde{\chi}_i^0$ (where the index $i$ 
lies in the range 1 to 4, ordered by increasing neutralino mass). 
The charginos and neutralinos are collectively referred to as electroweakinos. 

Since no statistically significant signals consistent with supersymmetry 
have yet been observed at the LHC, 
searches have been used to constrain the allowed 
supersymmetric model space. In the case of searches for supersymmetry, this
typically results in setting lower limits on the masses of the pertinent 
supersymmetric partner particles (sparticles). 

This paper presents the combined sensitivity and constraints from
\nanalyses{} separate ATLAS analyses of the \runone{} LHC dataset,
using centre-of-mass energies of 7 and \SI{8}{\TeV} 
and an integrated luminosity of up to \maxlumi{}.
Direct searches for the decay products of the 
sparticles listed above are considered together with  
searches for disappearing tracks, long-lived charged particles, 
monojet signatures, 
and a dedicated search for the heavier neutral Higgs bosons,
also expected in the MSSM.
More details about the searches can be found in 
Section~\ref{sec:analyses}.

The impact on the space of SUSY models has traditionally been 
presented in rather constrained frameworks, which
have particular limitations when considering 
large numbers of analyses.
One frequently used strategy for interpretation is in terms of 
models motivated by a particular mechanism of SUSY-breaking, 
for example via gravitational 
or gauge interactions.
While such models can have theoretically appealing features, 
they assert relationships between SUSY-breaking parameters  
that may not be realised in nature,
and they sample only a small part of the parameter space of the MSSM.
SUSY searches at the LHC have also been interpreted using `simplified' models.
Such models attempt to capture the behaviour of a small number of kinematically accessible sparticles,
assuming all others play no role.
The simplest case corresponds to one specific SUSY production process with a
fixed decay chain. 
Such models provide insight into the experimental constraints on the 
individual sparticle and decay mode,
but fail to capture the complex effects that can result from large numbers of 
competing production and decay processes.

The MSSM has over a hundred parameters that 
describe the pattern of sparticle masses and their decays. 
This parameter space is too large to be scanned exhaustively and compared to ATLAS data.
By applying a series of assumptions motivated by either experimental
constraints or general features of possible SUSY breaking mechanisms, the
number of parameters can be reduced to 19.
This is known as the phenomenological MSSM (pMSSM)~\cite{Djouadi:2002ze,Berger:2008cq,CahillRowley:2012cb}.
This model is assumed to conserve $R$ parity,\footnote{$R=(-1)^{3B+L+2S}$ where $B$ is baryon number, 
$L$ is lepton number and $S$ is spin.}
which ensures that sparticles are produced in pairs and the lightest 
supersymmetric particle (LSP) is stable. 
The parameters are assumed to be real so that new CP violation does not occur 
in the sparticle sector. Parameters that 
would give rise to additional flavour-changing neutral currents are absent.  
The LSP provides a dark matter candidate if colourless and electrically neutral. 
In this paper, the LSP is required to be the lightest neutralino.
Its production at the LHC gives rise to missing transverse momentum (whose magnitude is denoted \met), 
which is required by most of the ATLAS searches considered in this \paper. 
There is no theoretical upper bound on the parameters characterising 
the sparticle masses.
However, since the experiments have no sensitivity to sparticles with very large
masses, the following additional restriction is applied before a specific set
of parameters is considered: all sparticle masses must be less than
\SI{4}{\TeV}.
A specific set of the 19 parameters is referred to as 
a model point in parameter space;
\numTotal{} such model points, each consistent
with a range of previous experimental results, are considered.
More details about the selection of pMSSM points can be found in Section~\ref{sec:models}.

Several groups have advocated the use of the pMSSM for interpretation of LHC 
results~\cite{deVries:2015hva,Strege:2014ija,HenrotVersille:2014sca,Cahill-Rowley:2014wba,Cahill-Rowley:2014boa,Chakraborti:2014gea,CahillRowley:2012kx,AbdusSalam:2012ir,AbdusSalam:2012sy,CahillRowley:2012rv,CahillRowley:2012cb,Carena:2012he,Arbey:2012na,Strubig:2012qd,Arbey:2011un,Sekmen:2011cz,Allanach:2011ej,Conley:2010du,AbdusSalam:2009qd,Berger:2008cq}.
Most of these studies use estimated experimental efficiencies and acceptances
for pMSSM points,
and compare them to the model-independent limits from a selection of LHC searches 
to constrain the pMSSM parameter space.
Previous ATLAS analyses have also used the pMSSM for interpretation of 
individual searches~\cite{Aad:2014nua,Aad:2014vma,ATLAS-SUSY-2013-14,Aad:2014kra,Aad:2015pfx}
by fixing most of the parameters,
and varying just two or three: they therefore
explore only a small part of the parameter space.
This \paper{} makes full use of the 
ATLAS experimental simulation, reconstruction and analysis tools. 
It represents the most comprehensive assessment of the ATLAS constraints on supersymmetry models to date.

The \paper{} is organised as follows. 
The relevant ATLAS \runone{} analyses are summarised in Section~\ref{sec:analyses}.
A description of the pMSSM parameter space  
can be found in Section~\ref{sec:models}, along with the direct and
indirect constraints applied prior to the generation of the \numTotal{} model points.
Monte Carlo simulation of those model points is described in Section~\ref{sec:simulation}.
The effect of the ATLAS searches on this pMSSM space is described in Section~\ref{sec:results}.
Discussion and conclusions can be found in Section~\ref{sec:conclusion}.

\section{ATLAS searches}
\label{sec:analyses}

\begin{table}
\renewcommand\arraystretch{1.1}
\centering
\begin{tabular}{lrr}
{\bf Analysis}              & {\bf Ref.} & {\bf Category} \\ \hline
\ZeroLepton                 & \cite{ATLAS-SUSY-2013-02} & \multirow{7}{*}{Inclusive} \\
\Multijet                   & \cite{ATLAS-SUSY-2013-04} &            \\
\OneLeptonStrong             & \cite{Aad:2015mia} &            \\
\TauStrong                  & \cite{Aad:2014mra} &            \\
\SSThreeLepton              & \cite{Aad:2014pda} &            \\
\ThreeBjets                 & \cite{ATLAS-SUSY-2013-18} &            \\
\Monojet                    & \cite{Aad:2015zva} &            \\ \hline
\ZeroLeptonStop             & \cite{ATLAS-SUSY-2013-16} & \multirow{7}{*}{Third generation} \\
\OneLeptonStop              & \cite{Aad:2014kra} &            \\
\TwoLeptonStop              & \cite{Aad:2014qaa} &            \\
\StopMonojet                & \cite{Aad:2014stoptocharm} &            \\
\StopZ                      & \cite{Aad:2014mha} &            \\
\TwoBjet                    & \cite{ATLAS-SUSY-2013-05} &            \\
\TBmet                      & \cite{Aad:2015pfx}       &            \\ \hline
\OneLeptonHiggs             & \cite{Aad:2015jqa} & \multirow{6}{*}{Electroweak} \\
\TwoLepton                  & \cite{Aad:2014vma} &            \\
\TwoTau                     & \cite{ATLAS-SUSY-2013-14} &            \\
\ThreeLepton                & \cite{Aad:2014nua} &            \\
\FourLepton                 & \cite{Aad:2014iza} &            \\
\DisappearingTrack          & \cite{Aad:2013yna} &            \\
\hline
\LLSparticles               & \cite{Aad:2012pra,ATLAS:2014fka} &  \multirow{2}{*}{Other} \\
\HiggsToTauTau              & \cite{Aad:2014vgg} &            \\ \hline
\end{tabular}

\caption{The \nanalyses{} different ATLAS searches considered in this summary \note.
The term `lepton' ($\ell$) refers specifically to $e^\pm$ and $\mu^\pm$ states,
except in the cases of the electroweak \ThreeLepton{} and \FourLepton{}
analyses where $\tau$ leptons are also included.
}
\label{tab:susySearches}
\end{table}

A total of \nanalyses{} distinct ATLAS analyses are considered,
spanning a wide range of different search strategies and final states,
as listed in Table~\ref{tab:susySearches}.  Each analysis 
has several signal regions --- for
example the analysis requiring events with zero isolated electrons and muons
and a minimum of 2--6 jets in association with large
$\met$~\cite{ATLAS-SUSY-2013-02} has 15 different signal regions, each
with different requirements on kinematic parameters and/or
multiplicities of jets.  For each of the \nanalyses{} analyses, most 
of the signal regions from the original analysis are
considered. However, in some cases, for
practical reasons it was necessary to leave out some specialised signal
regions or more complex combined fits. This leads to a slight
underestimate of the full reach of the search.  In total, almost
200 distinct signal regions are considered.

The analyses are classified into the four broad categories
shown in Table~\ref{tab:susySearches}. 
`Inclusive' searches are those primarily targeting decays, including
cascade decays, initiated by production of squarks of the first two
generations or gluinos. `Third-generation' searches are those
targeted particularly at the production of top 
and bottom squarks,
known as the stop ($\tilde{t}$) and sbottom ($\tilde{b}$).
`Electroweak' searches include those for 
direct production of electroweakinos and
sleptons. Since each search involves multiple
signal regions, and since different SUSY production and decay processes can
contribute to each of those, this categorisation can only be
considered to be a rough guide when interpreting the type of
sparticles to which the analysis might show sensitivity.  `Other'
searches are those for heavy, long-lived particles (which are
only considered for a small subset of the model points)
and the search for heavy Higgs bosons.
The details of the analyses can be found in the corresponding papers 
(listed in Table~\ref{tab:susySearches}) and a brief summary for each is given below.

In what follows the term `lepton' ($\ell$) 
is used to refer specifically to the charged leptons $e^\pm$ and $\mu^\pm$
of the first two generations.
Where $\tau^\pm$ leptons are also included ---
for the \ThreeLepton{} and \FourLepton{} electroweak searches
--- this is indicated explicitly. 

\subsection{Inclusive searches}
\label{sec:analyses:inclusive}
The inclusive searches are designed to be sensitive to prompt decays of squarks, 
particularly those of the first two generations, and gluinos.
Strongly interacting sparticles may decay directly to the LSP,
via the decay $\squark \to q + \neutralino$ 
for the squark and via
$\gluino \to q + \bar{q} + \neutralino$ for the gluino.
Alternatively, cascade decays may also occur involving one or more additional sparticles
yielding final states with additional jets, 
large $\met$ and possibly leptons, including $\tau$ leptons.
The ATLAS searches targeting these final states are
classified according to the different dominant signal signatures, as follows.

The \ZeroLepton{} analysis~\cite{ATLAS-SUSY-2013-02}
has wide-ranging sensitivity to strongly interacting sparticle production.
It vetoes events with leptons in order to suppress the background from 
$W$ boson and $t\bar{t}$ decays.
Depending on the signal region, final states with a minimum jet requirement 
of 2--6 jets with large transverse momenta (denoted $\pt$ in the following)  
are considered, each in association with large $\met$.
Signal regions with small numbers of jets provide sensitivity 
to the direct production and decay of squarks,
while those with higher jet multiplicities are sensitive to the production and direct decay of gluinos,
and various cascade decays. The original analysis has two signal regions specifically targeting
hadronic decays of high-\pt{} $W$ bosons. These signal regions are
not considered in this paper.

The \Multijet{} analysis
~\cite{ATLAS-SUSY-2013-04} selects events with significant \met{} 
and with jet multiplicities ranging from 7 to 10 or more, depending on the signal region.
It was designed to target, amongst others, 
models where each gluino of a produced pair decays through a (possibly virtual) 
top squark to $t + \bar{t} + \neutralino$.
The four top quarks produced generally lead to large jet multiplicities in the final state.
This search also has sensitivity to other models in which cascade decays
generate large numbers of jets.
It has a looser requirement on the $\met$ than the 2--6 jet analysis 
described in the previous paragraph,
because the many possible intermediate stages of the cascade decay
tend to reduce the \met.
Unlike the original analysis, where the disjoint signal regions
could be statistically combined to improve sensitivity,
in this \paper{} signal regions are considered individually when determining 
whether or not a model point is excluded.

The \OneLeptonStrong{} analysis~\cite{Aad:2015mia} explicitly requires
one isolated lepton, several jets and high $\met$ in the selection.
Two sets of signal regions are used from this analysis -- one set with relatively
high $\pt$ leptons (sensitive to SUSY scenarios with larger mass
splittings between the produced sparticle and the LSP) and another set
using low-$\pt$ leptons (sensitive to smaller mass splittings). The signal regions
requiring two leptons are not considered.  Overall, it is sensitive to
decay chains where leptons can be produced through the cascade decay
of squarks and gluinos.

The \TauStrong{} search~\cite{Aad:2014mra} 
targets final states arising from cascades producing 
hadronically 
decaying $\tau$ leptons -- with signal regions requiring either one or two
$\tau$ leptons, and
 including large $\met$, jets and either
exactly zero or one additional light lepton. This search can
be sensitive to long decay chains in models with light staus.

Cascade decays of squark and gluino pairs can also lead to final states
 with multiple leptons, or with two leptons of the same electric charge,
known as same-sign (SS) leptons.
Those final states are addressed by the \SSThreeLepton{} 
analysis~\cite{Aad:2014pda}, which requires multiple jets in the final state, and 
either two SS leptons -- with or without 
jets containing $b$-hadrons (\emph{b}-jets) in the
final state -- or at least three leptons. 

For models where many $b$-jets are expected, the specially
designed \ThreeBjets{} analysis~\cite{ATLAS-SUSY-2013-18} is
sensitive.  This analysis
is designed around the definition of two sets of signal regions: one set with no
isolated leptons, and another with at least one isolated lepton.

The \Monojet{} analysis~\cite{Aad:2015zva} selects events where the
leading jet's \pt is as large as 50\% of the $\met$,
and there is large $\met$ and no leptons. 
The single jet can originate from initial-state QCD radiation (ISR),
providing sensitivity to collisions
in which no decay products from sparticle decays are observed.
This can occur either for direct pair production of invisible LSPs,
or if the produced sparticles are
only a little heavier than the LSP (up to a few \GeV{}) 
and their decays therefore produce SM particles of too low an energy
to be detected in the other searches.

\subsection{Third-generation searches}
This set of analyses is focused on searches for direct production of third 
generation squarks.
Their masses are generally expected to be at the \TeV{} scale or 
below if the Higgs boson(s) are to be protected from large unnatural
loop corrections. The decay of $\tilde{t}$ and $\tilde{b}$ squarks 
also leads to distinctive experimental signatures, typically involving
the production of $t$- or $b$-quarks in association with large \met.

The $0$-lepton stop  
search \cite{ATLAS-SUSY-2013-16} is optimised for the direct
production of pairs of top squarks decaying directly to a top quark and neutralino,
leading to an all-hadronic final state with at least two \emph{b}-jets and large $\met$. Most of the signal
regions rely on variables related to the reconstructed top quarks present in
the final state and on lepton vetoes,
but there are also signal regions that target the case where one of the pair of
top squarks decays to a top quark and a neutralino and the other decays to a bottom quark and a chargino. 

The 0-lepton stop search is complemented by the \OneLeptonStop{}  search \cite{Aad:2014kra}, 
in which all of the signal regions are characterised 
by exactly one isolated lepton, at least two 
jets and large $\met$.
The presence of a \emph{b}-jet is used in both signal regions targeting
$\tilde{t}_1\rightarrow b\chargino$ and those targeting  $\tilde{t}_1\rightarrow t\neutralino$, while the latter also use variables related to reconstructed top quarks.
The dedicated signal regions targeting top squark decays with soft leptons or boosted top quarks are
not included in this paper. 

The \TwoLeptonStop{} search \cite{Aad:2014qaa} is designed for final states
containing two isolated leptons and large $\met$, primarily targeting top squarks decaying
through an intermediate chargino. Only the so-called ``leptonic $m_{\rm T2}$'' signal regions, targeting
charginos decaying through on-shell $W$ bosons, are included in this paper.

The \StopMonojet{} analysis~\cite{Aad:2014stoptocharm} looks for final
states characterised by large $\met$, at least one high-$p_{\rm T}$ jet (vetoing
events with more than three jets), and no leptons. 
The signal regions of this search
were designed in the context of a search for top squarks, each decaying into an undetected charm
quark and a neutralino which is relevant for cases where the LSP mass is close
to the top squark mass. 
The \met{} requirement is less stringent than that of 
the \Monojet{} analysis described in Section~\ref{sec:analyses:inclusive}.
The signal regions targeting reconstructed and tagged charm jets
are not included in this paper.

The search for top squarks with a $Z$ boson in the final state \cite{Aad:2014mha} is
motivated by the decay of $\tilde{t}_2\rightarrow\tilde{t}_1Z$, which
can produce many leptons in the final state. The leptons are required to form a
pair with a mass consistent with the $Z$ boson, 
with at least one \emph{b}-jet and large $\met$.

The \TwoBjet{} analysis~\cite{ATLAS-SUSY-2013-05} searches for SUSY scenarios that
produce events containing exactly two $b$-jets, significant $\met$ and no
isolated leptons, for example those coming from decays of bottom squarks 
to a $b$-quark and the LSP and from top squark to a $b$-quark and chargino.

The \TBmet{} analysis~\cite{Aad:2015pfx} was designed for a mixed scenario:
direct production of pairs of top or bottom squarks 
each decaying (with various
branching ratios) to neutralinos or charginos, 
and yielding final states consisting of a
top quark, bottom quark and large $\met$.

\subsection{Electroweak searches}

This section details the analyses considered in this \paper{} which target sparticles produced via
electroweak interactions. This includes the production of pairs of
sleptons or electroweakinos which typically decay into final states containing
several high-\pt leptons and significant $\met$.

The lepton plus Higgs boson (\OneLeptonHiggs{}) 
analysis~\cite{Aad:2015jqa} is designed to search for direct pair production of a chargino and a
neutralino, which decay to final states with large $\met$, an isolated
lepton, and a Higgs boson $h$ which is identified by requiring either two
\emph{b}-jets, or two photons, or a second lepton with the same electric charge (targeting $h\rightarrow WW$ decays).
Only the signal regions for Higgs boson decays to bottom quarks are considered in this paper. 

The \TwoLepton{} analysis \cite{Aad:2014vma} targets electroweak
production of charginos and/or neutralinos, or sleptons in
events with exactly two leptons, large $\met$ and, for some signal regions, two or more jets in the
final state.

A complementary search targeting the third-generation leptons is the \TwoTau{} analysis
\cite{ATLAS-SUSY-2013-14} searching for SUSY in events with at least two
hadronically decaying $\tau$ leptons, large $\met$ and a jet veto.

The \ThreeLepton{} analysis~\cite{Aad:2014nua} is a search for the direct production of
charginos and neutralinos in final states with three leptons ---
which here may include up to two hadronically decaying $\tau$ leptons --- and large $\met$,
which can come through the decays via sneutrinos, sleptons or $W$, $Z$ or Higgs
bosons.

The \FourLepton{} analysis~\cite{Aad:2014iza} looks for SUSY in events with
four or more leptons, of which at least two must be electrons or muons. The
leptons may also include  hadronically decaying $\tau$ leptons in this case.
Such high lepton multiplicity final states
can occur if a degenerate $\tilde{\chi}_2^0$ $\tilde{\chi}_3^0$ pair is
produced which subsequently decay via sleptons, staus or $Z$ bosons to $\tilde{\chi}_1^0$ and many leptons.

The \DisappearingTrack{} analysis~\cite{Aad:2013yna} is motivated by
scenarios with a wino-like LSP in which the charged wino is typically only
$\sim\SI{160}{\MeV}$ heavier than the LSP.  In such models the
\chargino{} can have decay lengths of order a few tens of centimetres
before it decays to a \neutralino{} and a charged pion.  The
low-momentum pion track is typically not reconstructed, so the
distinctive signature is that of the high-\pt{} chargino track
apparently disappearing within the detector volume.

\subsection{Other searches}

The long-lived particle searches~\cite{Aad:2012pra,ATLAS:2014fka} are
designed to detect heavy long-lived particles by measuring their speed $\beta$
using the time-of-flight to the calorimeters and muon detectors and
$\beta\gamma$ (where $\gamma$ is the relativistic Lorentz
factor) from the specific ionisation energy loss in the pixel
detector. 
Only the direct production of pairs of long-lived 
top or bottom squarks, gluinos, staus or charginos
are considered in this paper. 
The search using 7\,\TeV{} data from 2011~\cite{Aad:2012pra} considered
sparticles as light as 200~\GeV, whereas in most cases the later analysis
\cite{ATLAS:2014fka} only
considered sparticles above 400~\GeV. Both searches are therefore included
for maximal sensitivity.

The \HiggsToTauTau{} search~\cite{Aad:2014vgg} is designed to detect
the heavy, neutral Higgs bosons predicted in the MSSM if they decay to
$\tau$-pairs.

\section{pMSSM points and indirect constraints}
\label{sec:models}
An overview of the method of selecting 
signal model points is provided in this section.
A summary of the scan of the pMSSM parameter space, the software employed, 
and the constraints applied to determine the final selection are described. 

\subsection{pMSSM points generation}
\label{sec:pMSSMpars}

The model set is generated
by selecting model points within the pMSSM 
using methods similar to those described in Ref.~\cite{CahillRowley:2012cb}, 
but with several important changes. 
The modifications are made after taking into account new
experimental results, updated calculational tools, knowledge gained
from the study described in Ref.~\cite{CahillRowley:2012cb},
and the improved capabilities of the ATLAS simulations.
The full details of the method by which the model points are selected, 
including the sampling procedure, the codes employed, 
and the constraints applied to determine the final selection
of `surviving' model points may be found in the Appendices.

The model points are selected after making the following
assumptions about the MSSM. They are motivated both by 
constraints from experimental observations and 
a desire for theoretical simplicity:
\begin{description}
\item[($i$)] $R$-parity is exactly conserved.
\item[($ii$)] The soft parameters are real, so that no new sources of CP violation exist beyond that present in the CKM matrix.
\item[($iii$)] Minimal Flavour Violation~\cite{DAmbrosio:2002ex} is imposed at the electroweak scale.
\item[($iv$)] The first two generations of squarks and sleptons with the same quantum numbers
are mass degenerate, and their Yukawa couplings are too small to affect sparticle production or precision observables.
\item[($v$)] The LSP is the lightest neutralino.
\end{description}

\noindent

This approach remains agnostic about the presence of
non-minimal particle content at higher scales, the mechanism of SUSY
breaking, and the unification of sparticle masses. 
Assumptions $ii$--$iv$ are motivated by the necessity of imposing 
some organising principle on SUSY flavour-violating parameters to allow
\TeV{}-scale masses for the squarks and sleptons. Combining assumptions
$i$--$v$ reduces the large MSSM parameter space to the 19-dimensional
subspace considered here.
The parameters and the ranges used to sample them are listed in
Table~\ref{tab:scanranges}. 
The 4\,\tev{} upper bound on most of the mass parameters
is chosen to make all states kinematically accessible at the LHC.
As might be expected, decreasing the value of this upper limit 
restricts the space, resulting in an increase in the apparent 
fraction of the pMSSM space to which ATLAS analyses are sensitive.
Further increasing any physical mass above 4\,\tev{}
has little effect on the LHC phenomenology in most cases. 
An exception is when decays proceed via virtual heavy sparticles, 
when increasing that sparticle mass would lead to further suppression 
of those decays.
A larger range is permitted for $|A_t|$, 
a parameter which affects loop corrections to the mass of the 
the Higgs boson. 
The larger range increases the fraction of 
model points having the mass of the lightest Higgs boson 
close to the measured value.

Given the large dimensionality of the pMSSM, 
a grid sampling technique at regular intervals is impractical. The space is
therefore sampled by choosing random values for each parameter.  It should be
noted that in many cases only some of the parameters are relevant for a
given observable, in which case the scan is effectively more
comprehensive within the subspace of relevant parameters. The
value of each parameter is chosen from a flat probability
distribution, with lower and upper bounds given in
Table~\ref{tab:scanranges}. The lower and upper limits of the parameter
ranges are chosen to avoid experimental constraints and to give a high
density of model points with masses at scales accessible by the LHC
experiments, respectively.

\begin{table}
\centering
\renewcommand\arraystretch{1.2}
\begin{tabular}{lrrl}
\hline
{\bf Parameter} & {\bf Min value} & {\bf Max value} & {\bf Note}\\
\hline
$m_{\tilde L_1}(=m_{\tilde L_2})$    & 90\,\GeV    & 4\,\TeV & Left-handed slepton (first two gens.) mass \\
$m_{\tilde e_{1}} (= m_{\tilde e_{2}})$    & 90\,\GeV    & 4\,\TeV & Right-handed slepton (first two gens.) mass\\
$m_{\tilde L_{3}}$    & 90\,\GeV    & 4\,\TeV & Left-handed stau doublet mass\\
$m_{\tilde e_{3}}$    & 90\,\GeV    & 4\,\TeV & Right-handed stau mass\\
\hline
$m_{\tilde Q_{1}} (= m_{\tilde Q_{2}})$     & 200\,\GeV   & 4\,\TeV & Left-handed squark (first two gens.) mass\\
$m_{\tilde u_{1}} (= m_{\tilde u_{2}})$     & 200\,\GeV   & 4\,\TeV & Right-handed up-type squark (first two gens.) mass\\
$m_{\tilde d_{1}} (= m_{\tilde d_{2}})$     & 200\,\GeV   & 4\,\TeV & Right-handed down-type squark (first two gens.) mass\\
$m_{\tilde Q_{3}}$      & 100\,\GeV   & 4\,\TeV  & Left-handed squark (third gen.) mass\\
$m_{\tilde u_{3}}$      & 100\,\GeV   & 4\,\TeV  & Right-handed top squark mass\\
$m_{\tilde d_{3}}$      & 100\,\GeV   & 4\,\TeV  & Right-handed bottom squark mass\\
\hline
$|M_1|$                & 0\,\GeV     & 4\,\TeV & Bino mass parameter   \\
$|M_2|$              & 70\,\GeV    & 4\,\TeV & Wino mass parameter  \\
$|\mu|$             & 80\,\GeV    & 4\,\TeV & Bilinear Higgs mass parameter \\
$M_3 $              & 200\,\GeV   & 4\,\TeV & Gluino mass parameter \\
\hline
$|A_{t}|$            & 0\,\GeV &   8\,\TeV & Trilinear top coupling\\
$|A_{b}|$           & 0\,\GeV &  4\,\TeV & Trilinear bottom coupling\\
$|A_{\tau}|$         & 0\,\GeV &  4\,\TeV & Trilinear $\tau$ lepton coupling \\
$M_A$                & 100\,\GeV  & 4\,\TeV & Pseudoscalar Higgs boson mass\\
$\tan \beta$       & 1 &  60 & Ratio of the Higgs vacuum expectation values\\
\hline
\end{tabular}
\caption{
Scan ranges used for each of the 19 pMSSM parameters.
Where the parameter is written with a modulus sign 
both the positive and negative values are permitted. 
In the above, ``gen(s)'' refers to generation(s). 
\label{tab:scanranges}
}
\end{table}

Condition $iv$ imposes the constraints that the soft mass terms
for the second generation are equal to those in the first,
as shown in Table~\ref{tab:scanranges}.
This means, for example, that $\tilde u_{\rm L}$ and  $\tilde c_{\rm L}$ have the same soft mass 
term in the Lagrangian so that their physical masses are very close.  
Furthermore the scalar partners of the left-handed fermions, such as $\tilde e_{\rm L}$ and 
$\tilde \nu_{e_{\rm L}}$, have the same soft mass due to $SU(2)_{\rm L}$ invariance, 
but \emph{D}-terms related to electroweak symmetry breaking split their mass-squared values 
by $\mathcal{O}(m_W^2)$.

Once each of the 19 parameters has been chosen, a variety
of publicly available software packages are used to calculate the
properties of each model point, % WJF end edit
as described in Appendix~\ref{sec:modelcalculation}. 
In some cases the software is modified to produce accurate results for the wide
range of models found in the pMSSM scan. 
The sparticle decays are calculated, again using a variety of codes and analytical
techniques, as described in Appendix~\ref{sec:decays}.

\subsection{pMSSM point selection}
\label{sec:pMSSMselection}

Acceptable model points are furthermore 
required to have consistent electroweak symmetry breaking, 
a scalar potential that does not break colour or electric charge, 
and all particles' mass-squared values must be positive.
Model points with theoretical pathologies, described in more detail in 
Appendix~\ref{sec:pathologies}, are discarded. 
Further experimental constraints, shown in Table~\ref{tab:constraints}, which indirectly affect the parameter space are applied and described below.

\begin{table}
\centering
\renewcommand\arraystretch{1.3}
\begin{tabular}{lcc}
\hline
{\bf Parameter} & {\bf Minimum value} & {\bf Maximum value} \\
\hline
$\Delta{\rho}$ & $-0.0005$ & 0.0017\\
$\Delta(g - 2)_{\mu}$ &  $-17.7 \times 10^{-10}$ & $43.8 \times 10^{-10}$ \\
BR($b \to s \gamma$) & $2.69 \times 10^{-4}$ &  $3.87 \times 10^{-4}$ \\
BR($B_s \to \mu^+ \mu^-$) & $1.6 \times 10^{-9}$ & $4.2 \times 10^{-9}$ \\
BR($B^+ \to \tau^+ \nu_{\tau}$) & $66 \times 10^{-6}$ & $161 \times 10^{-6}$ \\
\hline
$\Omega_{\tilde{\chi}_1^0} h^2$ & --- &  0.1208 \\
\hline
$\Gamma_{\rm invisible(SUSY)}(Z)$ & --- & \SI{2}{\MeV}\\
Masses of charged sparticles & 100\,\GeV & --- \\
$m(\chargino)$ & 103\,\GeV & --- \\
$m(\tilde{u}_{1,2},\,\tilde{d}_{1,2},\,\tilde{c}_{1,2},\,\tilde{s}_{1,2})$ & 200\,\GeV & --- \\
$m(h)$ & 124\,\GeV & 128\,\GeV \\
\hline
\end{tabular}
\caption{Constraints on acceptable pMSSM points from considerations of 
precision electroweak and flavour results, dark matter relic density,
and other collider measurements.
A long dash (---) indicates that no requirement is made.
Further details may be found in the text.
}
\label{tab:constraints}
\end{table}

\subsubsection{Precision electroweak and flavour constraints}
\label{par:precision}
Unless specified otherwise, the relevant observables are calculated
using \program{micrOMEGAs 3.5.5}~\cite{Belanger:2006is,Belanger:2010pz}. 
The constraint on the electroweak parameter
$\Delta{\rho}$ uses the limit on $\Delta T$ (the parameter describing the 
radiative corrections to the total $Z$ boson coupling strength, 
the effective weak mixing angle, and the $W$ boson mass) 
in Ref.~\cite{Baak:2012kk} and
$\Delta {\rho}$ = $\alpha \Delta T$ with $\alpha = 1/128$. 
The allowed branching ratio (BR) of $b \to s \gamma$ 
is the union of the two standard deviation ($2\sigma$) intervals
around the theoretical prediction and the experimental measurement 
from Ref.~\cite{Amhis:2012bh}. For
the branching ratio of $B_s \to \mu^+ \mu^-$, the value calculated by {\tt micrOMEGAs} is
scaled by $1 / (1 - 0.088)$ as proposed in Ref.~\cite{DeBruyn:2012wk} for
comparison with experiment. The scaled value is required to lie within the 
$2\sigma$ interval around the combined result from the LHCb and CMS
Collaborations~\cite{CMS:2014xfa}.  The $2\sigma$ theoretical prediction for
the SM $(3.20 $ to $ 4.12)\times 10^{-9}$ lies within this interval.
The branching ratio $B^+ \to \tau^+ \nu_{\tau}$ is
calculated using Ref.~\cite{Mahmoudi:2008tp}, which
includes $\tan \beta$-enhanced corrections. The allowed range is 
the union of the $2\sigma$ intervals around the experimental 
results~\cite{Aubert:2009wt,Hara:2010dk,Adachi:2012mm,Lees:2012ju}
and the SM prediction \cite{Charles:2004jd}.
Finally, for the SUSY contribution to the anomalous magnetic moment of the
muon, $\Delta(g-2)_\mu$,
 a very large range is allowed. 
This range in $\Delta(g-2)_\mu$ is the union of the $3\sigma$
intervals around the SM value, $(0.0 \pm 5.9) \times 10^{-10}$ from
combining \cite{Aoyama:2012wk} the results of
Refs.\cite{Hagiwara:2011af,Nyffeler:2009tw,Czarnecki:2002nt} and the
experimental measurement
\cite{PhysRevLett.92.161802,PhysRevD.73.072003} corrected to an
updated value of the muon to proton magnetic moment ratio
\cite{RevModPhys.80.633,1674-1137-34-6-021} giving
$\Delta(g-2)_{\mu,\mathrm{exp}}=(24.9\pm 6.3) \times
10^{-10}$. Three-sigma intervals are used to obtain a continuous range
from the union.

\subsubsection{Dark matter constraints}\label{sec:dmconstraint}
Since $R$-parity is assumed to be exactly conserved, the LSP is stable
and as a consequence has a non-zero cosmological abundance. It is assumed that the
LSP abundance is determined thermally and is not diluted by other processes
e.g.~late-time entropy addition. No assumption is made about whether
the LSP is the sole constituent of dark matter. As a result, the total
cold dark matter energy density is used as an upper limit on the LSP
abundance. The limit is based on the latest combined measurement from
the Planck Collaboration of $\Omega_{\mathrm{CDM}} h^2=0.1188\pm 0.0010$ (Table~4 of
Ref.~\cite{Ade:2015xua}).\footnote{It should be noted that in the context of the dark matter relic density the symbol $h$ corresponds to the normalised Hubble constant, rather than the Higgs boson.} The upper limit is set to the observed central value plus double the experimental uncertaintity.
The limit on the spin-independent cross-section is that
for the interaction of a
neutralino with a nucleus derived by the LUX experiment~\cite{Akerib:2013tjd}.
In the case of the LSP mass versus proton
spin-dependent cross-section plane the limit is from the COUPP 
Collaboration~\cite{Behnke:2012ys}, while in the LSP mass versus neutron
spin-dependent cross-section plane, 
the XENON100 Collaboration~\cite{Aprile:2013doa} limit is applied. 
\program{MicrOMEGAs 3.5.5} is used to
calculate the neutralino--nucleon cross-sections. These are scaled down by the ratio of
the expected relic density from the LSP to the observed relic
density to obtain the effective dark matter cross-sections, assuming
the remaining non-LSP dark matter is invisible to the direct
detection experiments. When accepting or rejecting models, the calculated value is allowed to be up to a factor of four higher
than the limits obtained by the experiments, to
account for nucleon form-factor uncertainties~\cite{Berger:2008cq}.
 
\subsubsection{Collider constraints}
\label{par:collider}
Finally, constraints from LEP and from the measurement of the Higgs boson
mass at the LHC are applied. To ensure consistency with LEP, model points
are discarded if their additional contribution to the 
invisible width of the $Z$ boson is above 2\,\MeV~\cite{ALEPH:2005ab},
or where any charged sparticle is lighter than 100\,\GeV. For
charginos, the bound is increased to 103\,\GeV, provided that all
sneutrinos are heavier than 160\,\GeV{} and the mass splitting
between the chargino and the LSP is at least 2\,\GeV. This constraint
comes from the combined LEP search~\cite{LEP_Chargino}.
First- and second-generation squarks are
required to be heavier than 200\,\GeV, although this has only a
very small effect given the scan range and the assumption of negligible
first- and second-generation squark mixing. The lightest Higgs boson mass, as
calculated by \program{FeynHiggs 2.10.0}~\cite{Heinemeyer:1998yj,Hahn:2013ria},
is required to be in the range 124 to 
128\,\GeV. This range is set around the central value of the Higgs mass at the
time of generation, 126\,\GeV, and with a 2\,\GeV~window that mainly reflects the typical theoretical uncertainty of the
\program{FeynHiggs} calculation. 
The results are found not to depend 
on the exact value of the Higgs mass within this interval
(as shown later in Section~\ref{sec:results:precision}).

\begin{table}
\centering\renewcommand\arraystretch{1.4}
\begin{tabular}{lccrcc}
\hline
\multirow{2}{*}{\bf LSP type} &  
\multirow{2}{*}{\bf Definition} & \multirow{2}{*}{\bf Sampled} & 
    \multicolumn{2}{c}{\bf Simulated}    & \multirow{2}{*}{\bf Weight}\\
 & & & {\bf Number} & {\bf Fraction} &  \\
\hline
`Bino-like'      & $N_{11}^2 > {\rm max} (N_{12}^2,\,N_{13}^2+ N_{14}^2)$ &  $480\times 10^6$ & \numBinoOnly & 35\% & $1/24$ \\
`Wino-like'      & $N_{12}^2 > {\rm max} (N_{11}^2,\,N_{13}^2 + N_{14}^2)$ &   \multirow{2}{*}{{ \Big\}} $20\times 10^6$ {\Big\{} }& \numWinoOnly & 26\% & 1 \\
`Higgsino-like'  & $(N_{13}^2 +N_{14}^2 ) > {\rm max} (N_{11}^2,\,N_{12}^2 )$ & & \numHiggsinoOnly & 39\% & 1 \\
\hline
{\bf Total}  & & $500\times10^6$ & \numTotal & & \\
\hline
\end{tabular}
\caption{\label{table:lsptype}
Categorisation of the \numTotal model points by the
type of the LSP (assumed to be the $\neutralino$)
according to the neutralino mixing matrix parameters $N_{ij}$, 
where the first index indicates the neutralino mass eigenstate
and the second indicates its nature
in the lexicographical order $(\tilde{B},\tilde{W},\tilde{H_1},\tilde{H_2})$.
For example, $N_{1,2}$ is the amplitude for the LSP 
to be $\tilde{W}$.
The final two columns indicate the fraction of model points in that category that
are sampled, and their weighted fraction after importance sampling.
}
\end{table}

\subsubsection{Importance sampling by LSP type}
\label{sec:importanceSampling}
Since low-mass SUSY models typically over-produce dark matter, 
the relic density constraint
in Table~\ref{tab:constraints} sculpts the distribution of the allowed model points.
The constraint depends strongly on the nature of the LSP.
Except where particularly effective neutralino annihilation mechanisms are available, 
model points with a bino-like LSP generally tend to produce too much dark
matter~\cite{ArkaniHamed:2006mb}, meaning that such models are infrequently sampled and accepted  
in a random scan employing flat priors.
The model points are therefore partitioned into three categories,
bino-like, wino-like and Higgsino-like.
The categorisation is made according to the dominant contribution
to the LSP within the neutralino mixing matrix $N_{ij}$
as shown in Table~\ref{table:lsptype}.
Model points are therefore selected, by importance sampling, in such a way that
approximately equal numbers are obtained for each LSP type.
In total 500 million model points are sampled randomly within the
ranges listed in Table~\ref{tab:scanranges}. From the first 20 million sampled, 
\numWinoPlusHiggsino{} model points had a wino-like or Higgsino-like LSP
and satisfied all of the constraints of Table~\ref{tab:constraints}.
To obtain a sufficiently high number with bino-like LSP,
the remaining $480$ million model points are used to find the \numBinoOnly{} 
which had a bino-like LSP and satisfied the Table~\ref{tab:constraints} constraints.
Generally models have a LSP dominated by one particular type, with over 87\% of
models having a LSP which is at least 90\% pure.
The phenomenology of each LSP type can be explored separately due 
to the large number of model points in each category. 
In the following plots, where all LSP types are shown together 
the contribution from each LSP
type is scaled according to the weights shown in Table~\ref{table:lsptype}.
\subsection{Properties of model points (before applying ATLAS constraints)}
\label{sec:models:properties}
\begin{figure}
\centering
\includegraphics[width=0.45\textwidth]{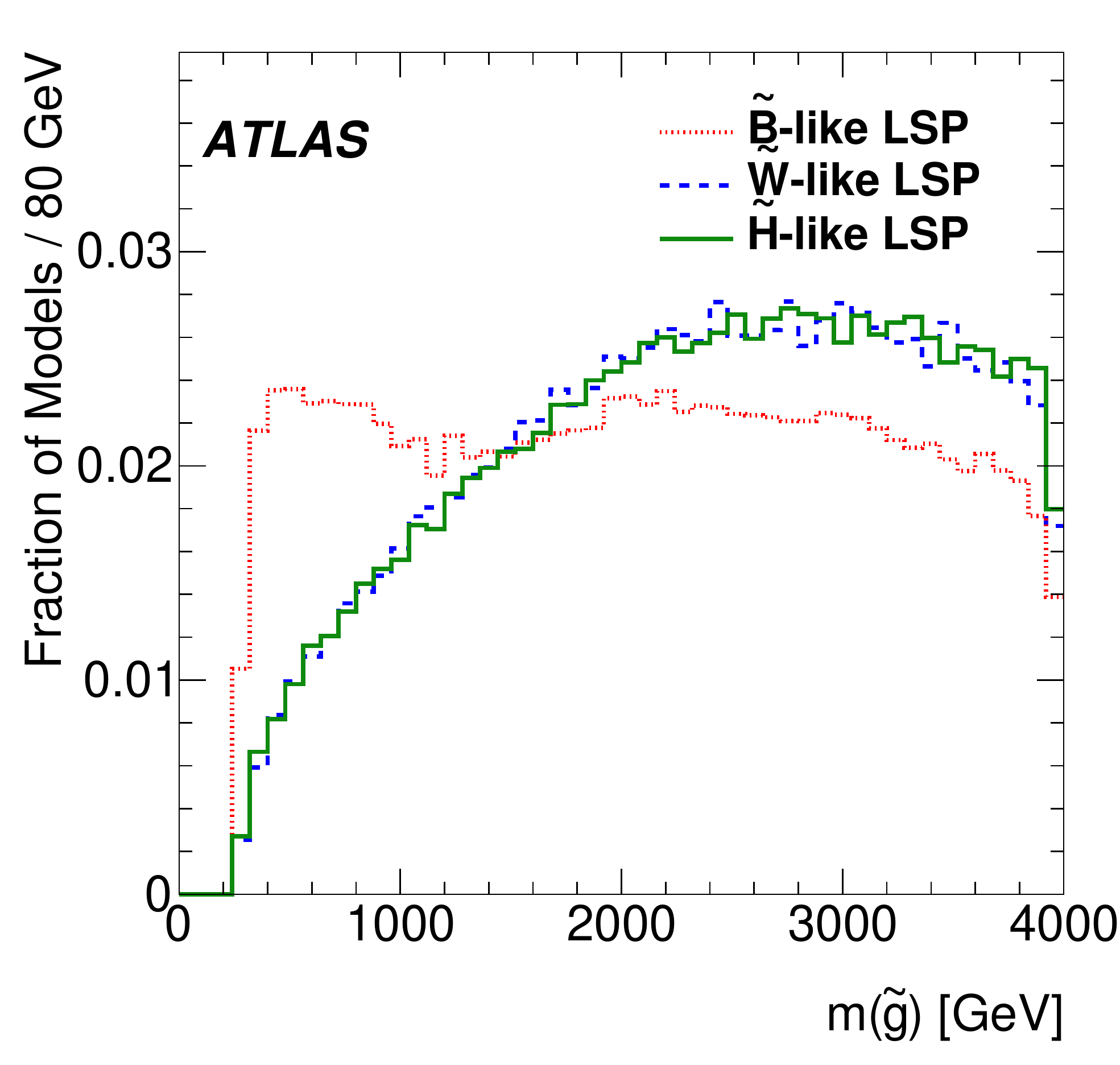}
\includegraphics[width=0.45\textwidth]{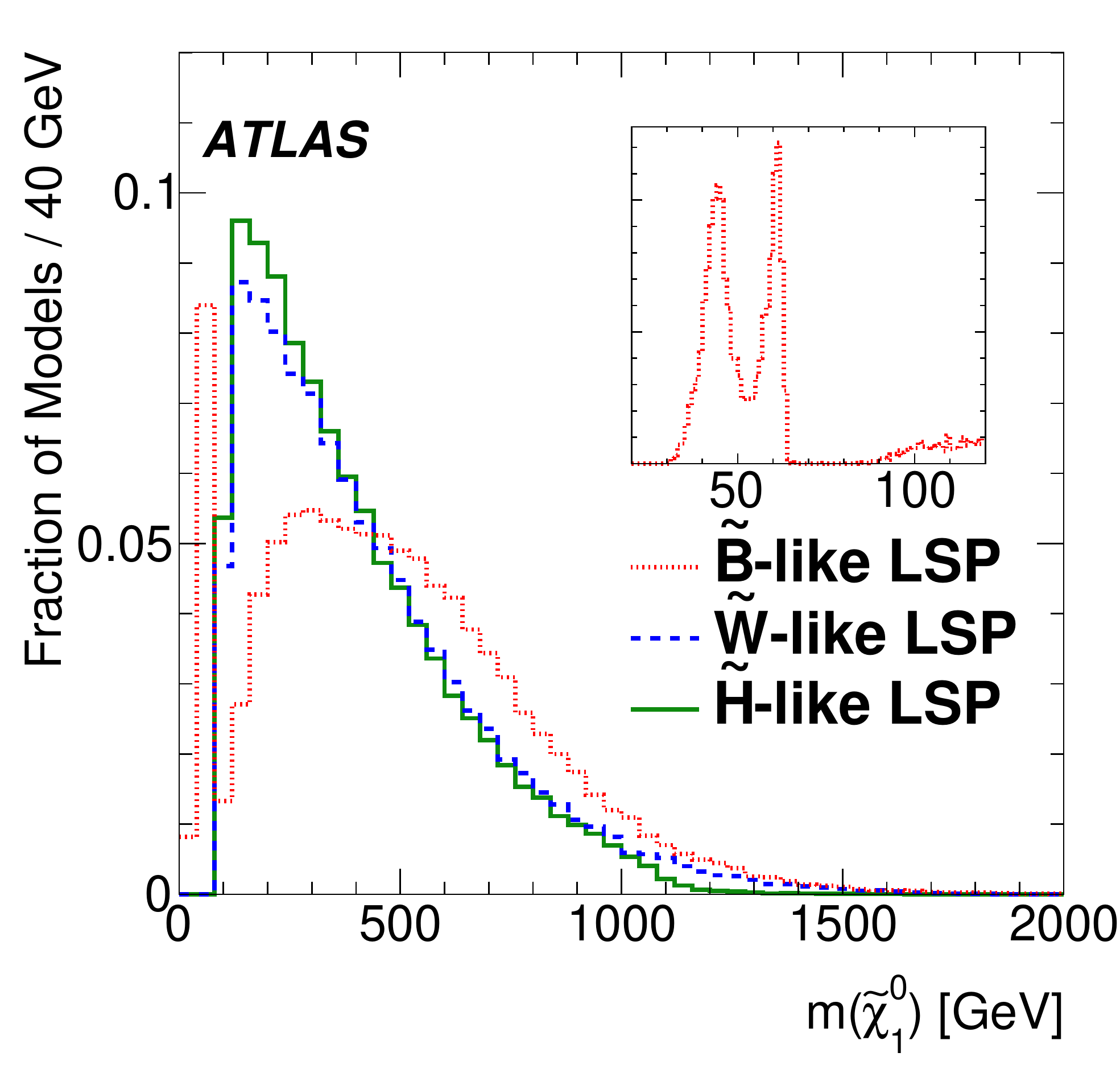}
\caption{Distributions of the gluino and LSP masses, shown separately 
for models with a bino-like (dotted red), wino-like (dashed blue) or Higgsino-like (solid green) LSP. 
The constraints listed in Table~\ref{tab:constraints} have
been applied, but not the constraints from the ATLAS searches.
The distributions are normalised to unit area.
The inset in the plot on the right shows in more detail the region of low neutralino mass for the models with bino-like LSP.
}
\label{fig:basic-gl-chi10}
\end{figure}

\begin{figure}
\centering
\includegraphics[width=0.5\textwidth]{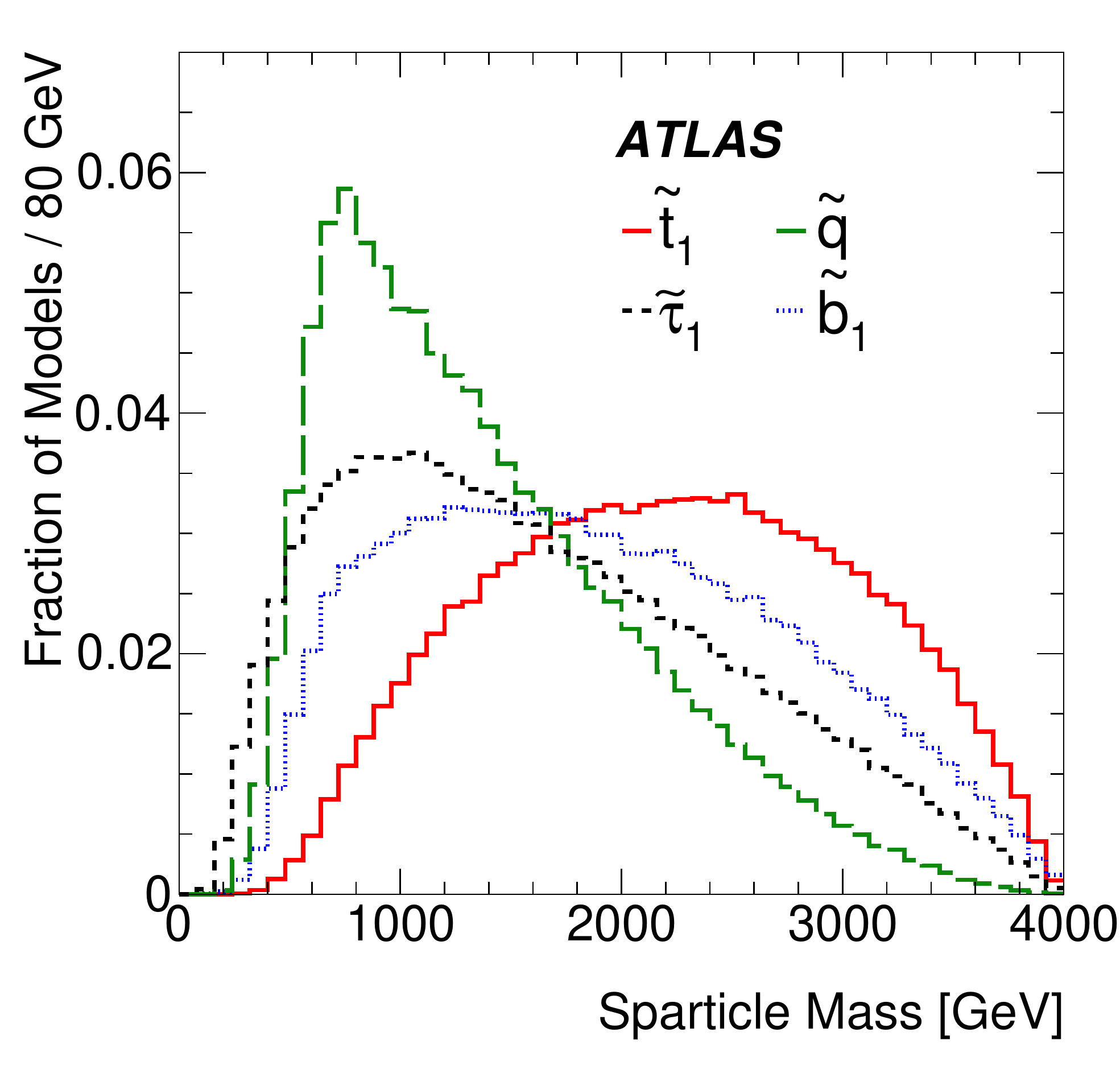}
\caption{Normalised distribution of sparticle masses for the lightest
  top (solid red), bottom (dashed blue), first- or second-generation squark
  (dashed green) and stau (dot-dashed black) for all
  LSP types combined.  
  The constraints listed in Table~\ref{tab:constraints} have
been applied, but not the constraints from the ATLAS searches.
}
\label{fig:basic-t-b-q-chi}
\end{figure}
The distributions of the gluino and LSP masses for the model points satisfying
the constraints from  Table~\ref{tab:constraints} are shown in
Figure~\ref{fig:basic-gl-chi10}, separately for models with a bino-, wino- or Higgsino-like LSP. 
Light gluinos are more common among model points with bino-like LSP. 
Dark matter for model points with bino-like LSP is typically over-produced, 
so the presence of
a gluino state close in mass to the LSP enables them to 
act as coannihilators with the dark matter in the early universe, 
reducing the relic density 
to a level that satisfies the constraint on $\Omega_{\neutralino} h^2$.
The neutralino mass distribution for the bino-like LSP
model points shows a sharp concentration of model points with
$m(\tilde{\chi}_1^0) \lesssim 100$\,\GeV. 
This concentration corresponds to model points in 
which the dark matter relic density constraint can be satisfied due to  
enhanced neutralino pair annihilation via the $Z$ or Higgs boson poles.
A plot of the low-mass region for
bino-LSP model points (shown in the inset to Figure~\ref{fig:basic-gl-chi10}(a)) 
confirms this
interpretation, showing two individual peaks corresponding to the 
separate $Z$ and Higgs poles. 

The identity of the next-to-lightest supersymmetric particle (NLSP)
can be important for the phenomenology of a model point, and is strongly
influenced by the LSP type. The NLSP is
nearly always a chargino or neutralino for wino-like or Higgsino-like
LSPs, as is expected given the small splittings between the
different components of the wino or Higgsino multiplets. In
particular, over 99\% of wino-like LSP model points have a chargino NLSP. 
The small mass difference between the \chargino and the \neutralino
can lead to long \chargino lifetimes for wino-like and Higgsino-like
LSPs, which result in the types of disappearing tracks
searched for in the analysis described in Ref.~\cite{Aad:2013yna}.

Bino-like LSP model points, by contrast, exhibit a much wider range of
NLSP types, the distribution of which is determined by their effect on
the LSP annihilation rate. Even for models with a bino-like LSP,
charginos and neutralinos are still the
most common NLSP type, since LSP--Higgsino mixing is important for many
of the possible annihilation mechanisms.
The remaining NLSP types are generally ordered by
their effectiveness as coannihilators, with coloured NLSPs being the
most prevalent and neutral NLSPs (sneutrinos) being the least so. 
The NLSP--LSP mass splitting for bino-like LSP model points is frequently
less than 50\,\GeV, supporting the assertion that light 
sparticles beyond the LSP are typically required to avoid
over-production of bino-like LSPs. 

In cases where a squark, gluino or slepton is almost mass degenerate
with the LSP, the small available phase-space in the decay can lead to
sparticles that are long-lived or even stable on the time scale for
traversal of the ATLAS detector. In total \numLonglived{} model points 
contain squarks, gluinos or sleptons with $c\tau>1\,\mathrm{mm}$. The sensitivity of
the SUSY searches targeting prompt decays (the first three categories in Table~\ref{tab:susySearches},
other than the \DisappearingTrack{} analysis) 
to these model points is reduced and therefore only the
long-lived particle searches were considered for these model points.

The distributions of the mass of several other sparticles of phenomenological interest
at the LHC can be found in Figure~\ref{fig:basic-t-b-q-chi}.
Even before ATLAS analyses are considered, 
the model set is depleted in light top squarks, and to a slightly lesser extent 
in light bottom squarks, by the requirement (in Table~\ref{tab:constraints})
that the lightest Higgs boson mass be close to the experimentally observed value.

\section{Signal simulation and evaluation of searches}
\label{sec:simulation}
For each of the \numTotal model points passing the preselection
described in Table~\ref{tab:constraints}
it has to be determined which, if any, of the ATLAS 
searches are sensitive to it and whether it can be excluded or
not. Simulating and running the full set of ATLAS analyses on
these would be extremely time and resource consuming.
Each model point is therefore evaluated in three steps as described
below. A special procedure is applied to evaluate the searches for long-lived
particles and heavy Higgs bosons as described in the following sub-sections.

\subsection{Supersymmetry signals}

First, each model point is categorised based on its production cross-sections 
for SUSY particles, as 
calculated using \program{Prospino 2.1}~\cite{Beenakker:1996ch,Beenakker:1997ut,Beenakker:1999xh,Spira:2002rd,Plehn:2004rp}.
The production processes are split into three separate groups:
strong production, electroweak production (encompassing electroweakino and
slepton pair production) and finally mixed production  (e.g.~of an
electroweakino in association with a squark or gluino). Model points with cross
sections for any of those processes larger than  the minimum values in
Table~\ref{tab:minCrossSections} % and an LSP mass less than 1\,\tev{}
are subsequently retained, and any such processes are investigated in more detail as described below. 
For strong and mixed production, the minimum cross-section corresponds to just five signal
events produced in the full $\sqrt{s}=\SI{8}{\TeV}$ dataset. 
Sensitivity to such small cross-sections occurs only for model points
with a very high fraction of events with four leptons in the final state,
for which the \FourLepton{} analysis has a high acceptance (up to 50\%).
Production of electroweakinos is most effectively observed by using decays to leptons,
which are often suppressed by the leptonic branching fractions of $W$, $Z$ and $h$ bosons, 
explaining the higher cross-sections limits. 
ATLAS searches have a greater sensitivity to low cross-section slepton pair
production than to electroweakino production. Therefore, if the model point does
not satisfy the higher cross-section criterion of the electroweak production group, a fourth group
that allows for model points with lower cross-section slepton pair production
is considered. Fewer than 10\% of the model points have no process passing this
selection and are thus not considered to be excludable.

\begin{table}
\centering
\renewcommand\arraystretch{1.3}
\begin{tabular}{l|c|c|c|c}
\hline
                         & \textbf{Minimum cross} & \multicolumn{3}{c}{\textbf{Fraction of models generated}}\\
\cline{3-5}
\textbf{Production mode} & \textbf{section [fb]}  & \textbf{Bino LSP} & \textbf{Wino LSP} & \textbf{Higgsino LSP}\\ \hline 
Strong                      &       0.25 &      82.5\% &      74.9\% &      76.7\% \\
Mixed                       &       0.25 &      52.6\% &      42.1\% &      13.9\% \\
Electroweak                &        7.5 &      38.3\% &      72.5\% &      75.0\% \\
Slepton pair                &       0.75 &       9.6\% &       7.9\% &       9.5\% \\ \hline

\end{tabular}
\caption{Minimum cross-sections required to do particle-level event generation for the four different production modes
and the fraction of the models above this cross-section for each LSP type.}
\label{tab:minCrossSections}
\end{table}

For each of the model points satisfying one or more of the production mode
cross-section criteria, a large sample of events is generated 
using \program{MadGraph5~1.5.12}~\cite{Alwall:2011uj} with the \program{CTEQ~6L1} 
parton distribution functions (PDF)~\cite{Pumplin:2002vw} and \program{Pythia~6.427}~\cite{Sjostrand:2006za}
with the \program{AUET2B} set of parameters~\cite{ATL-PHYS-PUB-2011-009}. 
\program{MadGraph5} is used to generate the initial pair of
sparticles and up to one additional parton in the matrix element, while
\program{Pythia} is used for all sparticle decays and parton showering\footnote{Polarisation from the decay of the initial sparticles is not taken into account in this analysis.}
MLM matching \cite{Mangano:2006rw} is used with up to one additional jet in the \program{MadGraph} matrix element, a \program{MadGraph} kT measure of 100\,\GeV, and a \program{Pythia} jet measure cut off of 120\,\GeV. 
Both \program{Tauola~1.20}~\cite{Jadach:1993hs} and \program{Photos~2.15}~\cite{Golonka:2005pn} are
enabled to handle the decays of $\tau$ leptons and final-state radiation of photons, respectively.

\newcommand\nmax{\ensuremath{N_{\rm max}}^{95}\xspace}
\newcommand\nsig{\ensuremath{N_{\rm sig}}\xspace}

To reduce the amount of computationally expensive detector simulation
and reconstruction that is required, a Monte Carlo particle-level
selection corresponding to each of the SUSY searches in the first three categories in
Table~\ref{tab:susySearches} is used to process the generated
events. In this step, inefficiencies from the detector-level
reconstruction are parameterised using a single
efficiency factor for each signal region, determined from previously simulated signal samples.
Exceptions to this are made for the $\tau$ reconstruction efficiency
in the \TauStrong{} and \TwoTau{} searches,
for which \pt-dependent efficiencies are applied for each signal $\tau$. 
Similarly, the \DisappearingTrack{} search
applies the reconstruction efficiency for decaying charginos as a function of the 
distance from the centre of the ATLAS detector and the angular coordinates as published in Ref.~\cite{Aad:2013yna}.
The expected event yield in each signal region is calculated for each model
 point.
For most analyses the categorisation is performed by directly comparing the 
expected signal yield \nsig to the model-independent
95\% confidence level (CL) upper bound on the number of beyond-the-SM events $\nmax$
in each signal region of that analysis.
Model points are then partitioned into three categories, on the basis of that 
particle-level simulation,
using criteria determined to be appropriate for each individual analysis.
The first category comprised those already excluded at this stage 
on the basis that $\nsig$ is sufficiently larger than $\nmax$ 
for at least one signal region of one analysis.
The expected sensitivity of all other analyses 
to such model points is calculated using particle-level yields,
and using average reconstruction efficiencies.
The second category corresponds to those found not to be excludable,
consisting of points with $\nsig$ materially smaller than 
$\nmax$ for all analyses. 
The exact relationship between $\nsig{}$ and $\nmax{}$ 
for the categorisation is determined separately for each signal region and 
depends on the accuracy with which the particle-level evaluation
reproduces the results of a full simulation.
In total 35.9\% (44.7\%) of the model points fall in the first (second) category and are 
deemed to have been excluded 
(not excluded) at the 95\% CL.
The validity of this classification
was confirmed using the full simulation and reconstruction procedure
described below for approximately 5\% of the model points in the first category for each analysis.
A final category of model points -- those with $\nsig$ close to $\nmax$
(typically within a factor of a few) for the most sensitive analyses 
-- are subject to more detailed investigation, as follows.

For the \numFullSim{} model points for which the overall exclusion is uncertain
based on the particle-level simulation described above, 
the final step is a fast, \program{GEANT4}-based \cite{Agostinelli:2002hh}
simulation using a parameterisation of the performance of the ATLAS 
electromagnetic and hadronic calorimeters \cite{SOFT-2010-01}
and full event reconstruction.
The simulation includes a realistic description of multiple $pp$ interactions per bunch crossing,
and is corrected 
for identification efficiencies and resolution effects.
For each such model point, signal events are generated 
corresponding to four times the integrated luminosity recorded 
(i.e. \ensuremath{{\SI{81.2}{fb^{-1}}}}).
The simulation is limited to those production modes which could 
contribute to the analyses of interest.
For these processes, the nominal cross-section and the uncertainty are
taken from an envelope of cross-section predictions using different
parton distributions and factorisation and renormalisation scales, as described in
Ref.~\cite{Kramer:2012bx}.
The addition of the resummation of
soft gluon emission at next-to-leading-logarithm accuracy
(NLL)~\cite{Beenakker:1996ch,Kulesza:2008jb,Kulesza:2009kq,Beenakker:2009ha,Beenakker:2011fu}
is performed in the case of strong production of sparticle pairs.

The status of each previously inconclusive model point is then
determined for each of the analyses using the same
procedure \cite{Baak:2014wma} as used in the original analyses. In each analysis
the signal region with the best expected sensitivity 
is identified and the ``CLs method''~\cite{CLS} is used to determine
if the model point is excluded or not at 95\%~CL.  
It should be noted that for the exclusion
fits, the nominal signal cross-sections are used, without any
theoretical uncertainties on the signal, except for the \StopMonojet{}
and \Monojet{} analyses. These two analyses are
particularly sensitive to the modelling of ISR as they rely on
a high-\pt ISR jet in their event selection. Therefore an additional
25\% ISR signal uncertainty is applied in those cases,
based on the observed variance in acceptance in signal samples with modified parameters
for the ISR modelling \cite{Aad:2014stoptocharm}.
For the \Multijet{}, \ThreeBjets{} and \TwoLeptonStop{} analyses, it is not possible
to apply the full combined fit procedure of the original
analyses. Instead only the individual signal regions are considered,
resulting in somewhat conservative modelling of the sensitivity for those three analyses.
For the overall exclusion, no attempt is made to combine the
individual analyses.  Instead the analysis with the best expected
exclusion is used for each model point to determine its status.

\subsection{Long-lived particle search}
\label{sec:sim:longlived}

Model points with heavy long-lived particles require special treatment 
since such particles can traverse part or all of the ATLAS
detector leaving rather distinct signatures. The dominant types of
long-lived particles in the model points
are the $\tilde{\chi}_1^+$ and the $\tilde{\chi}_2^0$
when they are almost mass-degenerate with the LSP. 
The decay of such long-lived particles is included in the simulation procedure described
above and model points with such long-lived particles are considered using the same procedure.

Aside from the electroweakinos, \numLonglived{} of the model points 
contain squarks, gluinos or sleptons with $c\tau>1\,\mathrm{mm}$.
These model points have not been simulated.
Instead only the results from the \LLSparticles{} searches
are used to constrain these model points.

The long-lived particle searches in Refs.~\cite{Aad:2012pra} and~\cite{ATLAS:2014fka}
provide limits on the production cross-section at 7 and 8\,\tev{},
respectively, of bottom squarks, top squarks, gluinos, staus and charginos in the
case where these live long enough to traverse the complete
detector. 
Model points with 
bottom squarks, top squarks, gluinos, staus or charginos with a lifetime above
85\,ns and production cross-sections exceeding the corresponding cross
section limit from either the 7 or 8\,\tev{} result are considered to be
excluded. In all other cases --  where the lifetime is shorter 
or the production cross-sections lower --
the model point is considered not to be excluded.

\subsection{Heavy Higgs boson search}

Cross sections and branching ratios for heavy Higgs bosons
are calculated for
gluon fusion, or for production in association with 
$b$-quark(s)~\cite{Dittmaier:2011ti,Dittmaier:2012vm,Heinemeyer:2013tqa}.
The high-$m_A$ category 
($m_A > 200$~\GeV) of the ATLAS search~\cite{Aad:2014vgg} is used since 
this regime is relevant to all the model points in this study. 
It is assumed that $b$-quark associated production dominates and this calculation
is performed for each
model point in the pMSSM parameter space using the software
\program{SusHi~1.3.0}~\cite{Harlander:2012pb,Harlander:2002wh,Harlander:2003ai,Aglietti:2004nj,Bonciani:2010ms,Degrassi:2010eu,Degrassi:2011vq,Degrassi:2012vt,Heinemeyer:1998yj,Heinemeyer:1998np,Degrassi:2002fi,Frank:2006yh,Harlander:2005rq}.
The large value of $m_A$ in all the model points has the effect that 
the $A$ and $H$ bosons are nearly mass degenerate, so both must be simulated.
The quantity $\sigma(bbH) \times {\rm BR}(H\rightarrow\tau\tau) + \sigma(bbA) \times {\rm BR}(A\rightarrow\tau\tau)$ is calculated
and is compared to the ATLAS 95\% CL upper limits~\cite{Aad:2014vgg}
for a scalar particle produced in association with
$b$-quark(s) and decaying to $\tau\tau$. For the overall exclusion, this heavy Higgs boson search is considered only if none of the SUSY searches are expected to exclude the model point at the 95\%~CL.

\FloatBarrier
\section{ATLAS constraints from LHC \runone}
\label{sec:results}

\subsection{Impact of ATLAS searches on sparticle masses}\label{sec:results:masses}

The effect of the ATLAS search constraints are most easily presented
as projections onto one-dimensional or two-dimensional
subspaces of the full 19~parameter pMSSM space.\footnote{A full list of model parameters, observables and which analyses, if any, are excluding each model, are available from the 
ATLAS Collaboration website~\cite{pMSSM_plots}.}
The most relevant parameters onto which to project are typically the sparticle masses.
Production cross-sections for sparticles decrease rapidly 
when their masses are increased.
When those initial sparticles decay, 
the masses of other sparticles affect the 
types of visible decay products and their kinematics.
The mass of the LSP is particularly important 
since a decay to a high-mass LSP 
results in less energy being available for the observable decay products.

The fraction of surviving model points in the projected space
necessarily depends both on the prior distribution of model points in the 
parameters that have been projected out, and on experimental constraints on 
sparticle masses other than those plotted.
Thus, some care is needed in their interpretation.
In particular the fractions of model points 
excluded can depend, in some cases sensitively, 
on the non-collider constraints shown in Table~\ref{tab:constraints},
the choice of scan ranges shown in Table~\ref{tab:scanranges}, and on
the choice of a flat prior.
Nevertheless, some general features of the impact of the ATLAS \runone{} searches are clear.

The simplified-model limits shown on the plots throughout this section 
are the observed limits from the indicated analysis. In many cases there 
are several analyses interpreting their results in the same simplified models, 
and in this paper the observed limits from the most constraining analysis are shown.
It should be noted that there is no minimum number of model points required in 
each bin.

The results are shown first for squark masses (of the first two generations)
and the gluino mass
in Section~\ref{sec:squarkgluino},
then for third-generation squark masses in Section~\ref{sec:thirdgen}, 
and for the electroweak sparticles in Section~\ref{sec:electroweak}.
A small subset of the model points contain long-lived squarks, gluinos or sleptons.
These \numLonglived{} model points are treated separately in Section~\ref{sec:results:longlived} as only
the long-lived particle search is considered for these model points.
Section~\ref{sec:results:htautau} describes the effect of a search for
the decay of heavy neutral Higgs bosons to two $\tau$ leptons. %or $H\rightarrow\tau\tau$.
The complementarity between the different ATLAS searches
is described in Section~\ref{sec:combination}.

\subsubsection{Squarks and gluinos}
\label{sec:squarkgluino}

\begin{figure}
\centering
\begin{tabular}{cc}
\subfloat[Gluino / LSP]{\includegraphics[width=0.49\textwidth,trim=0 0 0 0,clip]{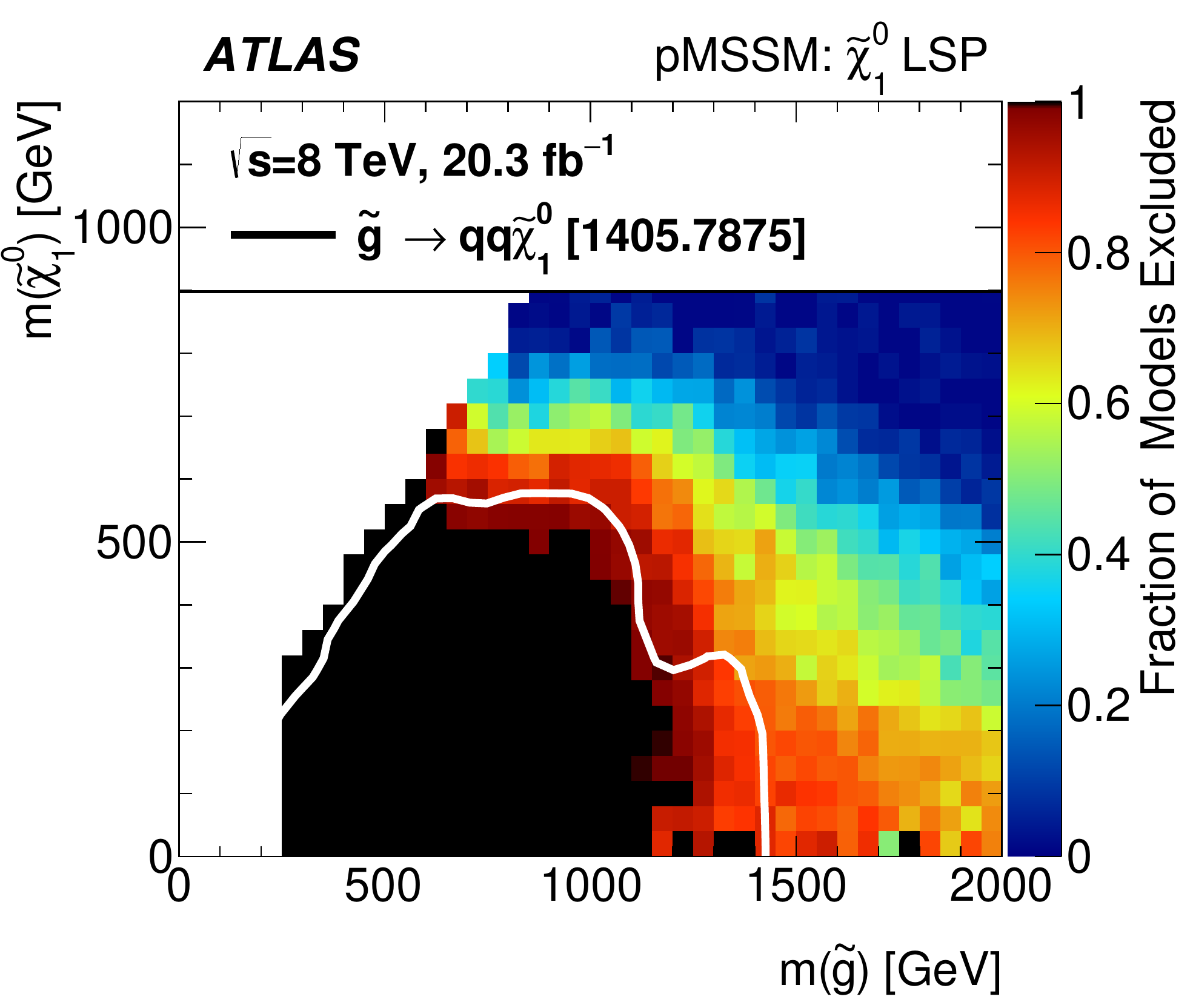}} &
\subfloat[Lightest (1$^{\rm st}$/2$^{\rm nd}$ gen) squark / LSP]{\includegraphics[width=0.49\textwidth,trim=0 0 0 0, clip]{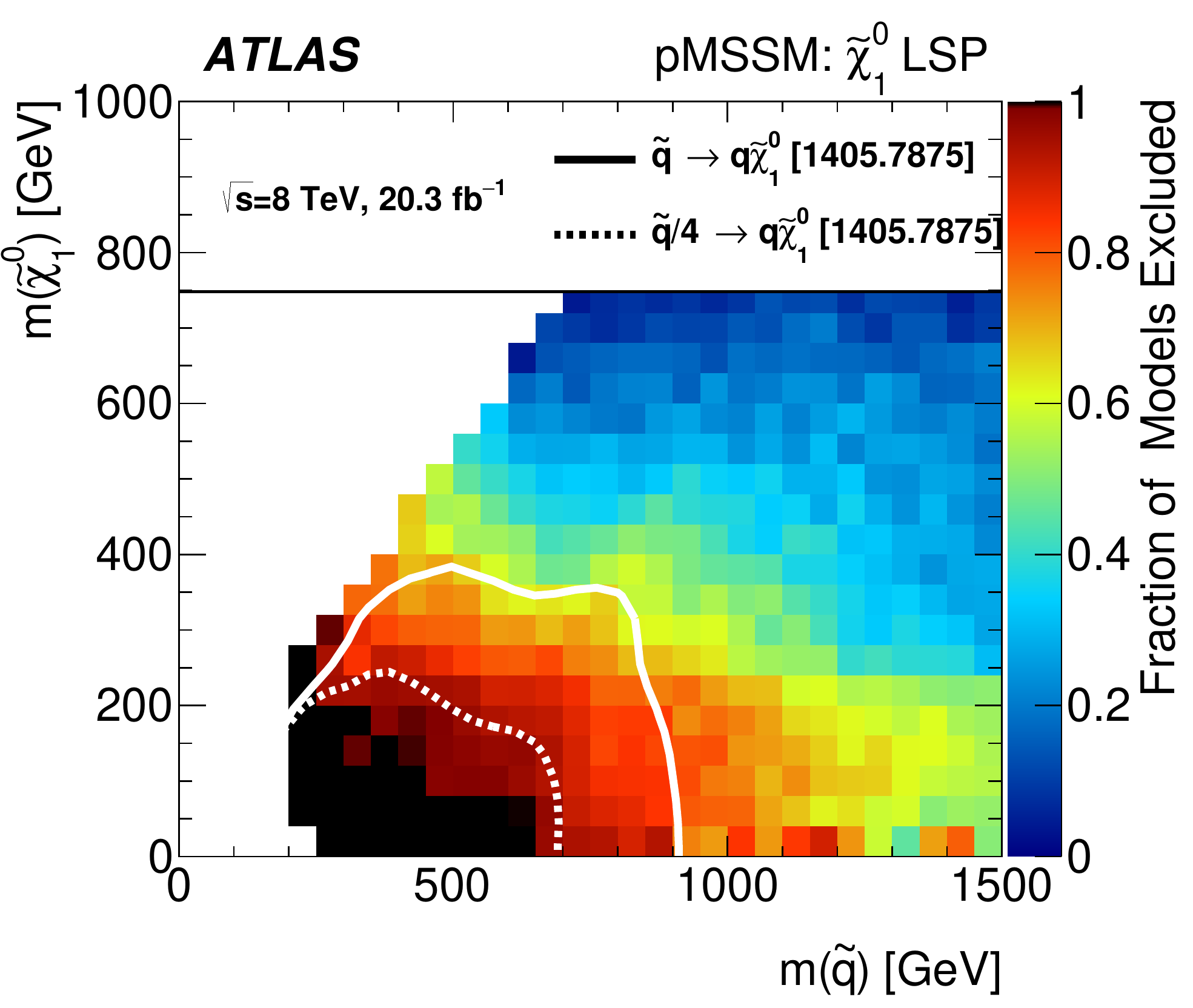}}
\end{tabular}
\caption{\label{fig:squark-lsp-gluino-lsp-2d-exclusions}
Fraction of pMSSM points excluded by the combination of 8\,\TeV{} ATLAS
searches in the (a) \gluino--\neutralino and (b) the
$\tilde{q}$--\neutralino mass planes. The colour scale indicates the
fraction of pMSSM points excluded in each mass bin, with black squares indicating
100\% of model points being excluded. 
The white regions indicate places where no model points were sampled which 
satisfied the constraints of Table~\ref{tab:constraints}.
In both cases, the solid white lines overlaid
are observed simplified-model limits
from the \ZeroLepton{} search \cite{ATLAS-SUSY-2013-02} at 95\%~CL. 
In the \gluino--\neutralino case, the simplified-model limit is set assuming 
direct production of gluino pairs and that the squarks are decoupled, 
with gluino decaying to quarks and a neutralino,
$\tilde g \rightarrow q + q + \neutralino$.
In the \squark--\neutralino plane, both lines are drawn 
assuming directly produced 
first/second-generation squark pairs, with each squark decaying to a
quark and a neutralino, $\squark \rightarrow q + \neutralino$. 
The solid line corresponds to the case where all eight squarks 
from the first two generations are assumed to be degenerate.
The dashed line has the squark production cross-section 
scaled down by a factor of four 
to emulate the effect of only two of those eight squarks 
being kinematically accessible.
}
\end{figure}

\newcommand\colourscale{{The colour scale is as described in Figure~\ref{fig:squark-lsp-gluino-lsp-2d-exclusions}.}}

Figure~\ref{fig:squark-lsp-gluino-lsp-2d-exclusions}(a) shows the fraction of model points
excluded by the ATLAS searches as projected onto the two-dimensional
space of the masses of the LSP and the gluino. 
As one would expect, light gluinos are robustly constrained by the ATLAS searches,
whereas at larger gluino masses the fraction of model points excluded
is reduced. 

It is instructive to compare these observed pMSSM exclusions to the observed limits previously presented for
simplified low-scale models.
Superimposed on Figure~\ref{fig:squark-lsp-gluino-lsp-2d-exclusions}(a)
is a line showing the 95\%~CL exclusion previously derived by the \ZeroLepton{} analysis 
from a simplified model in which only the gluino and the LSP are kinematically accessible~\cite{ATLAS-SUSY-2013-02}. 
It can be seen that there is generally good congruence between the region excluded
in the two different scenarios, demonstrating that the simplified model 
is successfully capturing the main pMSSM phenomenology in this case.
Nevertheless, the pMSSM sensitivity does differ in detail from that of the
simplified model. Not only is this because multiple analyses are considered
here but also because of residual dependence on the masses of other sparticles. 
For example, the pMSSM permits the existence of
additional particles with masses lying in between those of the gluino and the LSP,
which can lead to cascade decays. In general such cascade decays 
can be expected to yield different jet  $\pt$ and \met{} spectra,
and perhaps additional leptons, which have different experimental acceptances. 

Close to the diagonal line where $m({\gluino})$ is only a little larger than 
$m(\neutralino)$ (Figure~\ref{fig:squark-lsp-gluino-lsp-2d-exclusions}(a))
the simplified-model exclusion from Ref.~\cite{ATLAS-SUSY-2013-02}  
underestimates the ATLAS sensitivity. The $\gluino \to q \bar{q} \neutralino$ decays 
in this near-degenerate region produce low-energy
quarks which typically fail to meet the kinematic requirements
on jets.
That ATLAS does indeed show good sensitivity in this region 
is instead due to the monojet analyses~\cite{Aad:2014stoptocharm,Aad:2015zva}.
These analyses were designed to capture the recoil of LSPs 
(or other, slightly heavier, SUSY particles) against
initial-state QCD radiation.

\begin{figure}
\centering
\begin{tabular}{cc}
\subfloat[All LSP types]{\includegraphics[width=0.49\textwidth,trim=0 0 0 0, clip]{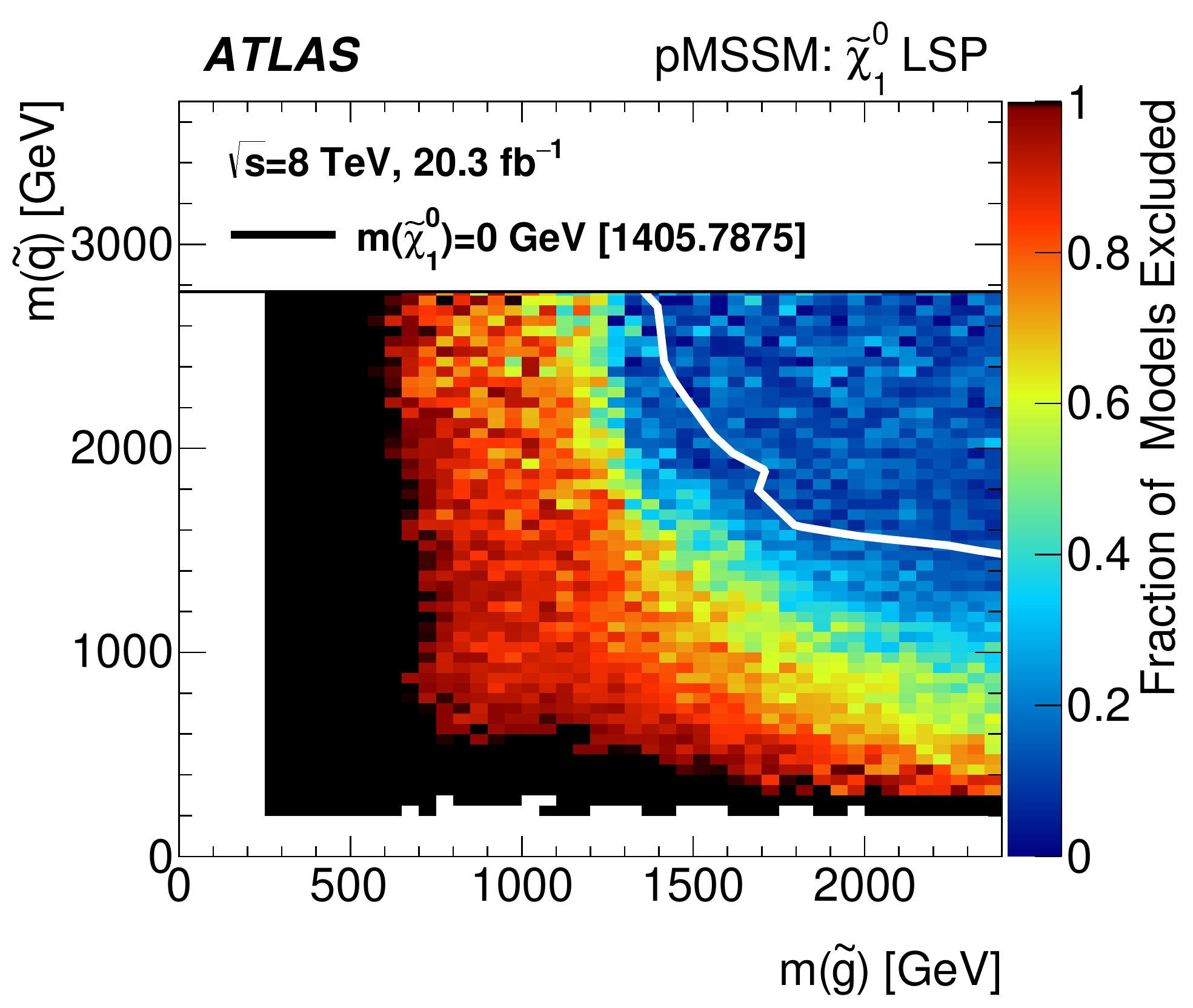}}&
\subfloat[Bino-like LSPs]{\includegraphics[width=0.49\textwidth,trim=0 0 0 0, clip]{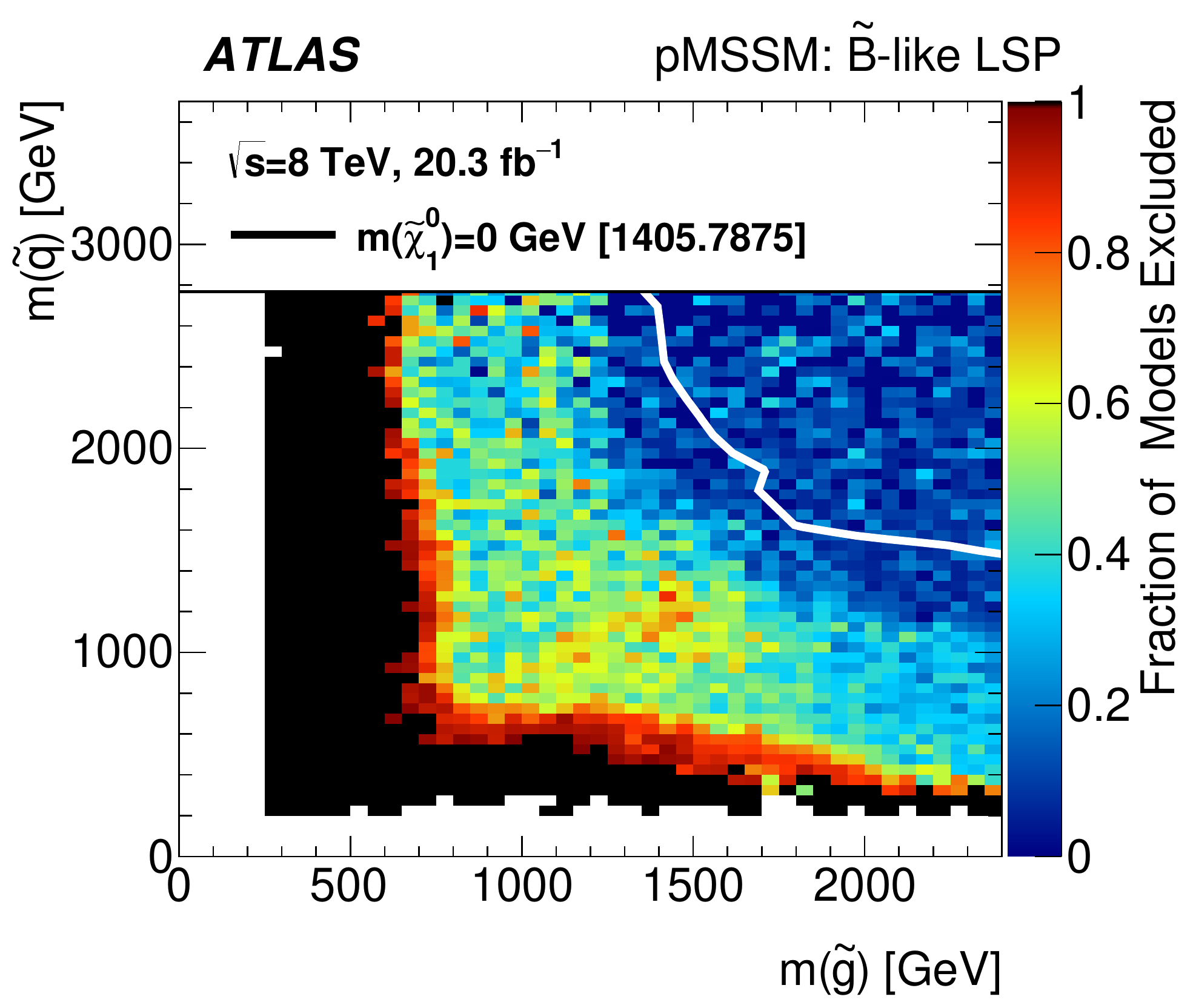}}\\
\end{tabular}
\caption{\label{fig:gluino-squark}
Fraction of pMSSM points excluded in the \gluino--\squark\ plane,
where $\squark$ represents the lightest squark from the first two generations. 
The overlaid line shows a limit for a simplified model from the \ZeroLepton{} search~\cite{ATLAS-SUSY-2013-02} 
which assumes strong production
of gluinos and eight-fold degenerate first- and second-generation squarks, with direct
decays to quarks and massless neutralinos. \colourscale}
\end{figure}

Figure~\ref{fig:squark-lsp-gluino-lsp-2d-exclusions}(b) shows a different projection,
in this case to the mass of the LSP versus the mass of the lightest
squark of the first two generations, $\tilde{q}_{\rm L,R}$ for $q \in \{u,\,d,\,s,\,c\}$, labelled here and in what follows as $\squark$. It can be observed that there is good sensitivity at low
squark mass and no models with a squark mass below 250~\GeV{}  
are allowed by the ATLAS analyses.
The solid line superimposed on
Figure~\ref{fig:squark-lsp-gluino-lsp-2d-exclusions}(b) 
shows the 95\%~CL exclusion obtained previously~\cite{ATLAS-SUSY-2013-02} 
for a simplified model in which the only kinematically accessible sparticles are the LSP
and the eight squark states of the first two generations,
where these squarks are all assumed to have the same mass.
It can be seen that the region within the solid simplified-model exclusion curve 
is only partially excluded within the pMSSM.
This is primarily because the pMSSM-19 parameter space
does not demand that the squarks be eight-fold degenerate,
reducing the cross-section.
There is a closer correspondence between the pMSSM sensitivity and 
that of an alternative simplified 
model (dashed line), in which the cross-section for direct (anti-)squark production
has been reduced by a factor of four, to model the effect of only two of those 
eight squarks being mass degenerate.\footnote{Reference~\cite{ATLAS-SUSY-2013-02} 
 emulates the effect of a single kinematically accessible squark by dividing the cross-section by a factor of eight rather than four.}

A noticeable excursion from the simplified-model lines, visible on
both plots in Figure~\ref{fig:squark-lsp-gluino-lsp-2d-exclusions} is a horizontal 
band of sensitivity to pMSSM points for LSP masses less than about \SI{200}{\GeV} 
stretching up to large gluino (or \squark) masses.
Since such high-mass strongly interacting sparticles have small production cross-sections, 
one would not expect sensitivity to their production.
Indeed these constraints are not the result of gluino or squark searches,
but instead of searches for disappearing tracks from long-lived charginos.
These long-lived chargino states are common for models with wino-like LSPs
with mass splittings between the charged NLSP and the neutral LSP
of less than about \SI{200}{\MeV}.
The NLSP, when it decays inside the detector volume, produces an invisible LSP
and a low-energy charged pion which itself often goes undetected.
The search for such `disappearing' charged-particle tracks is sensitive
even in the absence of direct squark or gluino production,
and hence sensitivity is observed for any mass of the strongly interacting sparticle.

\begin{figure}
\centering
\begin{tabular}{cc}
\subfloat[Left up squark]{\includegraphics[width=0.48\textwidth,trim=0 0 0 0, clip]{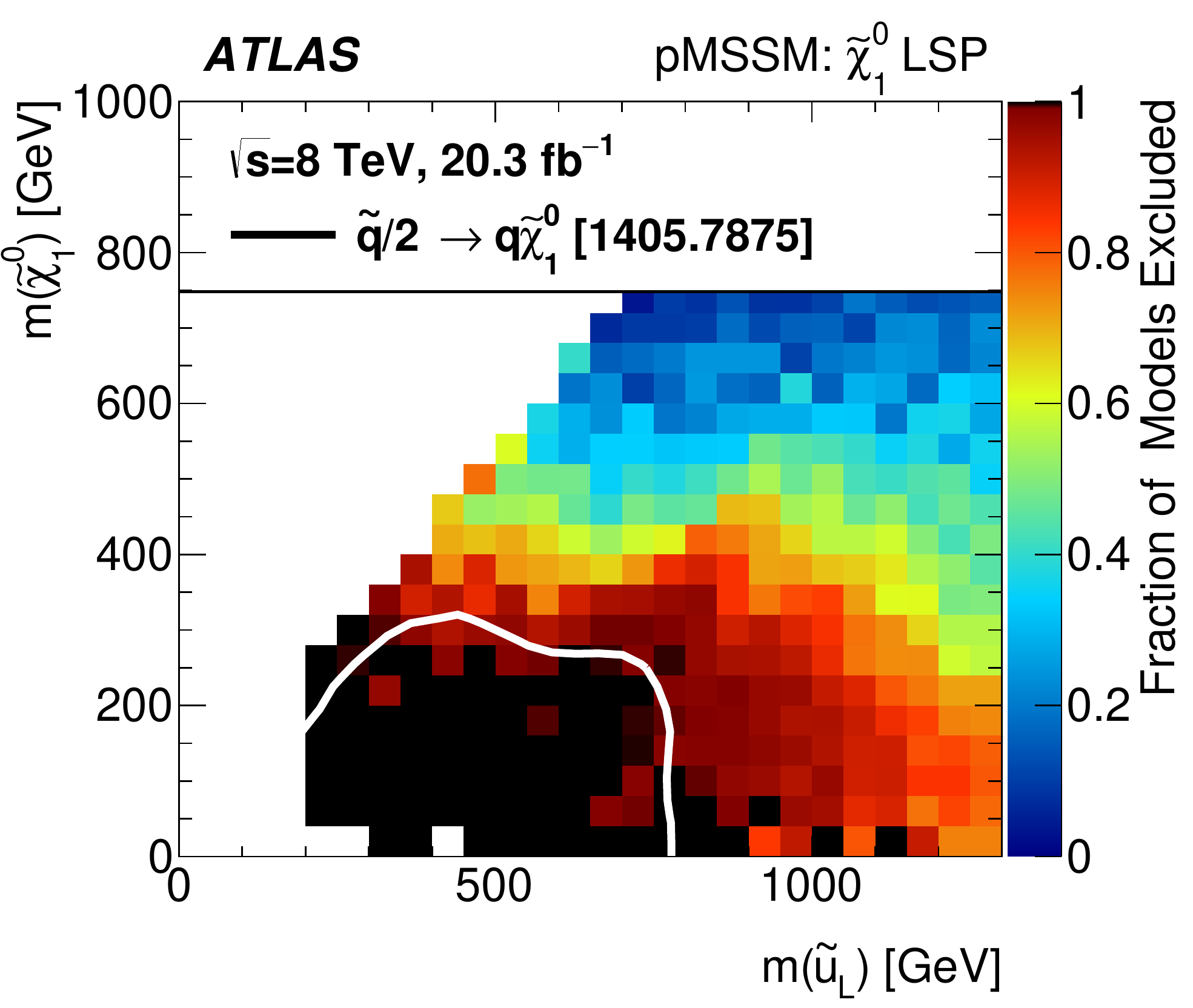}}&
\subfloat[Right up squark]{\includegraphics[width=0.48\textwidth,trim=0 0 0 0, clip]{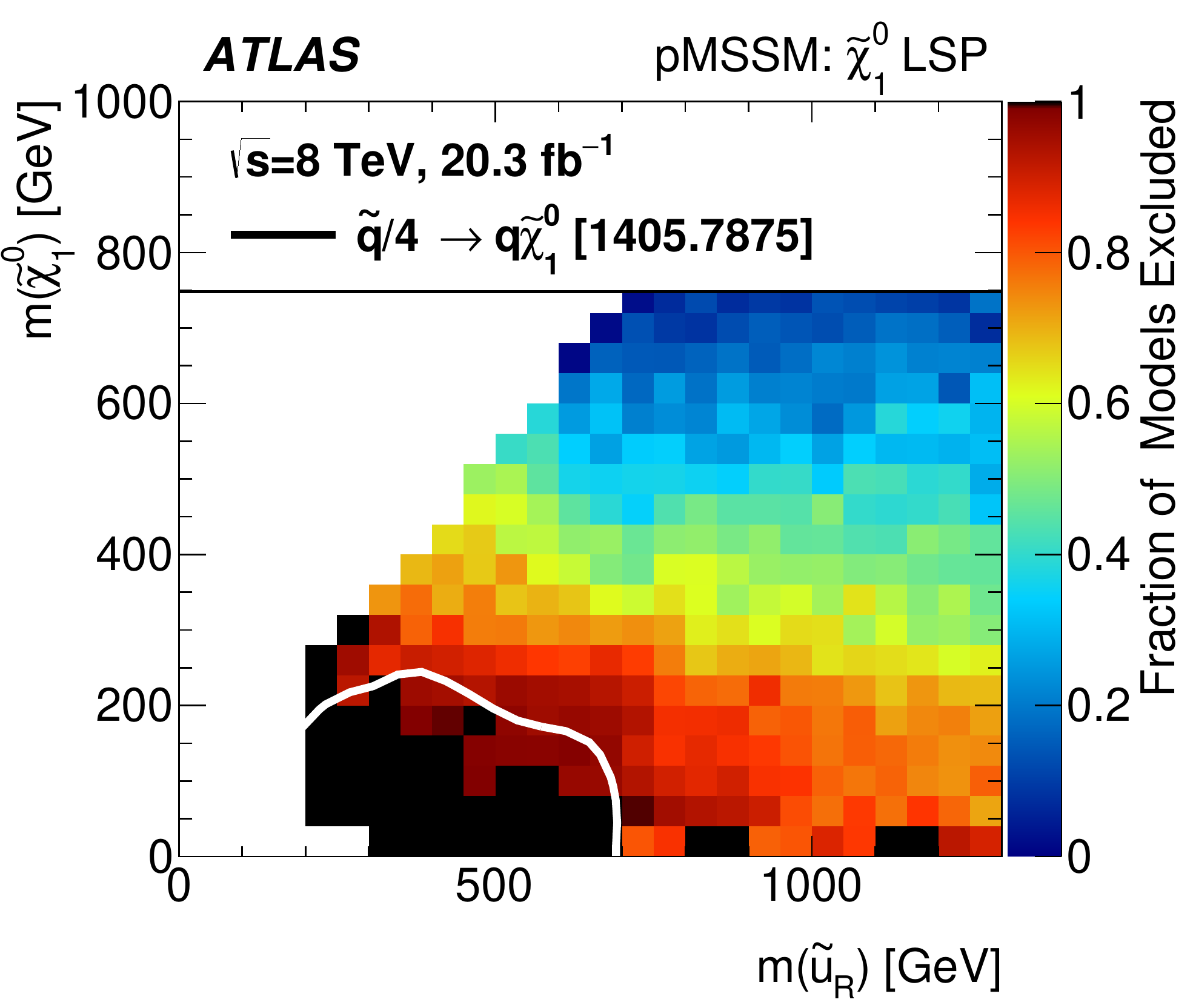}}\\
\subfloat[Left down squark]{\includegraphics[width=0.48\textwidth,trim=0 0 0 0, clip]{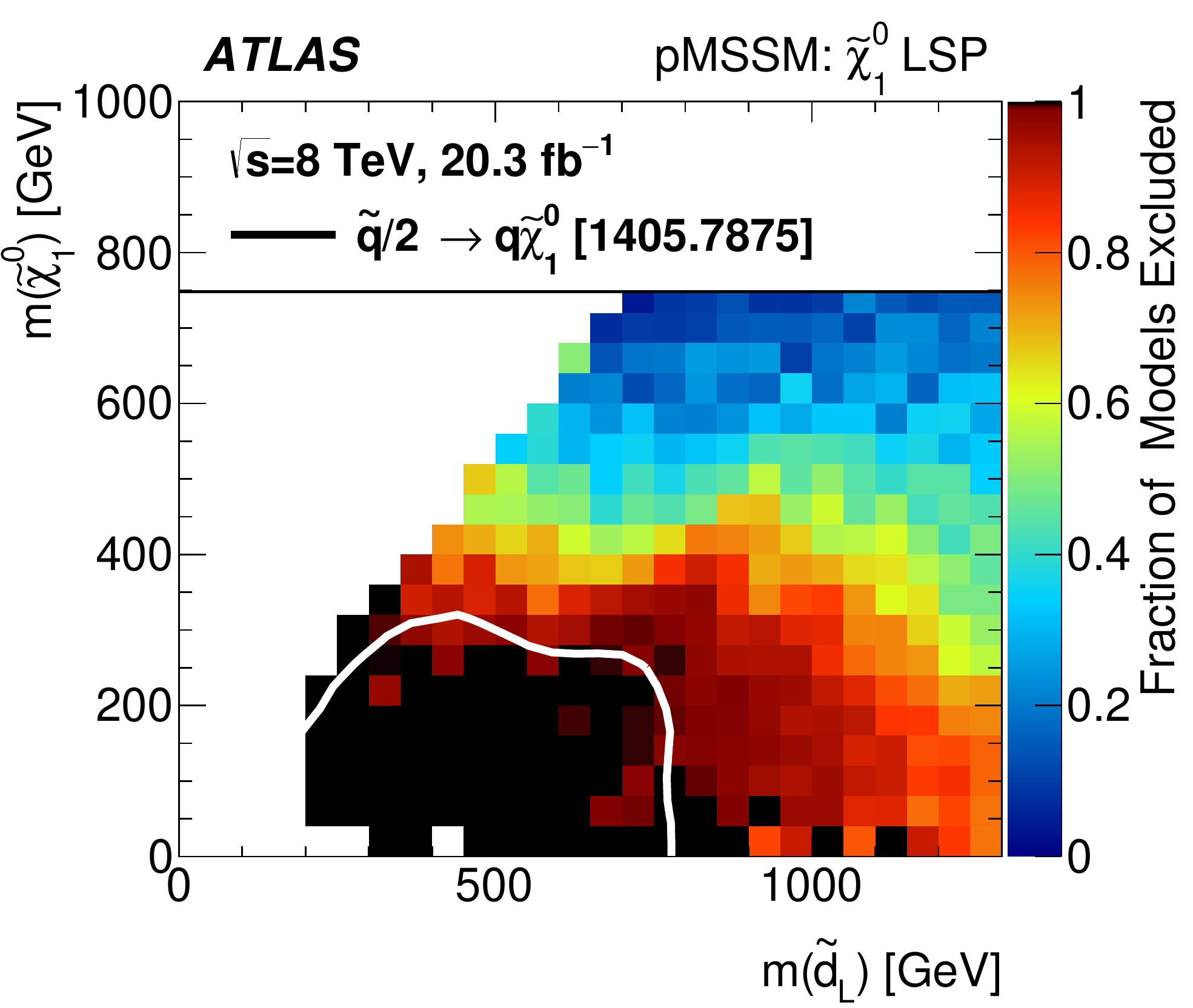}}&
\subfloat[Right down squark]{\includegraphics[width=0.48\textwidth,trim=0 0 0 0, clip]{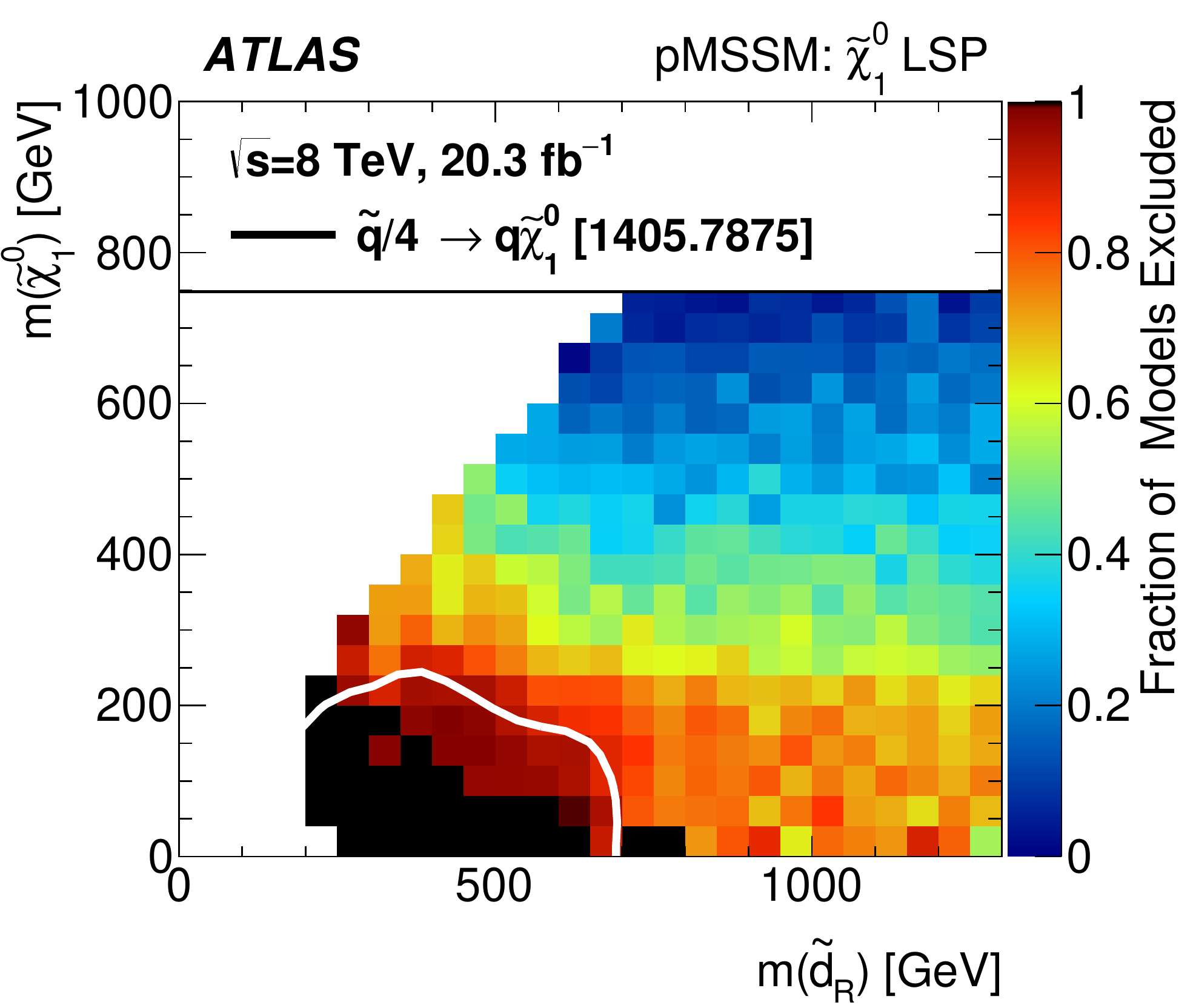}}\\
\end{tabular}
\caption{Fraction of model points excluded in the planes of the masses of the left-handed and right-handed
squarks (of the first two generations) versus the neutralino mass. 
Both simplified-model limit contours are 
taken from the \ZeroLepton{} analysis (Figure~10(c) of Ref.~\cite{ATLAS-SUSY-2013-02});
however, for the left (right) handed squarks the assumption of
four (two) degenerate squarks is emulated by dividing the cross-section
for production of the eight degenerate squark states 
by a factor of two (four). \colourscale}
\label{fig:ex-uRdR-chi10}
\end{figure}

Figure~\ref{fig:gluino-squark}(a) shows the sensitivity as projected onto the 
plane of the gluino and squark masses, where now the LSP mass may take any value.
One observes near-total exclusion by ATLAS analyses of gluinos with masses less than about
\SI{700}{\GeV}, with a high fraction of exclusion up to about \SI{1.2}{\TeV}, for all values of the lightest squark mass. 
Light squarks are also strongly constrained, although those constraints weaken as the gluino mass
increases, due to suppression both of direct squark-pair production via $t$-channel gluino exchange 
and of associated production of $\tilde{q} + \tilde{g}$.

The simplified model superimposed onto the squark--gluino plane 
is one that assumes an eight-fold degeneracy of squark masses in the first two generations
and a massless LSP~\cite{ATLAS-SUSY-2013-02}.
As one would expect, this simplified-model line lies close to the upper edge of the pMSSM
sensitivity, since the pMSSM permits non-degenerate squarks, and a non-zero LSP mass,
both of which reduce sensitivity, 
by reducing the signal cross-section and experimental acceptance respectively.
The reduction in sensitivity caused by a non-zero LSP mass is more pronounced in the case of 
model points with a bino-like LSP, Figure~\ref{fig:gluino-squark}(b).
These model points often have a small mass difference between the squark and the LSP
in order to satisfy the dark matter relic constraint, as discussed earlier in 
Section~\ref{sec:models:properties}.

The pMSSM also allows one to explore how the 
sensitivity to direct squark production depends on the nature of the 
lightest squark. Figure~\ref{fig:ex-uRdR-chi10}
shows that the search reach depends on whether the 
squarks are left- or right-handed, and whether they 
are up-type or down-type squarks.
Thus the assumption of equivalent sensitivity for all such squarks,
common in presentation of LHC SUSY searches, 
is a simplification that is not justified in the more general context of the pMSSM.
The production of down squarks is suppressed relative to up squarks
since there are fewer valence down quarks than up in the proton.
The results for left-handed squarks also differ from those for right-handed squarks.
The $\tilde{u}_{\rm L}$ and $\tilde{d}_{\rm L}$ squarks form a SU(2) doublet and so are degenerate in mass up to
electroweak symmetry breaking effects. This means that if the lightest squark
is a $\tilde{q}_{\rm L}$, there would be another squark with similar mass -- a statement not
usually true for $\tilde{q}_{\rm R}$, which have no similar constraint. For a particular value
of the lightest squark mass, the presence of a pair of left-handed squarks
effectively increases the overall squark production cross-section, leading to
larger apparent sensitivity.
The improved sensitivity to left-handed squarks is further enhanced by the dominant squark decay
modes. Right-handed squarks, which lack weak couplings and have small 
Yukawa couplings, have suppressed decays to the wino-like or Higgsino-like LSPs 
which dominate the model sample.
Instead right-handed squarks generally cascade decay via other electroweakino states
resulting in events with smaller \met,  a greater number of lower-\pt{} jets,
and generally a smaller experimental sensitivity. 

\subsubsection{Third-generation squarks}
\label{sec:thirdgen}

\begin{figure}
\centering
\begin{tabular}{cc}
\subfloat[All pMSSM points, all searches]{\includegraphics[width=0.49\textwidth]{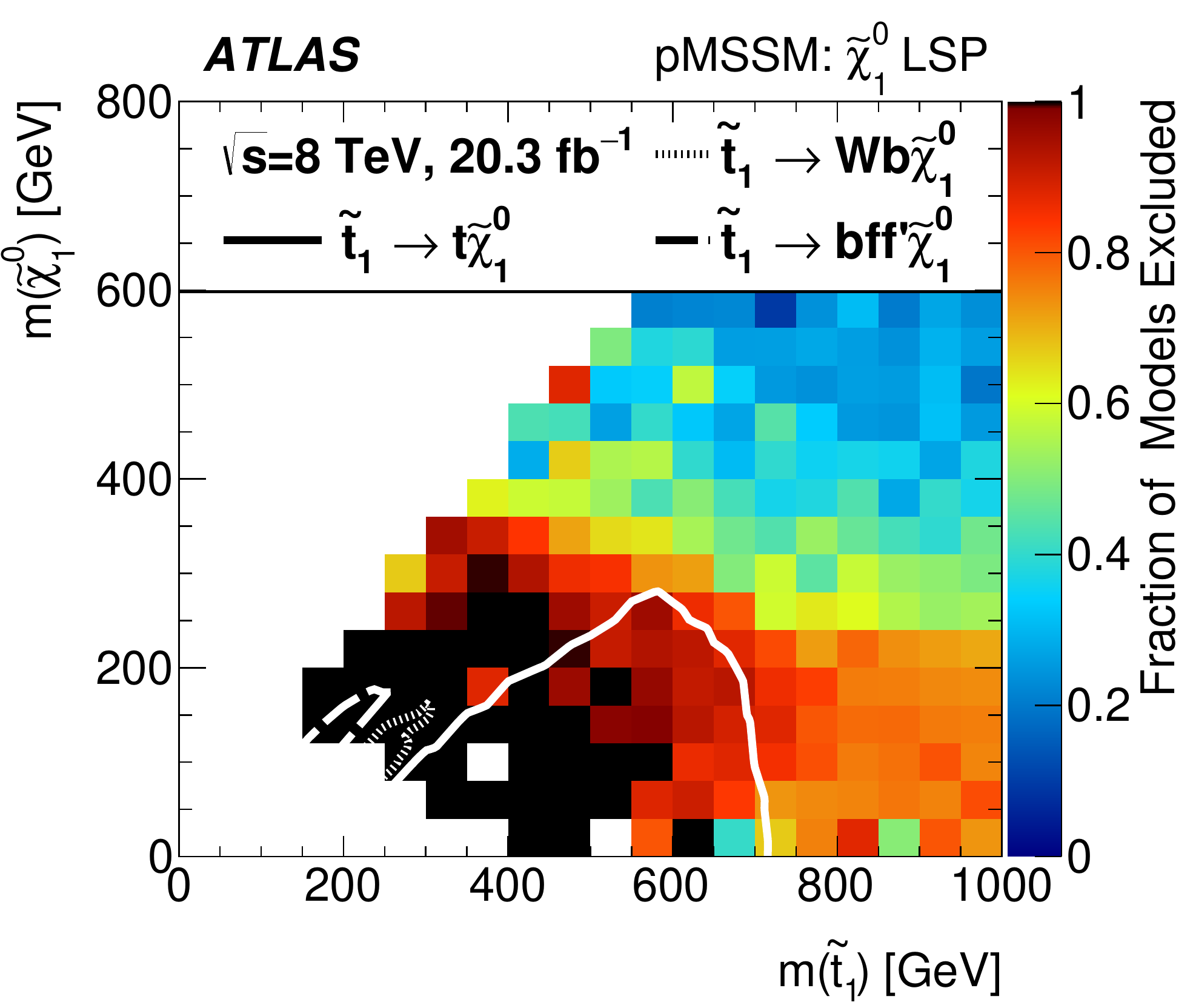}}&
\subfloat[All pMSSM points, $3{\mathrm{rd}}$ gen.~searches]{\includegraphics[width=0.49\textwidth]{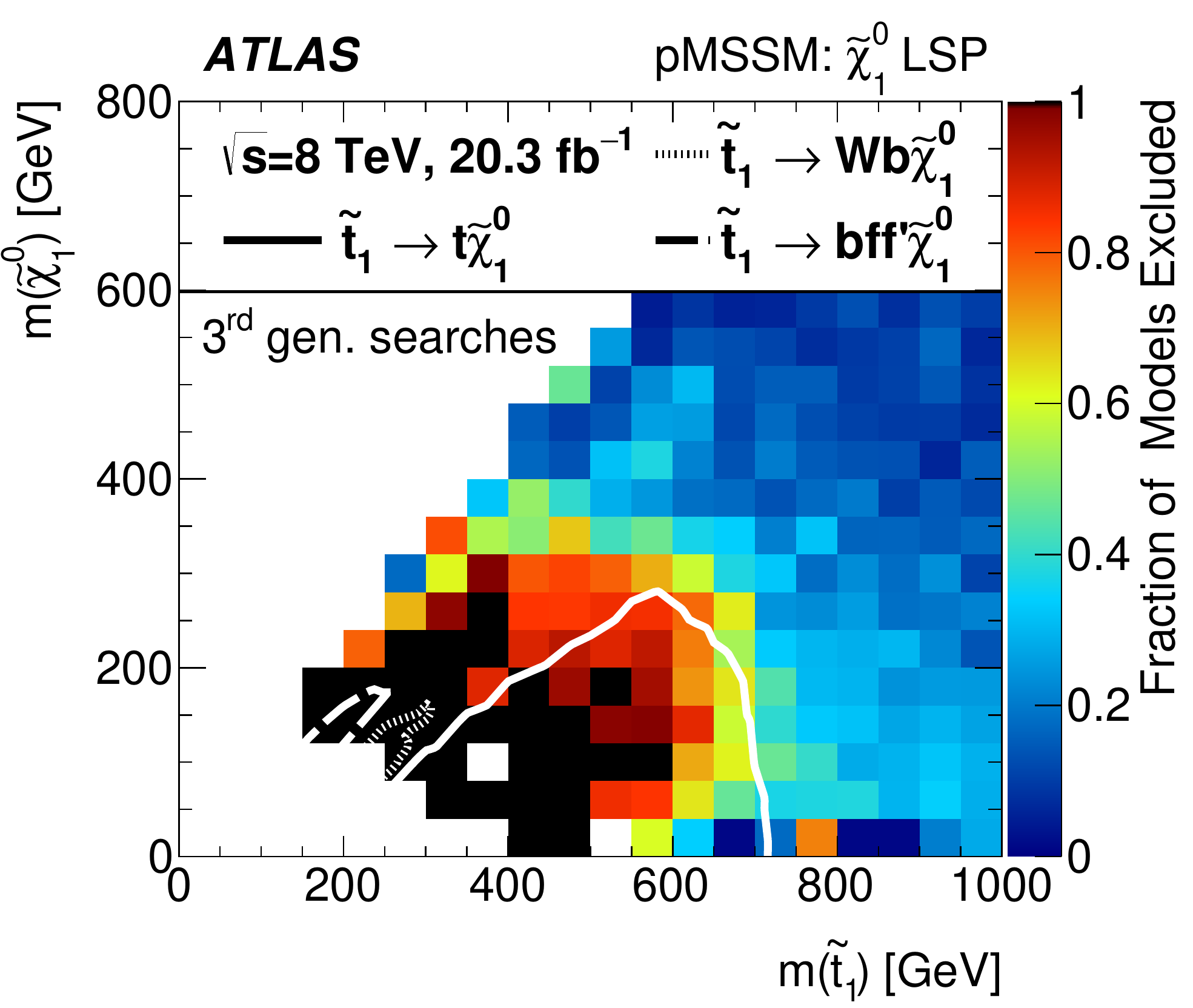}}\\
\subfloat[Left handed $\tilde{t}_1$, all searches]{\includegraphics[width=0.49\textwidth]{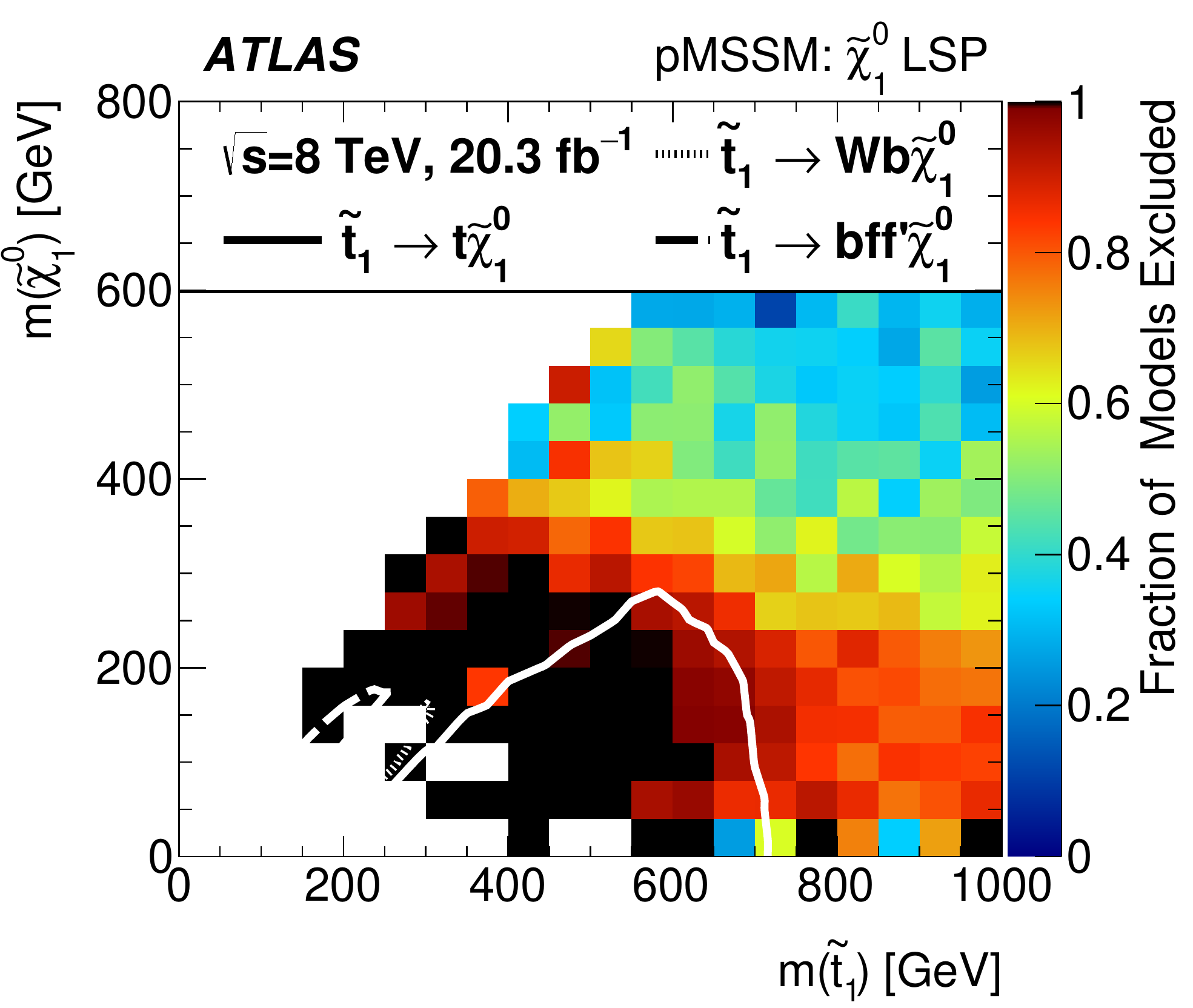}}&
\subfloat[Right handed $\tilde{t}_1$, all searches]{\includegraphics[width=0.49\textwidth]{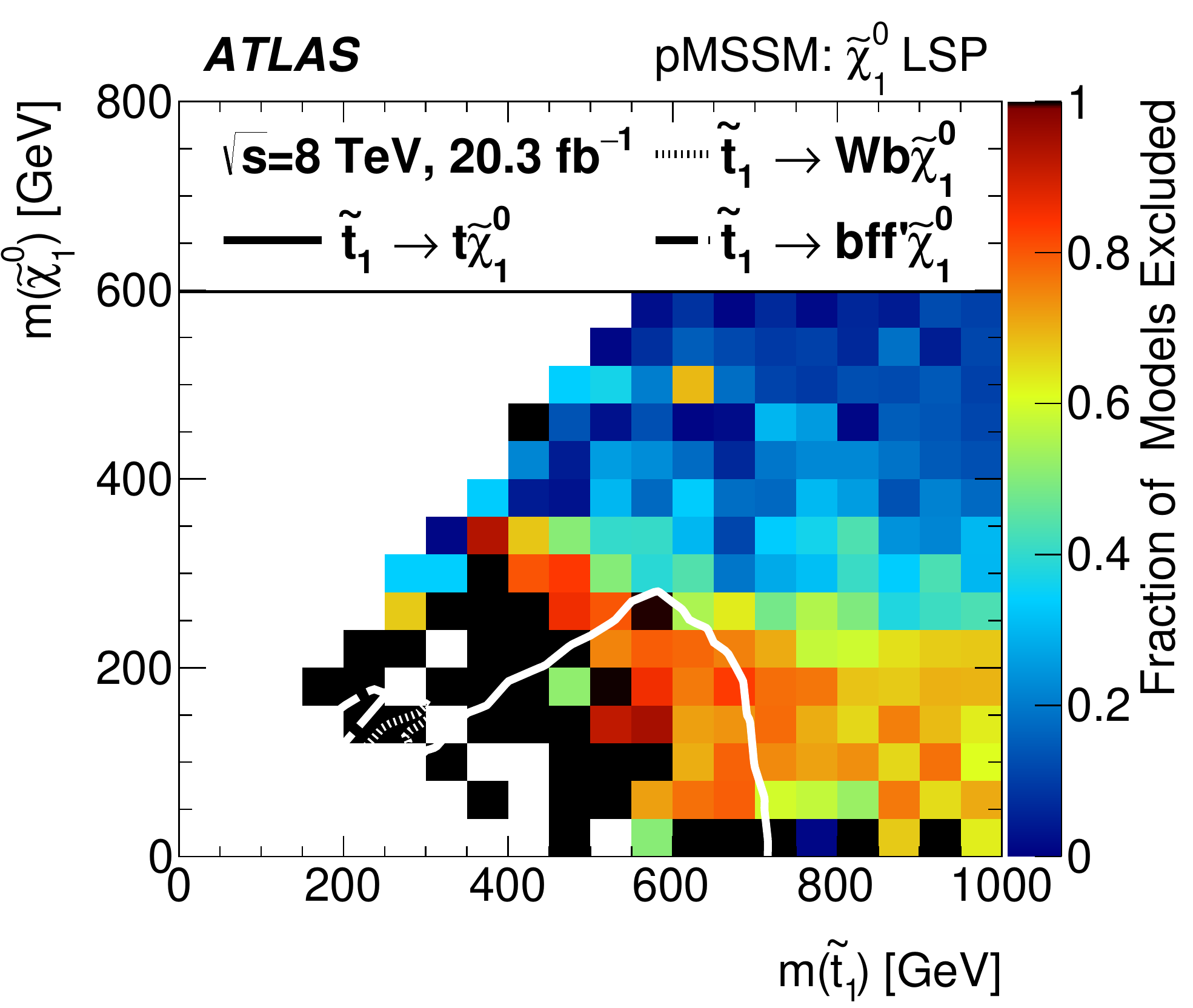}}\\
\end{tabular}
\caption{\label{fig:ex-stop-chi10}
Fraction of pMSSM points excluded in the $\tilde{t}_1$--\neutralino
plane in various cases. The top two plots show the full model set, with impact of all 
searches on the left and only the third-generation searches on the right. 
The bottom row of plots separates the models according to whether
the $\tilde{t}_1$ is either mostly (c) left handed or (d) right handed.
There are relatively few pMSSM points at low $\tilde{t}_1$ mass for the reasons
described in the text.
The simplified-model limit overlaid \cite{Aad:2014kra,Aad:2015pfx} is set 
assuming directly produced top squark pairs, with each decaying to a top quark
and neutralino, $\tilde{t}_1 \rightarrow t \neutralino$.
\colourscale
}
\end{figure}

The third-generation squarks are of particular phenomenological interest,
since light top squarks (and to a lesser extent bottom squarks) 
are usually required if SUSY is to solve the naturalness problem
associated with the Higgs boson mass. 
They have been the subject of several dedicated
searches, listed as `third-generation' in Table~\ref{tab:susySearches}. 
Within the 19-parameter pMSSM the masses of these third-generation squarks 
are controlled by different parameters than for the 
first two generations, allowing the $\tilde{t}_{1,2}$ and $\tilde{b}_{1,2}$
masses to differ substantially both from one another and from the
squarks of the first two generations.

Figure~\ref{fig:ex-stop-chi10}(a) shows the fractional exclusion of model points
as projected onto the plane of the mass of the lighter of the two top squarks
and that of the LSP. As discussed in Section~\ref{sec:models:properties}, there
are relatively few model points at low top squark mass. This is because a large
$m({\tilde{t}})$ (and/or a large trilinear coupling $A_t$) is required to
obtain the large quantum corrections needed to obtain the observed Higgs boson
mass. Despite there being relatively few points in the initial sample with
small top squark masses, one observes that when $m({\tilde{t}})$ is below about
\SI{600}{\GeV}, most points are excluded by ATLAS analyses.

The sensitivity to direct production of top squarks can be seen by considering the 
ATLAS exclusion using only those analyses from Table~\ref{tab:susySearches}
that target direct production of third-generation squarks.
The results, presented in Figure~\ref{fig:ex-stop-chi10}(b),
show that when only these third-generation ATLAS searches are considered,
good sensitivity continues to be 
observed for a lightest top squark mass up to about \SI{700}{\GeV}.

A reasonable correspondence is found between the 
sensitivity to the pMSSM points and those of 
the simplified-model decay considered in Ref.~\cite{Aad:2015pfx},
in which the top squark was assumed to decay
with certainty to $t + \neutralino$  (including off-shell top decays).
For larger top squark masses, in the range
\SI{600}{\GeV} to \SI{700}{\GeV},
this example simplified-model line extends into a 
region where the observed fraction of pMSSM points
that are excluded is less than 100\%.
This can be understood, since in Ref.~\cite{Aad:2015pfx} 
it was shown that the sensitivity of the $\tilde{t}$ search 
analyses depends on the branching ratio of the top squark to the LSP. 
When the decay proceeds via the two-step process
$\tilde{t}_1 \to b+\tilde{\chi}_1^\pm$ followed by
$\tilde{\chi}_1^\pm \to W^{(*)} + \neutralino$,
with the assumption that $m(\chargino)=m(\neutralino)$,
the 95\% CL exclusion limit did not extend beyond  $m(\tilde{t}_1) = \SI{540}{\GeV}$.
Since a range of such branching ratios is found in the pMSSM model sample, 
it is to be expected that there is partial sensitivity
in this intermediate region.

The lower two plots in Figure~\ref{fig:ex-stop-chi10} again show that the
sensitivity of the ATLAS analyses can depend in a non-trivial way on sparticle
properties that are not captured by simplified models. The dependence of the
sensitivity on the left versus right chirality of the top squark is caused in
part by different branching ratios for decays both to and from that squark.
The branching ratios of decays involving wino-like gauginos are affected by
the SU(2) coupling of the top squark, resulting in significant differences in
the sensitivity depending on whether the $\tilde{t}_1$  is dominantly left- or
right-handed. Furthermore, if the $\tilde{t}_1$ is mostly left-handed then 
the top squark's SU(2) partner, the $\tilde{b}$, would have a similar mass,
allowing analyses targeting sbottom production 
to become relevant in constraining the models.

\begin{figure}
\centering
\begin{tabular}{cc}
\subfloat[All searches]{\includegraphics[width=0.49\textwidth]{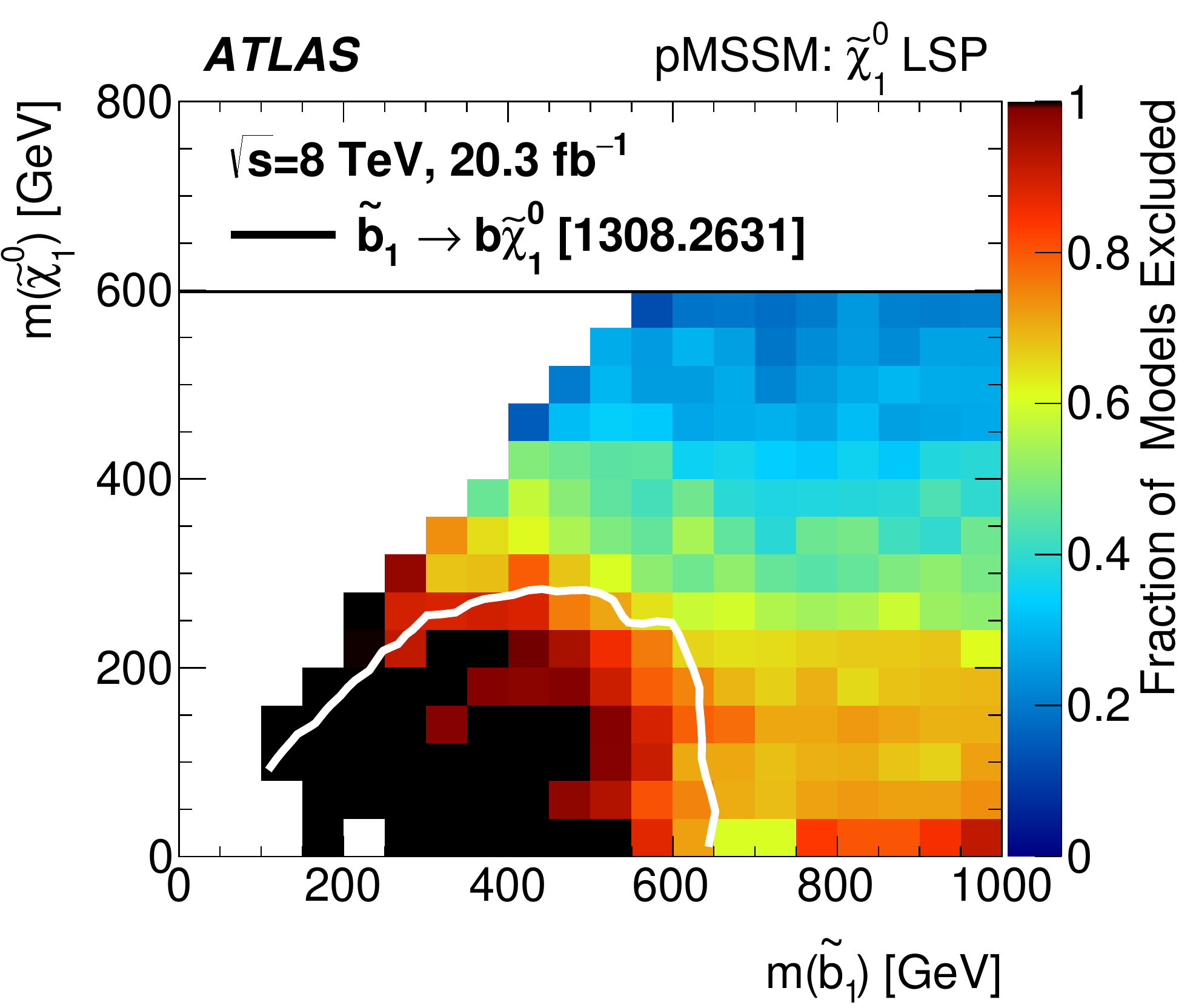}}&
\subfloat[Third-generation searches]{\includegraphics[width=0.49\textwidth]{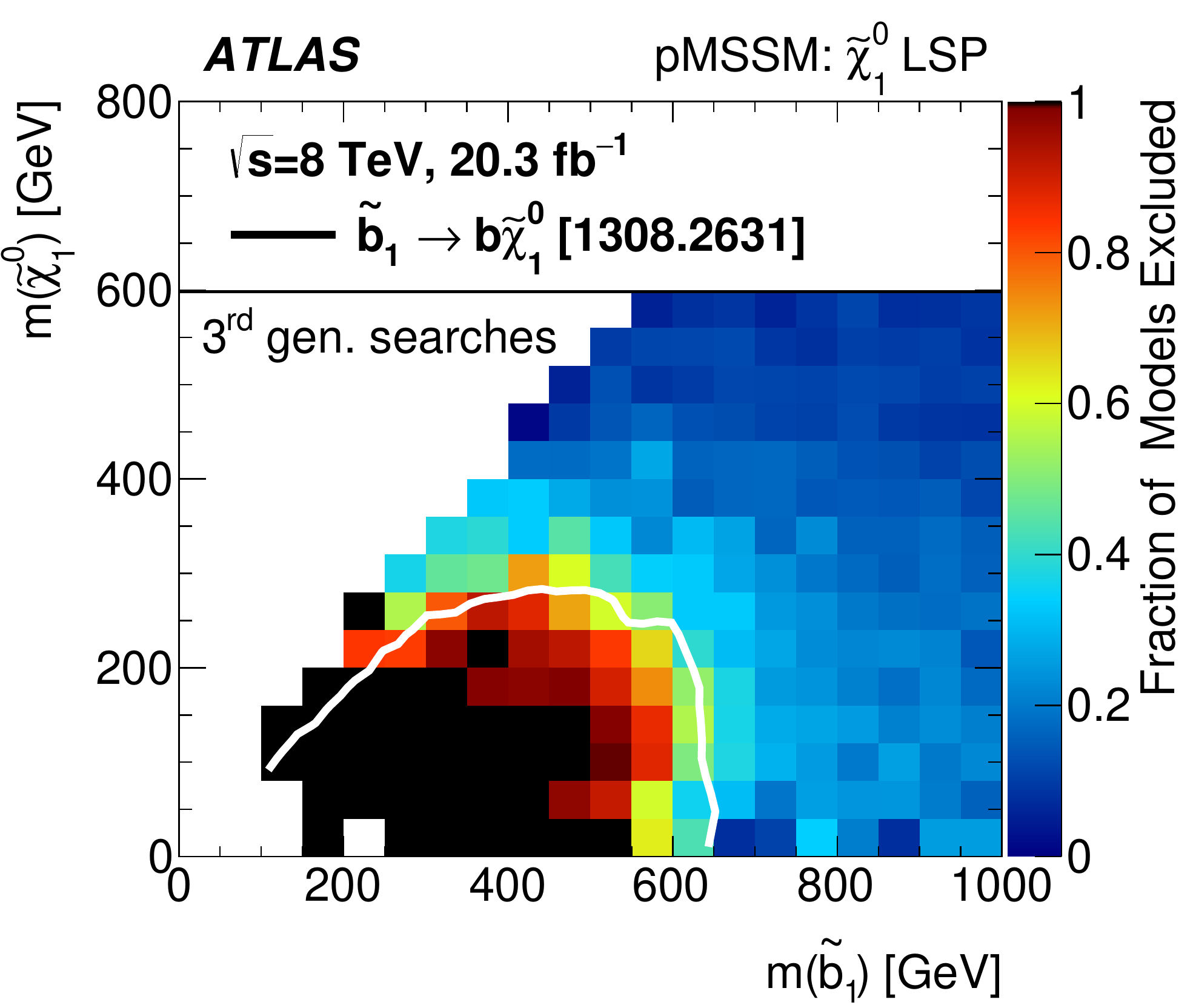}}\\
\end{tabular}
\caption{\label{fig:ex-sbottom-chi10}
The fraction of pMSSM points excluded by (a) all the listed ATLAS \runone{} searches  and (b) just the
third-generation searches in the \sbottom--\neutralino mass plane.
The white line shows a simplified-model limit
\cite{ATLAS-SUSY-2013-05} made assuming directly produced
bottom squark pairs, with bottom squarks decaying to a bottom quark and a neutralino,
$\sbottom \rightarrow b + \neutralino$. \colourscale
}
\end{figure}

Figure~\ref{fig:ex-sbottom-chi10}(a) shows the sensitivity of ATLAS analyses 
as projected onto the mass plane of the lightest bottom squark versus the LSP. 
The sensitivity is generally well captured by a simplified model
containing only an additional bottom squark and the LSP~\cite{ATLAS-SUSY-2013-05}.
When searches for sparticles other than those of the third generation are omitted
(Figure~\ref{fig:ex-sbottom-chi10}(b)) the similarity of the pMSSM 
with the simplified model becomes still clearer.
The sensitivity of the ATLAS searches was found to be similar regardless of 
whether the bottom squark was dominantly left- or right-handed. 

\subsubsection{Electroweak sparticles and sleptons}
\label{sec:electroweak}
\begin{figure}
\centering
\includegraphics[width=0.49\textwidth]{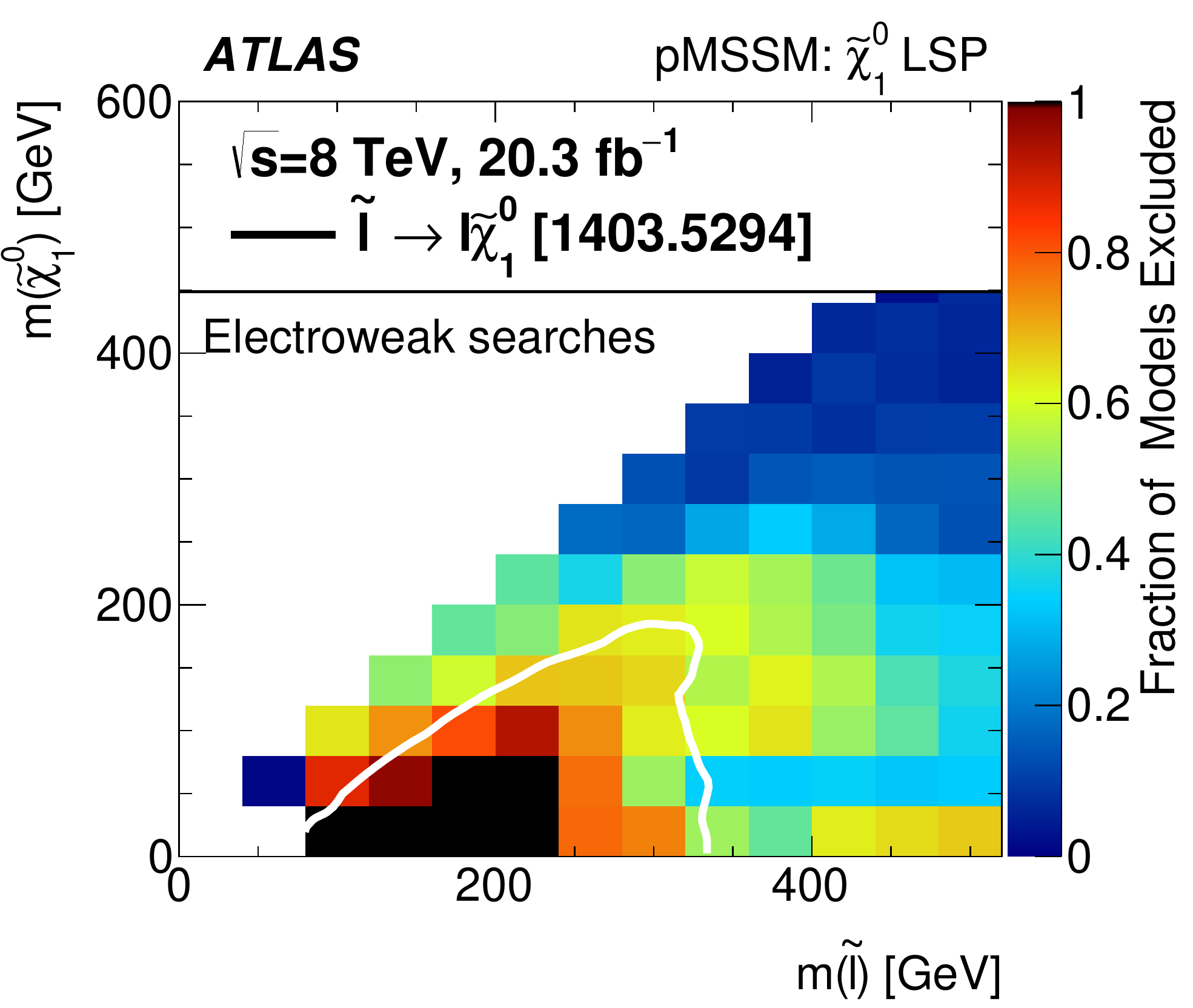}
\caption{\label{fig:ex-EWK-slepton-chi10}
The fraction of pMSSM points excluded by just the electroweak 
ATLAS searches listed in Table~\ref{tab:susySearches}, projected
onto the \slepton--\neutralino mass plane,
where \slepton is the lightest slepton of the first two generations.
The white line reflects  
one of the simplified-model limits (Figure~8(c) of Ref.~\cite{Aad:2014vma})
made for direct slepton pair production assuming that 
left- and right-handed selectrons (or smuons) are mass-degenerate
and that each decays via $\tilde{\ell}^\pm \to \ell^\pm + \neutralino$.
\colourscale
}
\end{figure}

The sensitivity to selectrons, smuons, and their sneutrino counterparts (here denoted collectively by $\tilde{\ell}$) is displayed in Figure~\ref{fig:ex-EWK-slepton-chi10}.
It can be seen that the ATLAS searches have 
good sensitivity to slepton masses up to about \SI{200}{\GeV},
particularly when the LSP mass is lighter than about \SI{75}{\GeV},
where only bino-like LSPs survived the preselection of Table~\ref{tab:constraints}.
Nevertheless, the reduced sensitivity in the near-degenerate region where
the slepton decay produces soft leptons and small \met{} 
means that existing ATLAS searches cannot place any lower bound
on the slepton mass.
The region of sensitivity in the pMSSM projected plane is found to 
have a degree of correspondence with one of 
the simplified models of Ref.~\cite{Aad:2014vma}.
Generally, reduced sensitivity is found in the pMSSM
when compared to the more-constrained simplified model.
This can be understood by recognising that this particular model presupposes that the left- and right-handed selectrons and smuons
are all mass degenerate, and that each has a 100\% branching ratio to a lepton and a LSP.
Breaking these assumptions reduces the number of signal events,
and hence allows more models to evade detection.

\begin{figure}
\centering
\begin{tabular}{cc}
\subfloat[All LSPs, $\tilde{\ell}_{\rm L}$]{\includegraphics[width=0.48\textwidth,trim=0 0 0 0,clip]{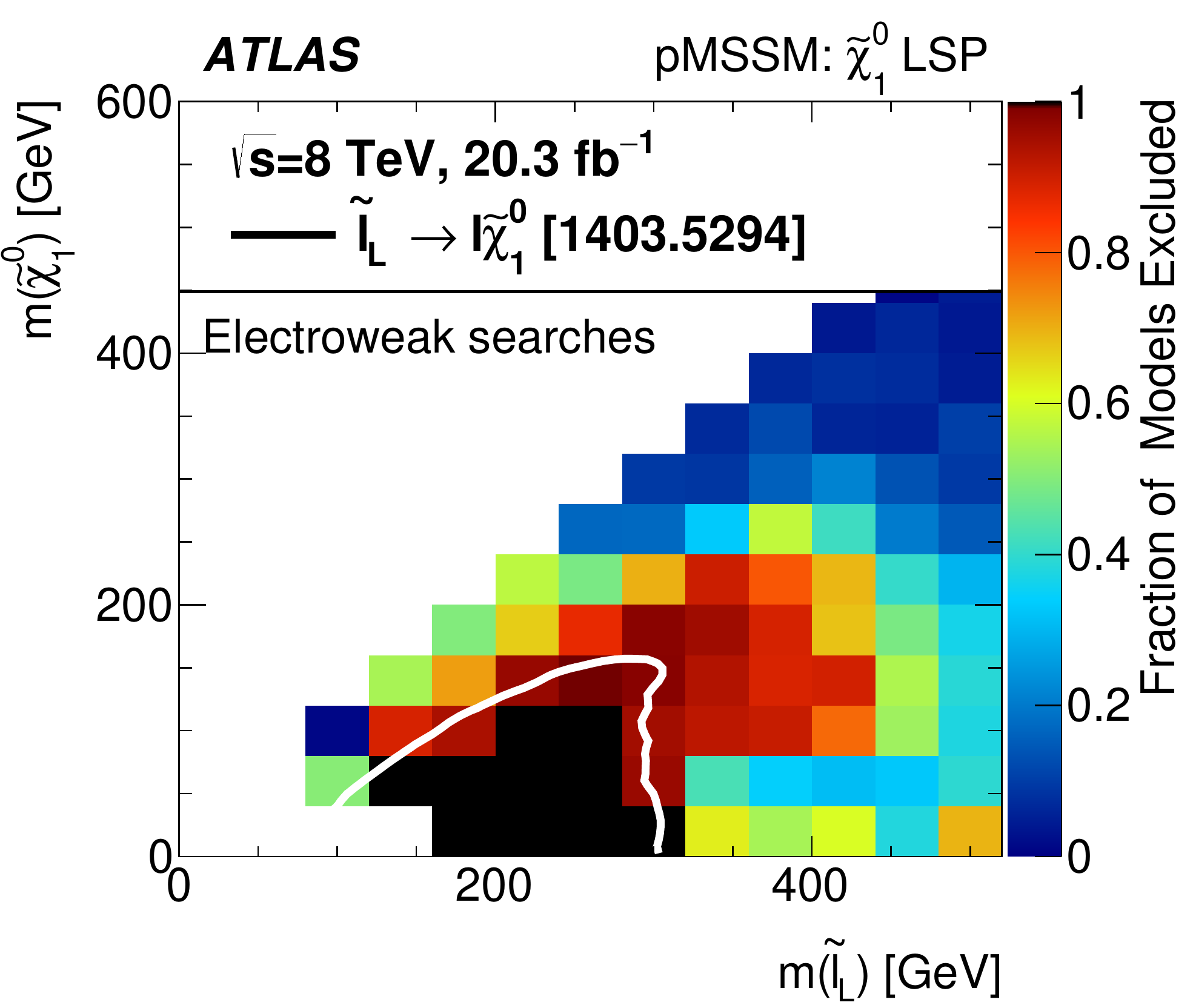}}&
\subfloat[All LSPs, $\tilde{\ell}_{\rm R}$]{\includegraphics[width=0.48\textwidth,trim=0 0 0 0,clip]{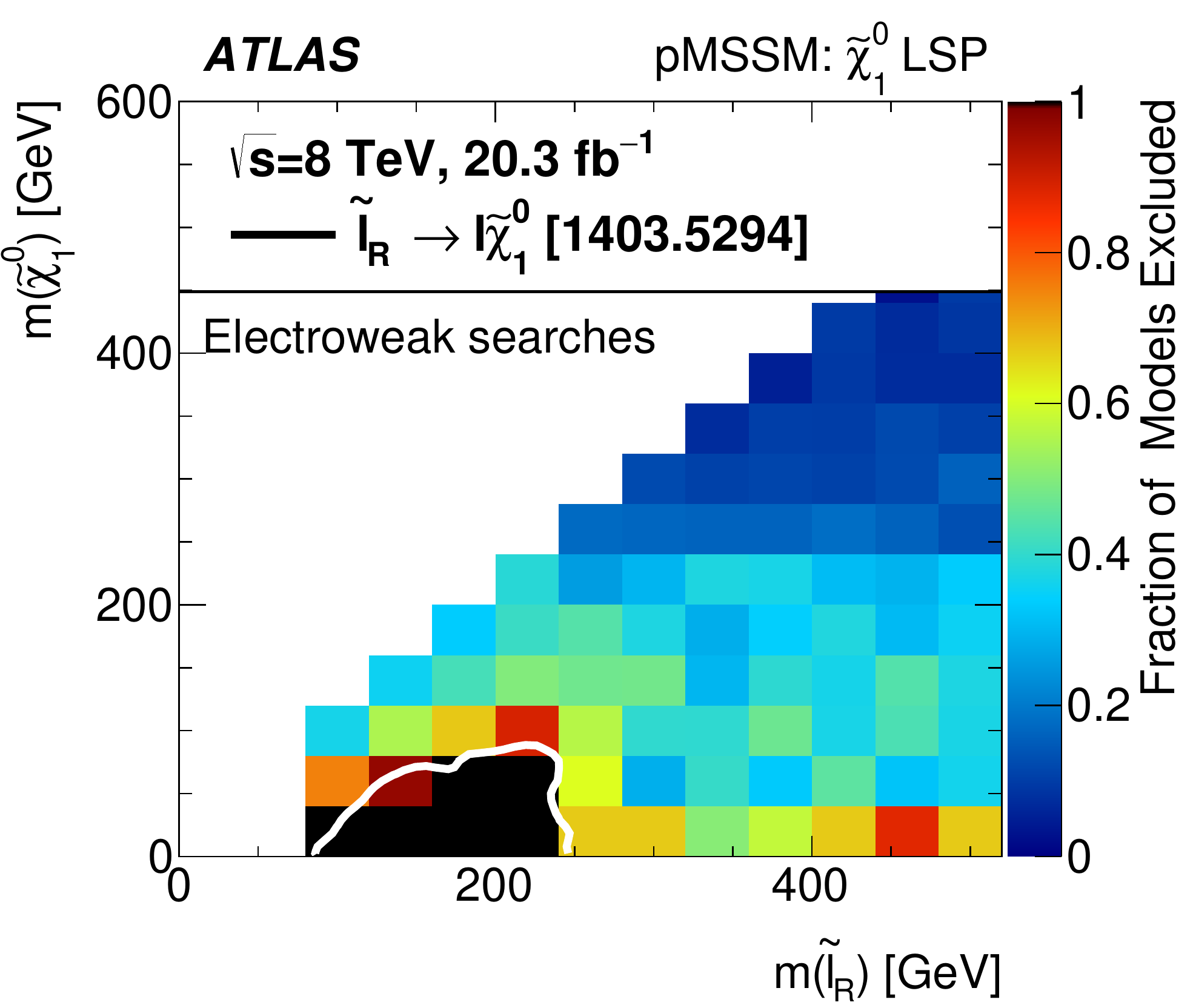}}
\end{tabular}
\caption{\label{fig:ex-EWK-e-chi10}
Impact of electroweak searches (as listed in Table~\ref{tab:susySearches}) 
in (a) the $\tilde{\ell}_{\rm L}$--\neutralino and (b)
$\tilde{\ell}_{\rm R}$--\neutralino projections.
It should be noted that in the 19-parameter pMSSM, 
the first- and second-generation sleptons of each handedness are required to be degenerate. 
The simplified-model limit in the $\tilde{\ell}_{\rm L}$ ($\tilde{\ell}_{\rm R}$)
case is set assuming directly pair-produced left (right)
handed selectrons/smuons, decaying to an electron/muon and neutralino.
The simplified-model limits are from Figures~8(a) and 8(b) of Ref.~\cite{Aad:2014vma}.
\colourscale
}
\end{figure}

When the assumption of degenerate left- and right-handed states is dropped from the simplified model,
the resulting limits are similar to those of the pMSSM.
This can be seen in Figure~\ref{fig:ex-EWK-e-chi10},  
showing the pMSSM space projected separately onto the mass of the left-handed or
right-handed slepton. The fraction of model points excluded is compared
to simplified models in which either only left- or right-handed sleptons ($\tilde{e}$ and
$\tilde{\mu}$) are produced.   
ATLAS searches have more sensitivity to the production of left-handed sleptons
since right-handed states lack SU(2) couplings
and so have a smaller $\tilde\ell^+\tilde\ell^-$ production cross-section.

\begin{figure}
\centering
\includegraphics[width=0.48\textwidth]{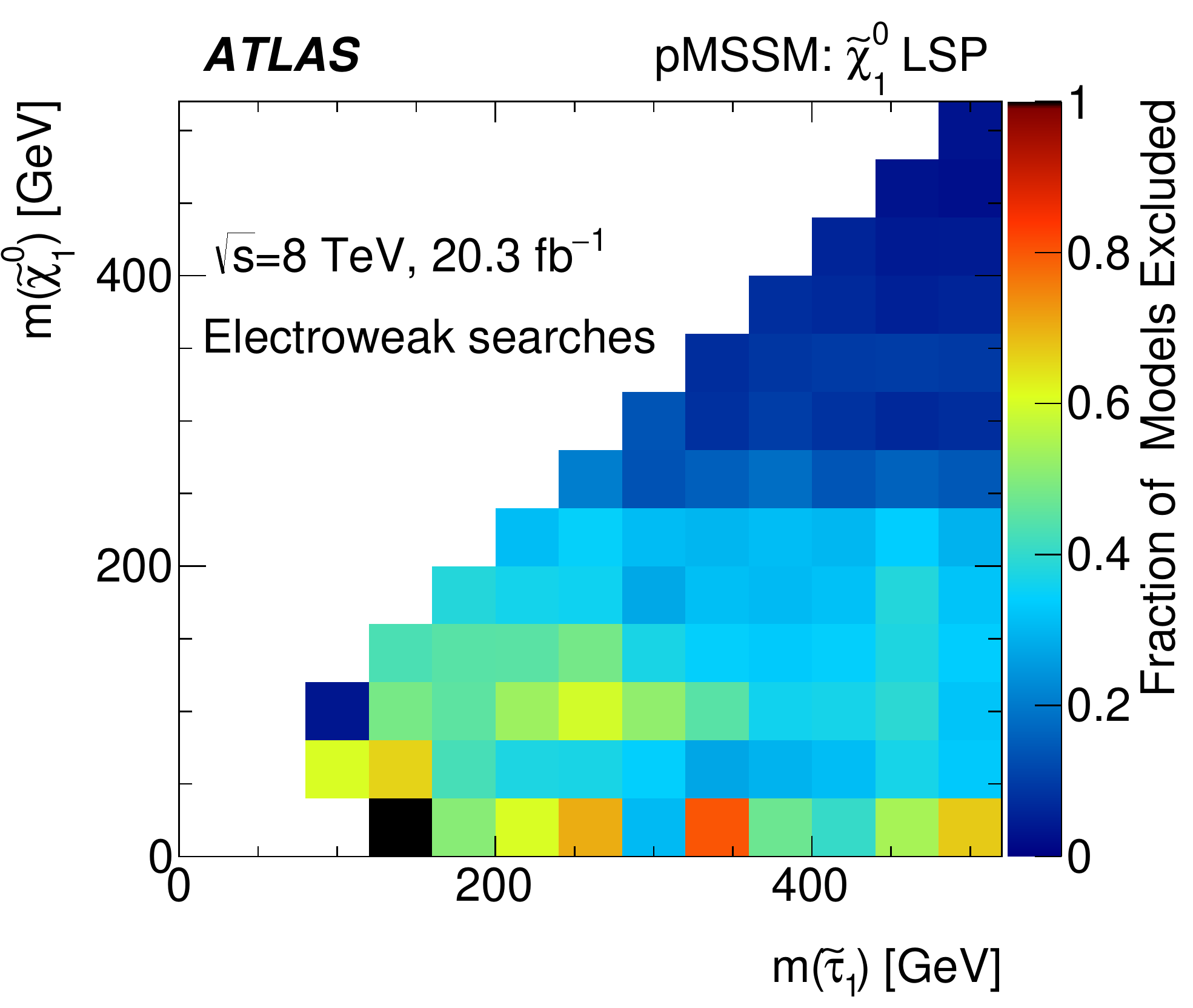}
\caption{\label{fig:ex-EWK-stau-chi10}
Impact of electroweak searches (as listed in Table~\ref{tab:susySearches})
on the \stau--\neutralino plane. \colourscale 
}
\end{figure}

Figure~\ref{fig:ex-EWK-stau-chi10} shows the fraction of model points excluded by just the electroweak
ATLAS searches listed in Table~\ref{tab:susySearches}, projected onto the plane of LSP and 
lightest stau mass.
It can be seen that the \runone{} sensitivity to staus is limited, 
with large fractions of model points
surviving even at the lowest stau masses. This is largely because it is
difficult to trigger on events resulting from direct stau production, and 
backgrounds to stau searches are much larger than for the equivalent 
search for sleptons of the first two generations.
No definitive lower bound can be placed on the stau mass by ATLAS after \runone. 

\begin{figure}
\centering
\begin{tabular}{cc}
\subfloat[Neutralinos]{\includegraphics[width=0.48\textwidth]{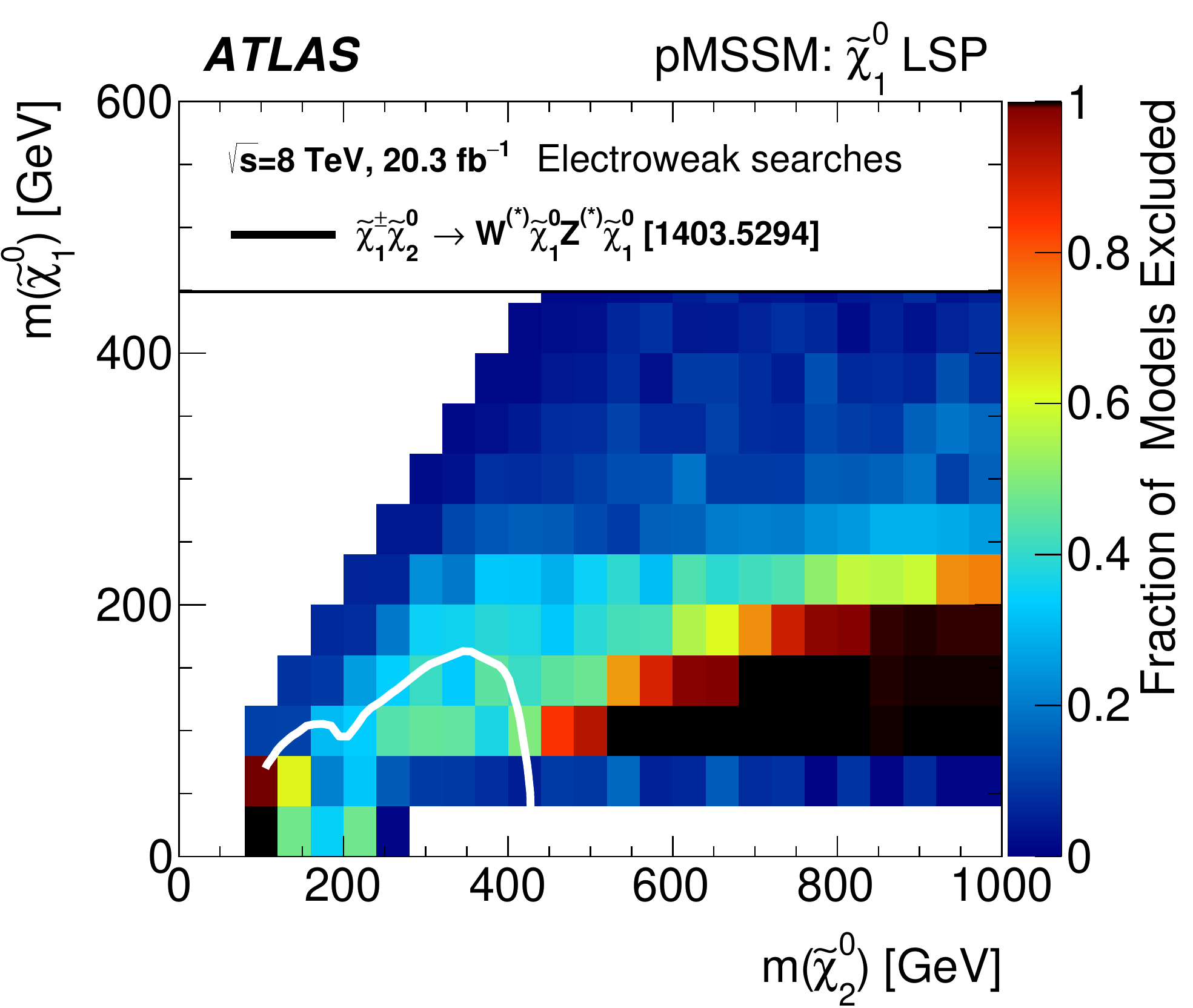}} & 
\subfloat[Chargino--neutralino]{\includegraphics[width=0.48\textwidth]{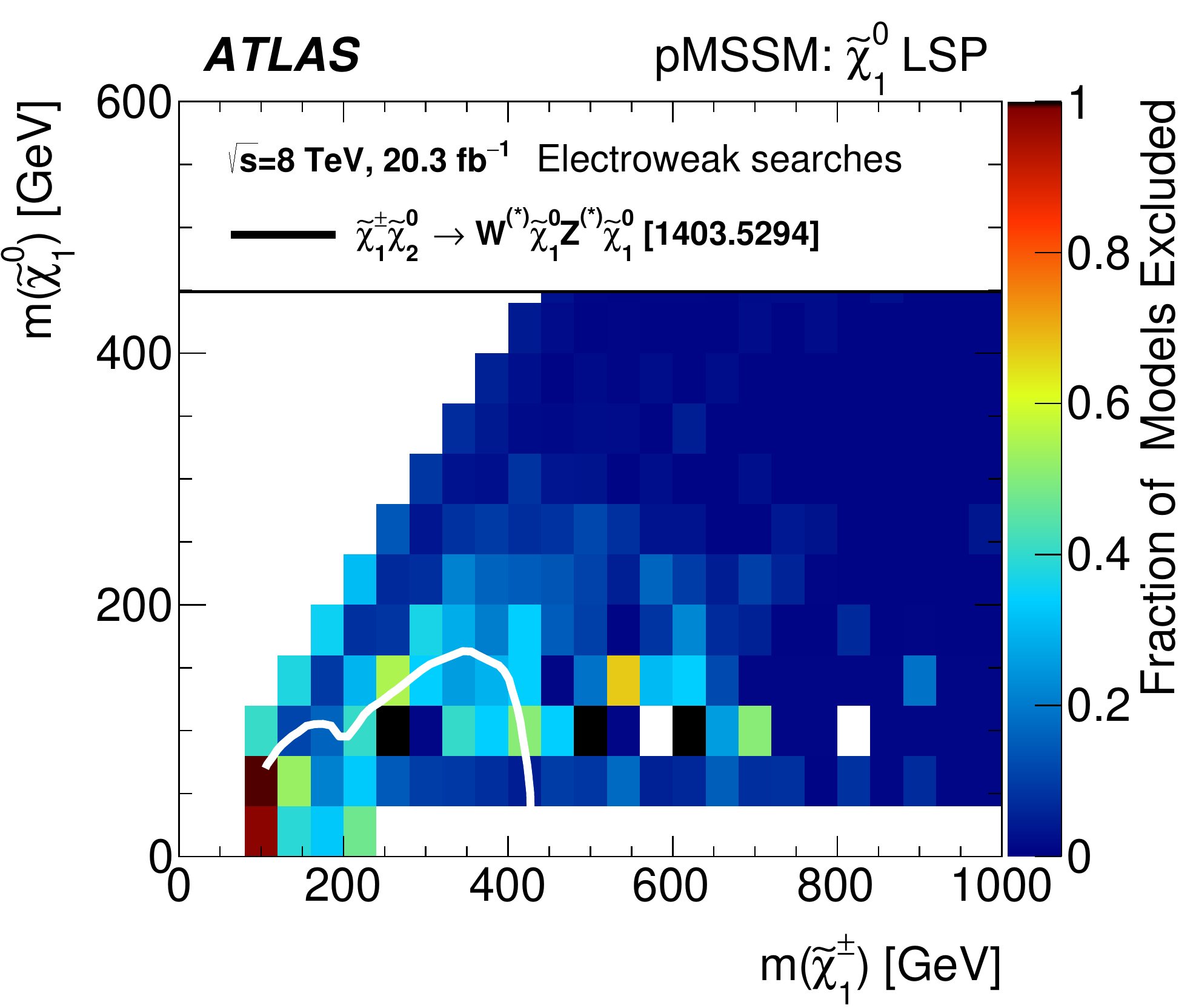}}
\end{tabular}
\caption{\label{fig:ex-EWK-electroweakinos}
Impact of electroweak searches (as listed in Table~\ref{tab:susySearches})
(a) on the \neutralinotwo--\neutralino plane and  
(b) on the \chargino--\neutralino{} plane.
The 95\% CL observed exclusion limit from Ref.~\cite{Aad:2014vma}
is for a simplified model that assumes pure-wino
$\chargino+\neutralinotwo$ production, followed
by the decays
$\chargino \neutralinotwo \to W^{*} \neutralino Z^{*} \neutralino$.
\colourscale 
}
\end{figure}

Figure~\ref{fig:ex-EWK-electroweakinos}(a) shows the fraction of models excluded
by only the electroweak ATLAS searches (see Table~\ref{tab:susySearches}), 
this time projected onto the plane of the masses of the lightest two neutralinos.
Two prominent features are visible.
For \neutralino{} masses lighter than about \SI{200}{\GeV},
a large fraction of models are excluded, particularly as $m(\neutralinotwo)$ 
becomes large. The dominant exclusion mechanism 
for large $m(\neutralinotwo)$ 
is due to the \DisappearingTrack{} analysis and is strongest when
the LSP is wino-dominated. 
The gaugino mass difference $\Delta m_\chi = m(\chargino)-m(\neutralino)$
is typically less than a few hundred \MeV{} for winos and
of order a few \GeV{} for Higgsinos.
As the \neutralinotwo{} mass decreases, approaching that of the 
\neutralino, there is more neutralino mixing, 
leading to a larger $\Delta m_\chi$, and a shorter \chargino lifetime,
hence the \DisappearingTrack{} analysis loses sensitivity.
The Figure~\ref{fig:ex-EWK-electroweakinos}(a) row 
in which $m(\neutralino)\sim \SI{50}{\GeV}$
has lower sensitivity for the \DisappearingTrack{} analysis.
This region is dominated by models for which the relic density
is controlled by the $Z$ and $h$~boson funnels, so has bino-like LSPs
with a Higgsino admixture.
Such models do not typically feature long-lived charginos.

For $m(\neutralinotwo)\lsim\SI{400}{\GeV}$ and
$m(\neutralino)\lsim\SI{200}{\GeV}$, 
direct production of \neutralinotwo (and/or \chargino) states
provides sensitivity via the 
\TwoLepton{}, \ThreeLepton{} and \FourLepton{}
analyses.
The sensitive region for these multi-lepton analyses 
is similar to that shown from the simplified
model of Ref.~\cite{Aad:2014vma}.
Nevertheless there remain many viable pMSSM points within 
the region excluded in the simplified-model scenario.
For example, many points in the $Z$ and $h$~boson funnel regions 
($m(\neutralino)\sim \SI{50}{\GeV}$)
have little sensitivity in the multi-lepton 
analyses as the \neutralinotwo{} is predominantly
Higgsino-like, leading to a lower production cross-section.

The equivalent plot for the projection onto the plane of the
lightest chargino and the LSP is shown in Figure~\ref{fig:ex-EWK-electroweakinos}(b),
again showing the fraction excluded by the electroweak ATLAS searches.
In this figure the \DisappearingTrack{} analysis
has sensitivity to models with wino-like LSPs 
which lie close to the leading diagonal where 
$m(\chargino)$ is only a little larger than $m(\neutralino)$.
Models with Higgsino-like LSPs also lie close to that 
diagonal, but have larger mass splittings and so
little sensitivity from the  \DisappearingTrack{} analysis.
Away from that diagonal only bino-dominated LSPs are found.
Here the best sensitivity is from the multi-lepton
electroweak search analyses (\TwoLepton{}, \ThreeLepton{} and \FourLepton{}),
particularly for  $m(\chargino)\lsim\SI{400}{\GeV}$ and
$m(\neutralino)\lsim\SI{200}{\GeV}$.
The region with sensitivity to the multi-lepton searches
again shows some similarity
with the simplified-model limit from Ref.~\cite{Aad:2014vma},
but again no region is totally excluded.

\subsubsection{Long-lived squarks, gluinos and sleptons}\label{sec:results:longlived}

As described in Section~\ref{sec:sim:longlived}, 
model points with long-lived squarks, gluinos and
sleptons are treated separately as the sensitivity 
to these models of the regular
searches for prompt decays is difficult to
assess. Instead only the \LLSparticles{} searches
\cite{ATLAS:2014fka,Aad:2012pra} are considered. These
analyses have very good sensitivity to these model points as can be seen
from Table~\ref{tab:longLivedExclusion}.
Particularly good sensitivity is found in cases where the bottom squarks are 
long lived.
The sensitivity to gluinos is quite poor as these are usually too
short-lived to be picked up by the long-lived particle searches.

\begin{table}[h]
\centering
\renewcommand\arraystretch{1.2}
\begin{tabular}{l|c|c|c|c|c|c}
\hline
\textbf{Long-lived}  & \multicolumn{2}{c|}{\textbf{Bino LSP}} & \multicolumn{2}{c|}{\textbf{Wino LSP}}& \multicolumn{2}{c}{\textbf{Higgsino LSP}}\\
\cline{2-7}
\textbf{Particle} & \textbf{Models} & \textbf{Excluded}  & \textbf{Models} & \textbf{Excluded} & \textbf{Models} & \textbf{Excluded} \\ \hline 
$\tilde{g}$                 & 899 (5.2\%) &       5.1\% &  58 (3.4\%) &       3.4\% &   9 (0.0\%) &       0.0\% \\
$\tilde{b}_1$               & 1252 (99.6\%) &      76.4\% & 51 (100.0\%) &      78.4\% & 67 (100.0\%) &      80.6\% \\
$\tilde{t}_1$               & 345 (56.8\%) &      36.5\% & 6 (100.0\%) &      66.7\% & 17 (82.4\%) &      47.1\% \\
$\tilde{\tau}_1$            & 406 (100.0\%) &      37.4\% & 2 (100.0\%) &       0.0\% & 41 (100.0\%) &      14.6\% \\

\hline
\end{tabular}
\caption{Number of model points with long-lived particles and their exclusion fraction. The percentages in parenthesis
are the fractions of these model points where the long-lived particle has a lifetime long enough to traverse the entire detector.}
\label{tab:longLivedExclusion}
\end{table}

\subsubsection{Heavy neutral Higgs bosons}\label{sec:results:htautau}

\begin{figure}
\centering
\includegraphics[width=0.48\textwidth, trim=0 0 0 0,clip]{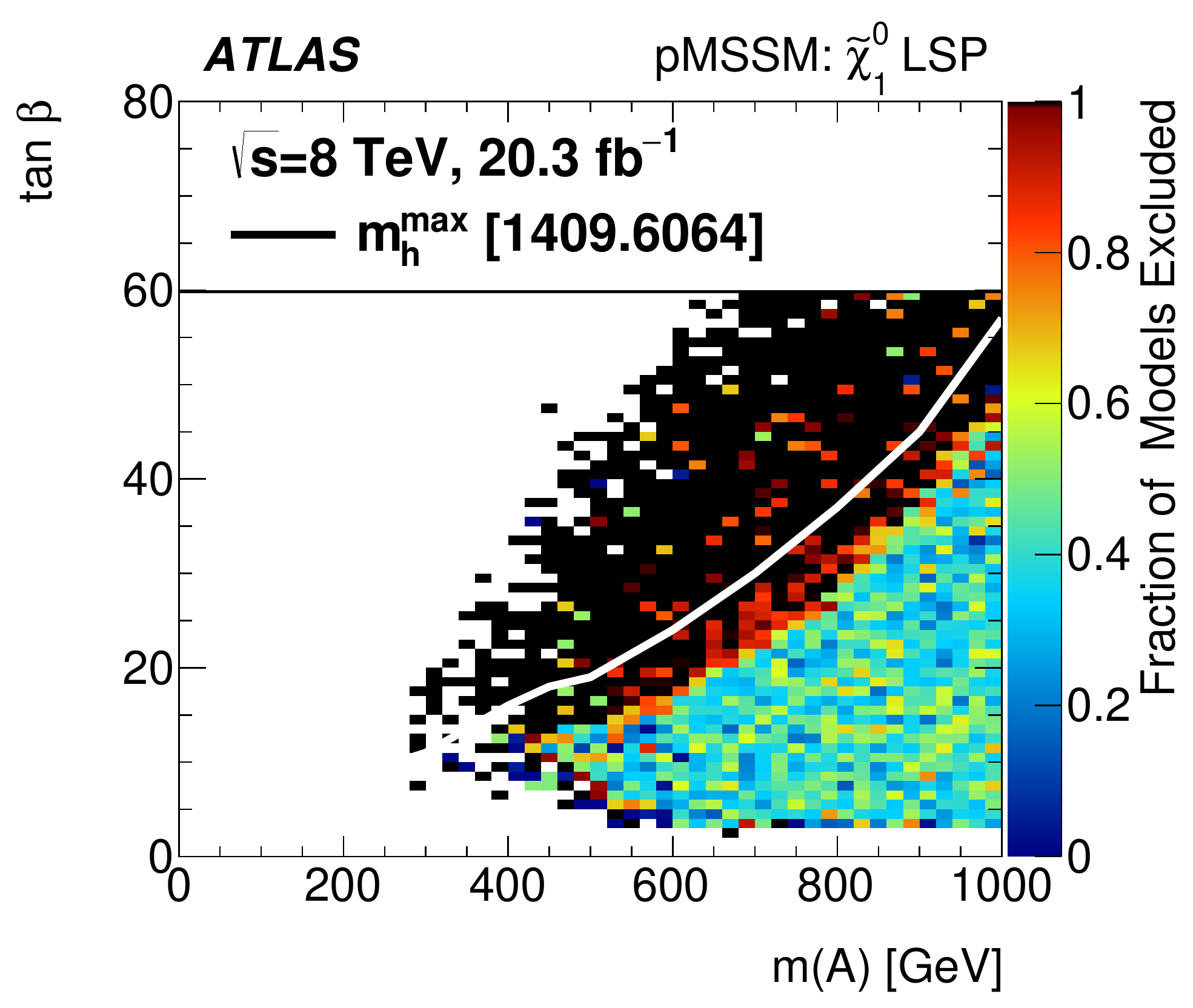}
\caption{\label{fig:higgs-impact}
The fraction of model points excluded by ATLAS \runone{} searches,
including those for heavy Higgs bosons, projected into the $m_A$--$\tan \beta$ plane.
The white line overlaid is the observed limit
for the \mhmax scenario from the $H/A \rightarrow \tau \tau$ search of Ref.~\cite{Aad:2014vgg} where
the region above the line is excluded.
}
\end{figure}

Figure~\ref{fig:higgs-impact} shows
the fraction of model points excluded by the
ATLAS \runone{} searches 
(which include searches for heavy Higgs bosons)
when projected into the $m_A$--$\tan \beta$ plane.
The white line overlaid is the observed limit
from the ATLAS search~\cite{Aad:2014vgg} 
for heavy neutral Higgs boson(s), $H$ or $A$, decaying
to $\tau\tau$, as interpreted in the \mhmax scenario.
A close correspondence can be observed between the 
limit obtained in that scenario and the region of the 
pMSSM space excluded by the \runone{} searches in the 
general pMSSM-19 space. The \mhmax scenario is seen to give
a slightly conservative limit compared to the pMSSM space as
it has light electroweakinos leading to a lower branching fraction
for $A/H\rightarrow \tau\tau$ than for most pMSSM points with 
similar $m_A,\tan \beta$ values.

\afterpage\clearpage
\subsubsection{Complementarity of searches}
\label{sec:combination}

\begin{figure}[htb]
\centering
\includegraphics[width=0.98\linewidth]{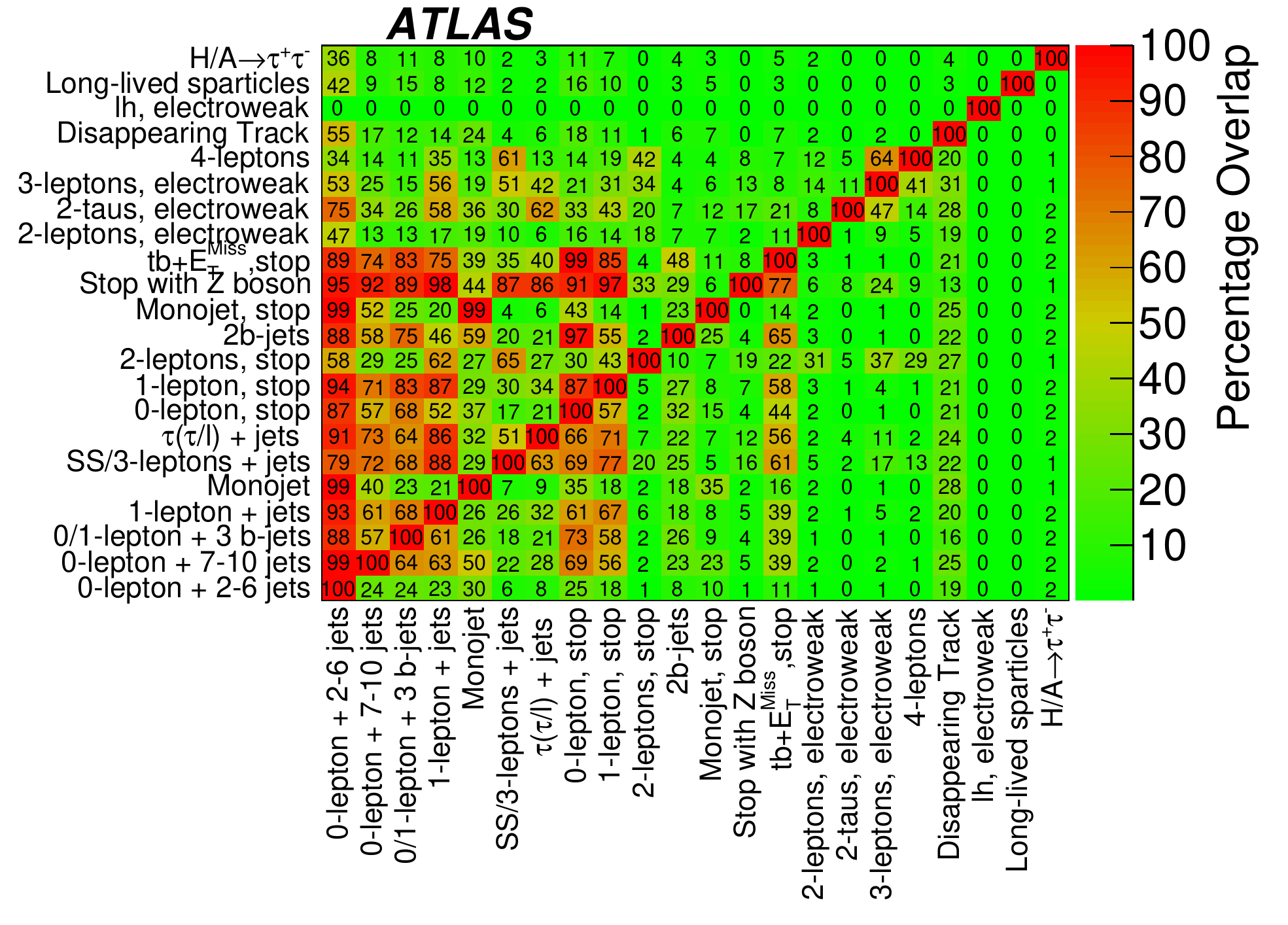}
\caption{\label{fig:overlap}
Complementarity of ATLAS searches: the figure shows the percentage of model
points excluded
by the analysis on the $y$-axis that were also excluded by
the analysis on the $x$-axis.
References for the individual analyses can be found in Table~\ref{tab:susySearches}.
As none of the models considered are excluded by the \OneLeptonHiggs{} analysis, it has no
overlap with any other search.
}
\end{figure}

The ATLAS searches are designed to be sensitive to different final 
states, or different kinematic regions.
Nevertheless it is often the case that a model which produces
an excess in one search can have a significant signal expectation in others,
since related cascade decays can lead to complementary final states.
The degree of complementarity is explored in 
Figure~\ref{fig:overlap}, which shows the fraction of model points
excluded (at 95\%~CL) by one analysis that are also excluded
by another. For example, one can observe 
that of the model points that were excluded by the
\OneLeptonStrong{} analysis (fourth row from bottom), 
93\% were also excluded by the \ZeroLepton{} analysis (first column).

\begin{table}[htb]
\centering\renewcommand\arraystretch{1.2}
\begin{tabular}{l|c|c|c|c}
\textbf{Analysis} & \textbf{All LSPs}  & \textbf{Bino-like} & \textbf{Wino-like} & \textbf{Higgsino-like}  \\ \hline 
\ZeroLepton{}               &      32.1\% &      35.8\% &      29.7\% &      33.5\% \\
\Multijet{}                 &       7.8\% &       5.5\% &       7.6\% &       8.0\% \\
\ThreeBjets{}               &       8.8\% &       5.4\% &       7.1\% &      10.1\% \\
\OneLeptonStrong{}          &       8.0\% &       5.4\% &       7.5\% &       8.4\% \\
\Monojet{}                  &       9.9\% &      16.7\% &       9.1\% &      10.1\% \\
\SSThreeLepton{}            &       2.4\% &       1.6\% &       2.4\% &       2.5\% \\
\TauStrong{}                &       3.0\% &       1.3\% &       2.9\% &       3.1\% \\
\ZeroLeptonStop{}           &       9.4\% &       7.8\% &       8.2\% &      10.2\% \\
\OneLeptonStop{}            &       6.2\% &       2.9\% &       5.4\% &       6.8\% \\
\TwoBjet{}                  &       3.1\% &       3.3\% &       2.3\% &       3.6\% \\
\TwoLeptonStop{}            &       0.8\% &       1.1\% &       0.8\% &       0.7\% \\
\StopMonojet{}              &       3.5\% &      11.3\% &       2.8\% &       3.6\% \\
\StopZ{}                    &       0.4\% &       1.0\% &       0.4\% &       0.5\% \\
\TBmet{}                    &       4.2\% &       1.9\% &       3.1\% &       5.0\% \\
\OneLeptonHiggs, electroweak       &       0 &       0 &       0 &       0 \\
\TwoLepton, electroweak      &       1.3\% &       2.2\% &       0.7\% &       1.6\% \\
\TwoTau, electroweak         &       0.2\% &       0.3\% &       0.2\% &       0.2\% \\
\ThreeLepton, electroweak    &       0.8\% &       3.8\% &       1.1\% &       0.6\% \\
\FourLepton                 &       0.5\% &       1.1\% &       0.6\% &       0.5\% \\
\DisappearingTrack{}        &      11.4\% &       0.4\% &      29.9\% &       0.1\% \\
\LLSparticles{}             &       0.1\% &       0.1\% &       0.0\% &       0.1\% \\
\HiggsToTauTau{}            &       1.8\% &       2.2\% &       0.9\% &       2.4\% \\ \hline
Total                       &      40.9\% &      40.2\% &      45.4\% &      38.1\% \\

\end{tabular}
\caption{Percentage of model points excluded by the individual analyses.
It should be noted that the fraction of model points that can be excluded will depend on the 
model employed and range of input masses initially generated.
The reader is reminded (Table~\ref{tab:scanranges}) that the
sparticle mass terms in this \paper{} extend to \SI{4}{\TeV}.
References for the individual analyses can be found in 
Table~\ref{tab:susySearches}.
}
\label{tab:overallExclusion}
\end{table}

Figure~\ref{fig:overlap} demonstrates that there is a good complementarity 
-- relatively small overlap in sensitivity -- between
searches for strongly interacting particles, which are characterised by 
final states with jets, and searches for electroweak production.
Further study shows that, for the sampling of pMSSM points made in this paper, 
the analyses with the largest regions of unique sensitivity are
the \ZeroLepton{} analysis~\cite{ATLAS-SUSY-2013-02}, 
and the \DisappearingTrack{} analysis~\cite{Aad:2013yna}.
Nevertheless some care is required in interpreting these results.
The degree of apparent overlap is subjective, 
in that it depends, in some cases sensitively, 
on the metric used when sampling the pMSSM space.
Even in cases where the apparent overlap appears to be large, 
for example between the 
\ZeroLepton{} and \Multijet{} analyses, 
both searches are found to have regions of pMSSM space in 
which they provide unique sensitivity.
The \DisappearingTrack{} analysis is mostly sensitive
to model points with a wino-like LSP,
so an alternative prior (or weighting by LSP type) 
of the sample model points would 
directly affect the apparent relative sensitivity of this analysis.

The overall fraction of model points within the pMSSM space
excluded by each analysis for each of the LSP types
is shown in Table~\ref{tab:overallExclusion}.
Only the \OneLeptonHiggs{} analysis is unable
to constrain the pMSSM set with the luminosity available.
The lack of sensitivity for that analysis
is not unexpected since for simplified models it 
excludes only points with very light LSPs~\cite{Aad:2015jqa}.
It should again be noted that the absolute values
of the fractions of model points excluded 
is strongly affected by the prior sampling, 
in particular by the upper mass bounds used for the scan in
selecting the pMSSM input parameters (see Table~\ref{tab:scanranges}).
The relative fractions of model points excluded by each analysis
are a little more informative, but again care is necessary
in their interpretation
since they too are sensitive to changes to the 
assumptions or constraints applied
to the initial model set.
Nevertheless, the high sensitivity of the 
\ZeroLepton{} analysis for all LSP types,
and the \DisappearingTrack{} analysis for models with a wino-like LSP
is unambiguous.

\subsection{Impact of ATLAS searches on dark matter}\label{sec:results:dm}

\begin{figure}
\begin{center}
\begin{tabular}{cc}
\subfloat[Before ATLAS \runone]{\includegraphics[width=0.49\textwidth,trim=30 20 40 5,clip]{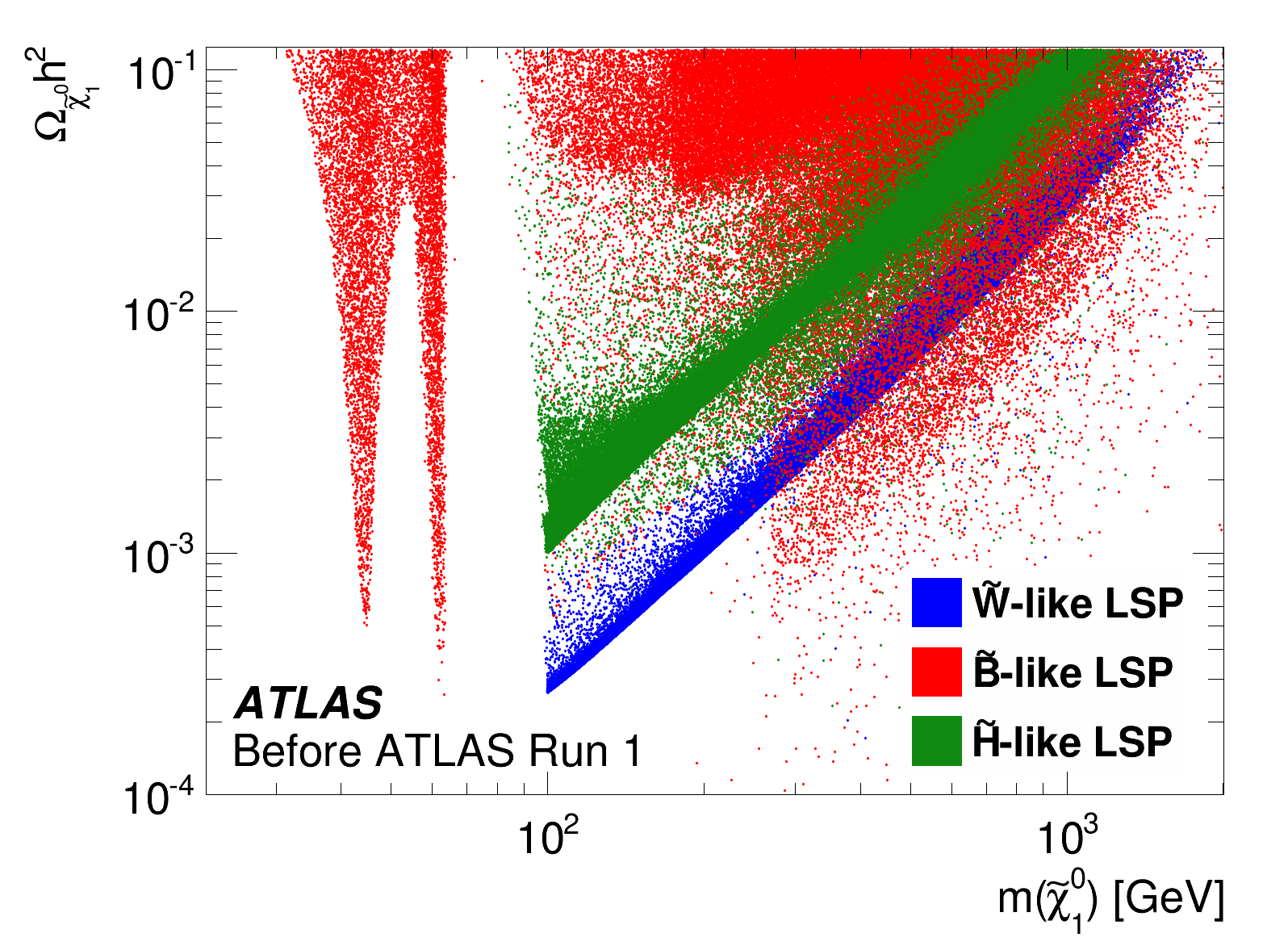}} &
\subfloat[After ATLAS \runone]{\includegraphics[width=0.49\textwidth,trim=30 20 40 5,clip]{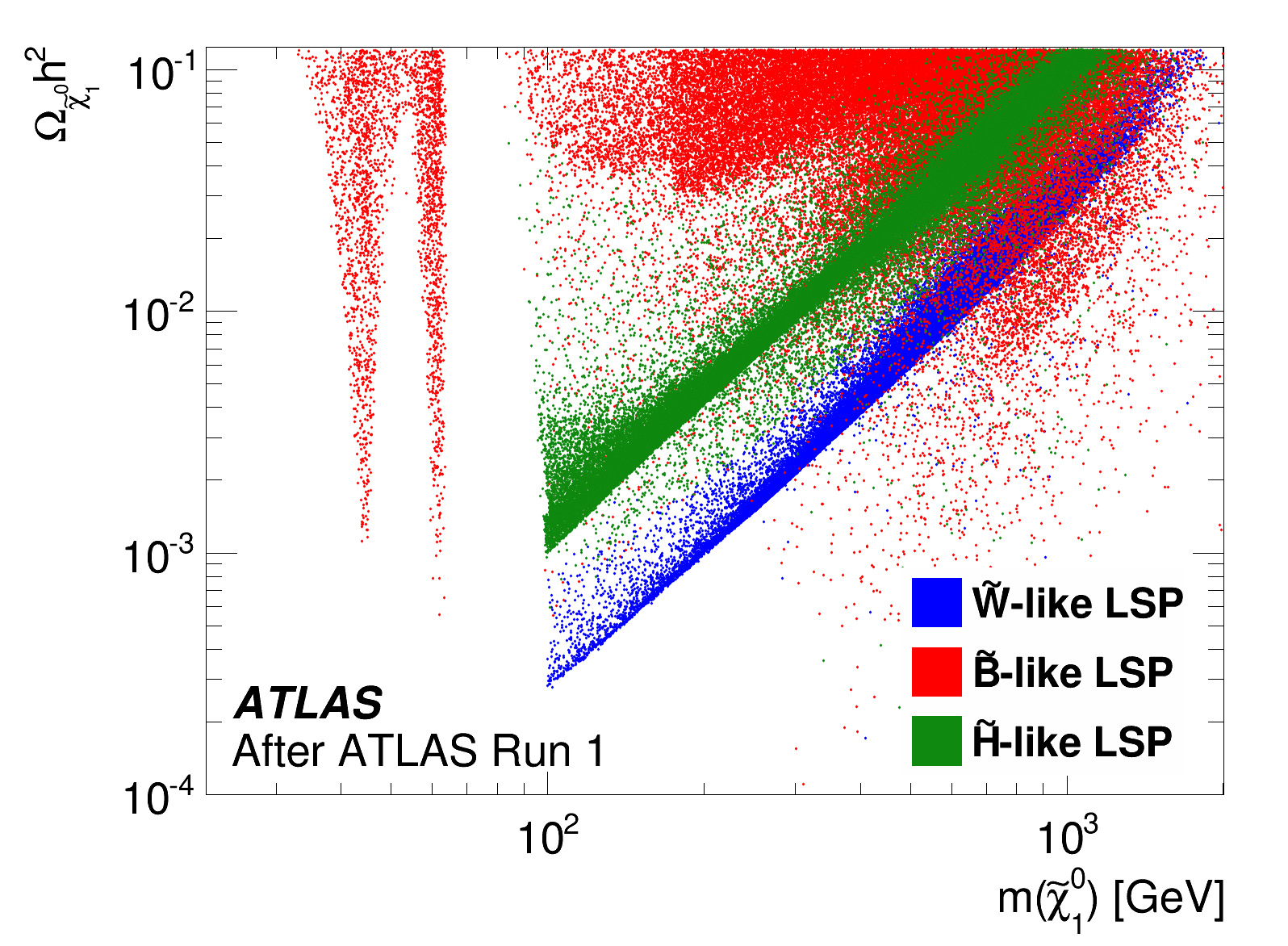}}\\
\end{tabular}
\end{center}
\caption{\label{fig:lsptype_omega_mlsp}
The density of pMSSM points projected onto the plane of dark matter 
relic density versus LSP mass, before and after the constraints from the search analyses.
The colours labelling the different LSP types, as defined in Table~\ref{table:lsptype}.
}
\end{figure} 

The nature of the LSP has a strong influence on the expected dark matter relic density. 
For the pMSSM points, the initial mass spectrum of the LSP --
before applying any ATLAS SUSY search constraints --
is sculpted by the requirement that the dark matter relic density,
as predicted from thermal production and annihilation calculations in the early universe,
should not be larger than that observed.
As discussed in Section~\ref{sec:dmconstraint}, the LSP is not required to
account for all of the dark matter density, since other particles may contribute.

The effect of the relic density preselection can be seen in 
Figure~\ref{fig:lsptype_omega_mlsp}(a), which shows
the density of pMSSM points in the plane of the $\tilde{\chi}_1^0$ relic density 
(\omegahsquared) versus the mass of the $\tilde{\chi}_1^0$.
The model points with a bino-like LSP are shown in red, while those with wino-like 
and Higgsino-like LSPs are in blue and green respectively.
The features at low LSP mass are due to the effective annihilation of LSPs through $s$-channel $Z$ or Higgs bosons
-- the so-called $Z$-funnel and Higgs-funnel regions.
There are few wino- or Higgsino-dominated LSPs at low mass
since in such cases the $\tilde{\chi}_1^0$ is expected to be 
accompanied by an almost degenerate chargino, which would have been observed at LEP~\cite{LEP_Chargino}.
Most of the models with wino- and Higgsino-dominated LSPs 
lie on bands which are almost straight lines on the logarithmic plot,
since the thermally averaged annihilation cross-section is expected to be 
proportional to the inverse square of the LSP mass, 
resulting in almost exact proportionality 
between $\omegahsquared$ and $m(\tilde{\chi}_1^0)^2$.

Figure~\ref{fig:lsptype_omega_mlsp}(b) shows the distribution of model points
in the same plane after applying the constraints from the ATLAS SUSY searches.
Considering first the bino-dominated LSP model points, one observes that 
model points are excluded by the ATLAS searches across a wide range of LSP masses
(and expected relic densities).
It is found that the ATLAS searches have some sensitivity up to $m(\neutralino)\lsim\SI{800}{\GeV}$,
while at low bino mass about two-thirds of the LSP model points in the $Z$- and Higgs-funnel
regions are excluded.
For wino-dominated LSPs, the overall range of sensitivity in $m(\neutralino)$ again extends
up to about \SI{800}{\GeV}.
Model points with $m(\neutralino)\lsim\SI{220}{\GeV}$ are particularly depleted by the ATLAS searches with 
about 80\% of model points with wino-like LSPs in this mass range being
excluded, mostly by the sensitivity of the \DisappearingTrack{} analysis to the charged wino.
For Higgsino-dominated LSPs, the ATLAS sensitivity is smaller as the $\chargino$--$\neutralino$ mass splittings
are mostly too large to have an observable $\chargino$ lifetime.

\begin{figure}
\begin{center}
\begin{tabular}{cc}
\subfloat[Before ATLAS \runone]{\includegraphics[width=0.49\textwidth]{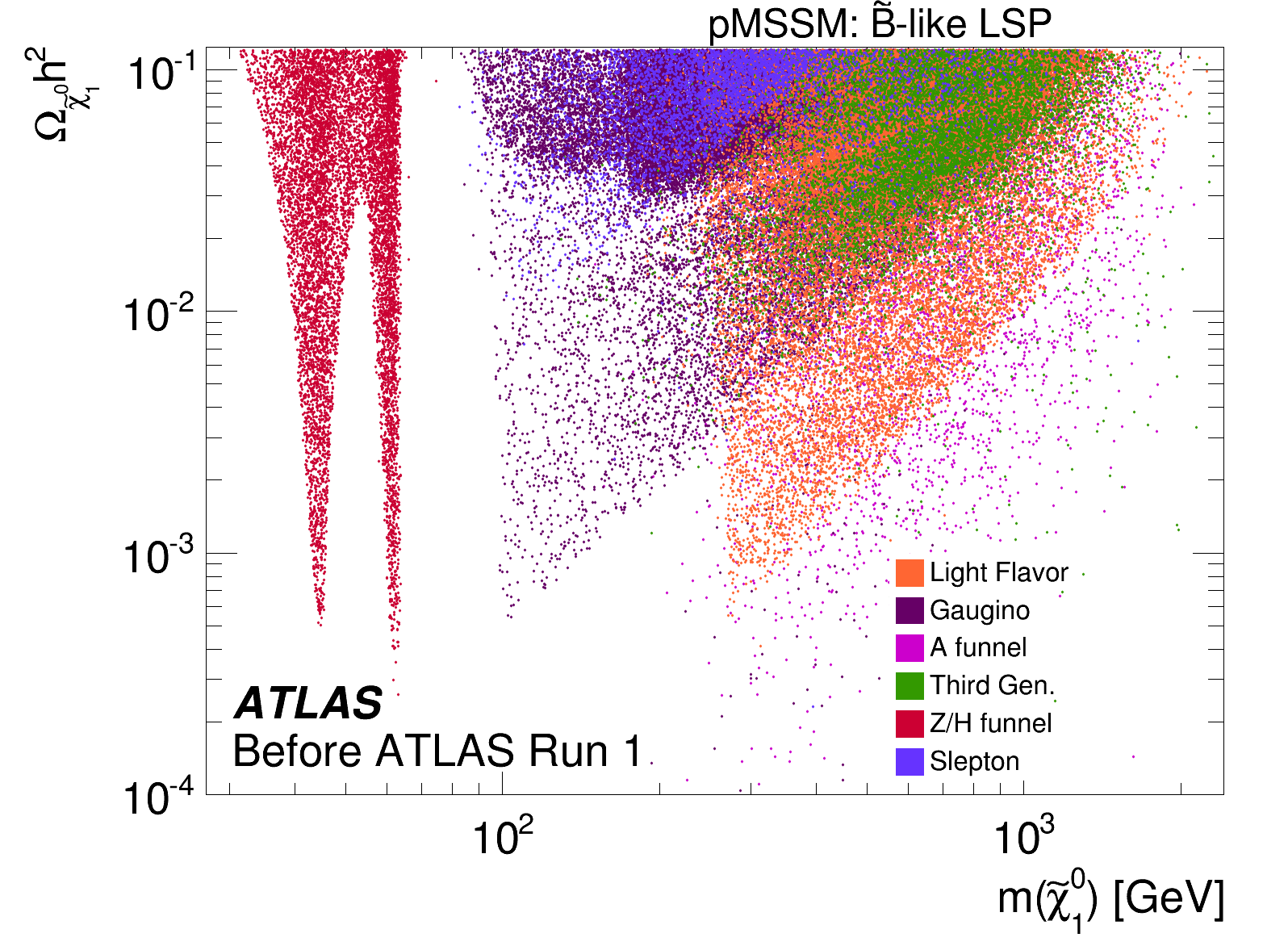}}&
\subfloat[After ATLAS \runone]{\includegraphics[width=0.49\textwidth]{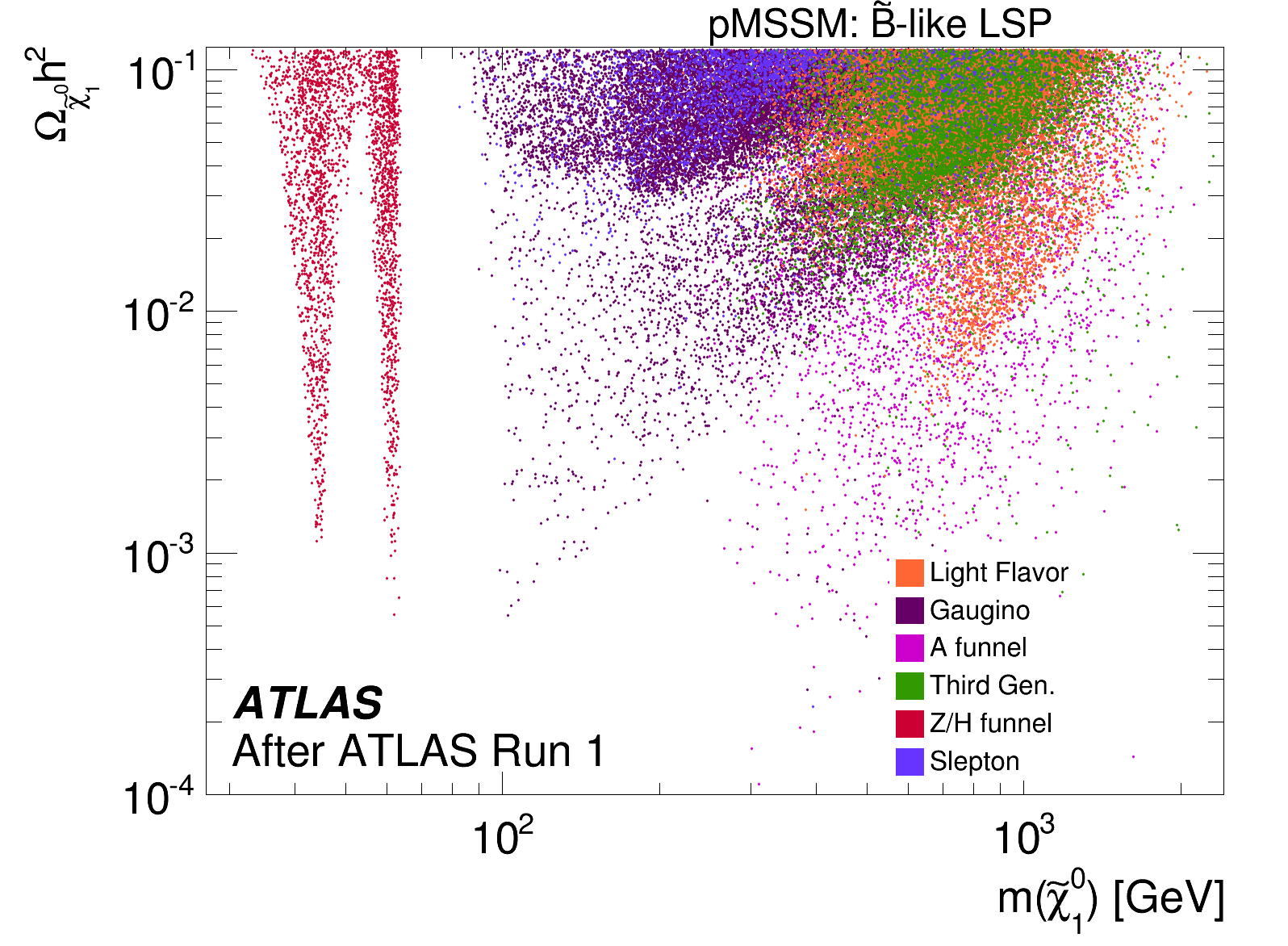}}\\
\end{tabular}
\end{center}
\caption{\label{fig:density_mlsp_bino_split}
The density of pMSSM points on the plane of relic density versus LSP mass, before and after the ATLAS constraints, 
for the bino-like LSP models.
The model points are colour coded by annihilation mechanism as described in the text.
`Gaugino' refers to electroweak gaugino coannihilation, while `Light Flavour' refers to coannihilation either against squarks of the first two generations or against gluinos.
}
\end{figure} 

Focusing on model points with bino-dominated LSPs, 
Figure~\ref{fig:density_mlsp_bino_split}
shows the model point density before and after ATLAS \runone{} searches.
In this figure the colour code now corresponds to the dominant annihilation mechanism of the 
dark matter -- for example an orange point has at least one squark of the first two generations, or the gluino, 
close in mass to the LSP, allowing annihilation mechanisms such as
$\squark + \neutralino \to q + \gamma$ to proceed effectively.
The vertical cut-offs for each coannihilator 
correspond to the preselection from 
previous experimental constraints -- 
for example charged sparticles lighter than \SI{100}{\GeV} are forbidden.
The different minima seen for the relic dark matter energy density for each coannihilator correspond to the
different coupling strengths.

The ATLAS searches are seen to be particularly 
sensitive to model points with light-flavour squark or gluino coannihilators.
One would expect such model points,
which have at least one coloured particle near-degenerate with the LSP, to produce
events with missing transverse momentum, 
and soft jets from the coloured sparticle decays.
Such model points are difficult to observe in the inclusive analyses,
but when produced in association with ISR can be 
observed through the monojet analyses~\cite{Aad:2014stoptocharm,Aad:2015zva}.

\begin{figure}
\vspace*{-0.5cm}
\begin{center}
\begin{tabular}{cc}
\subfloat[Spin independent, before ATLAS \runone]{\includegraphics[width=0.48\textwidth]{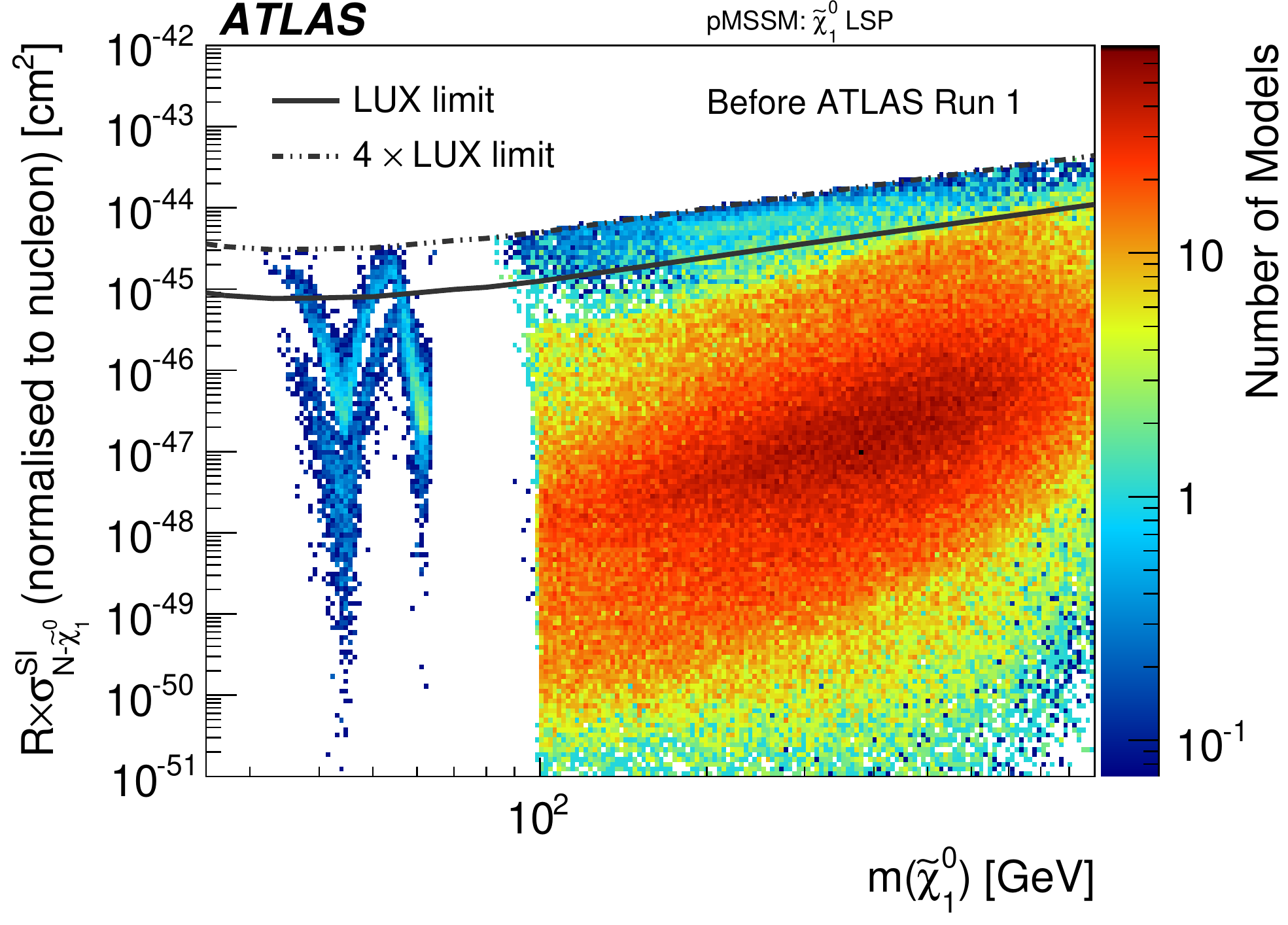}}&
\subfloat[Spin independent, after ATLAS \runone]{\includegraphics[width=0.48\textwidth]{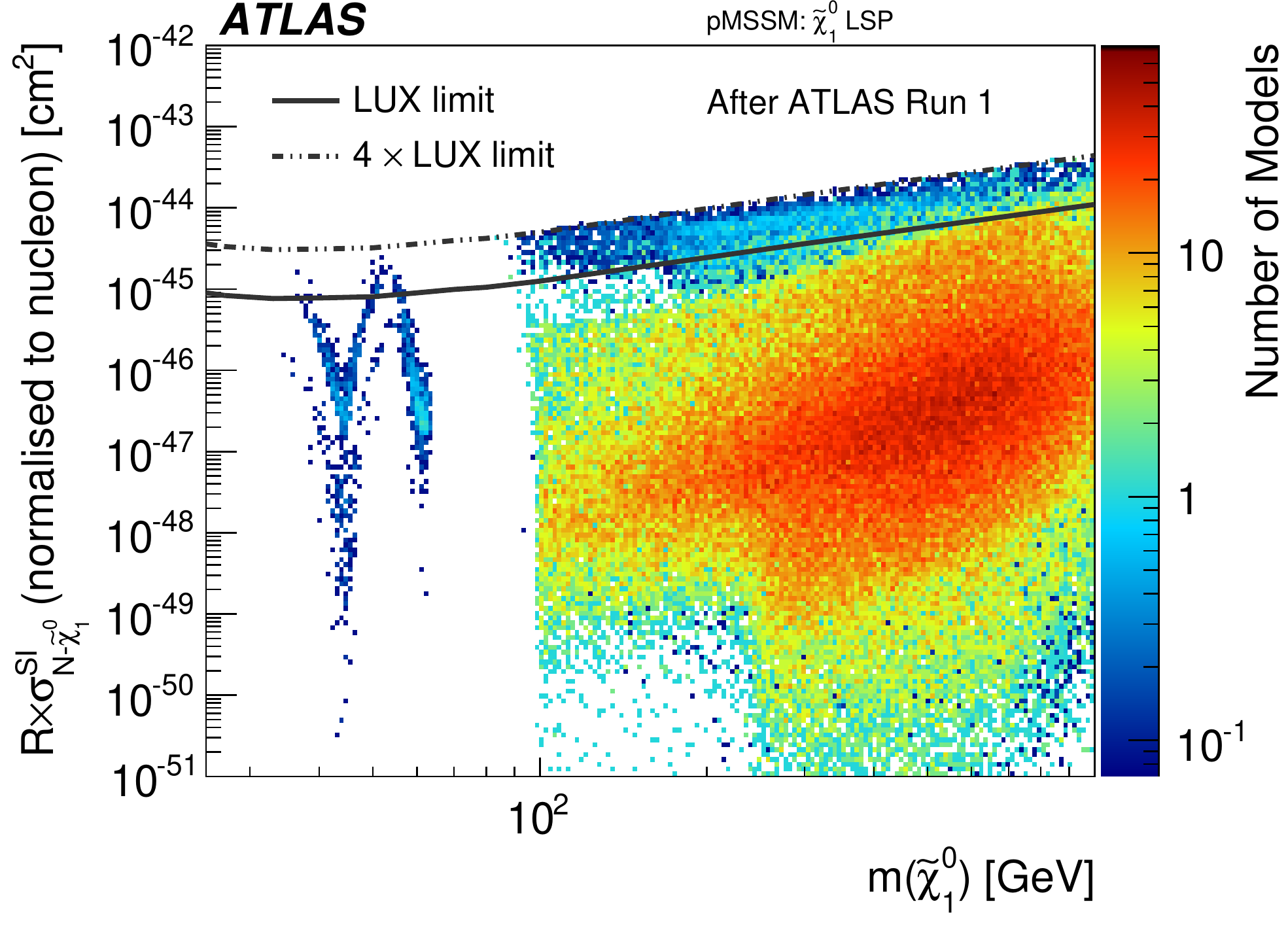}}\\
\subfloat[Spin dependent, before ATLAS \runone]{\includegraphics[width=0.48\textwidth]{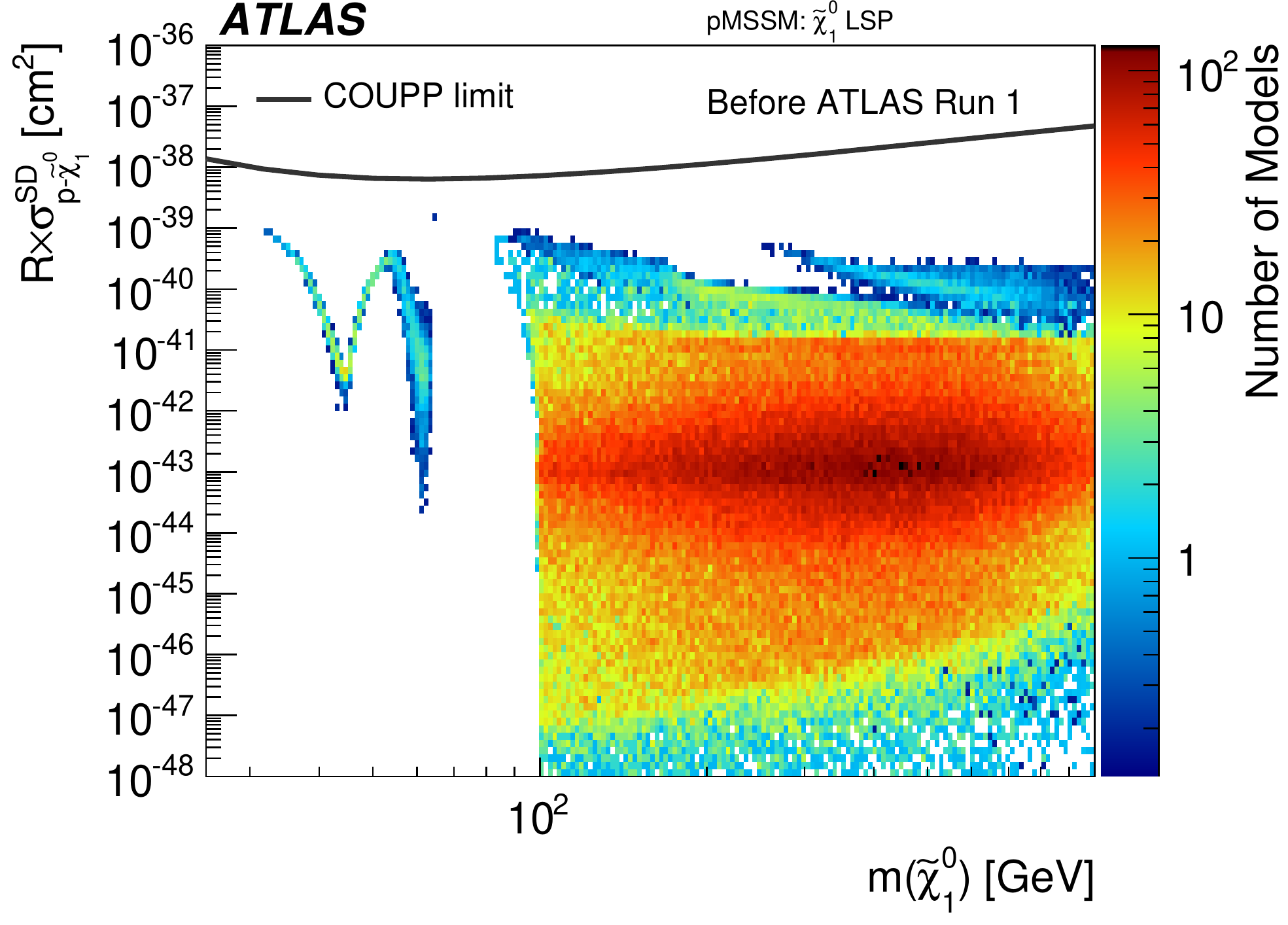}}&
\subfloat[Spin dependent, after ATLAS \runone]{\includegraphics[width=0.48\textwidth]{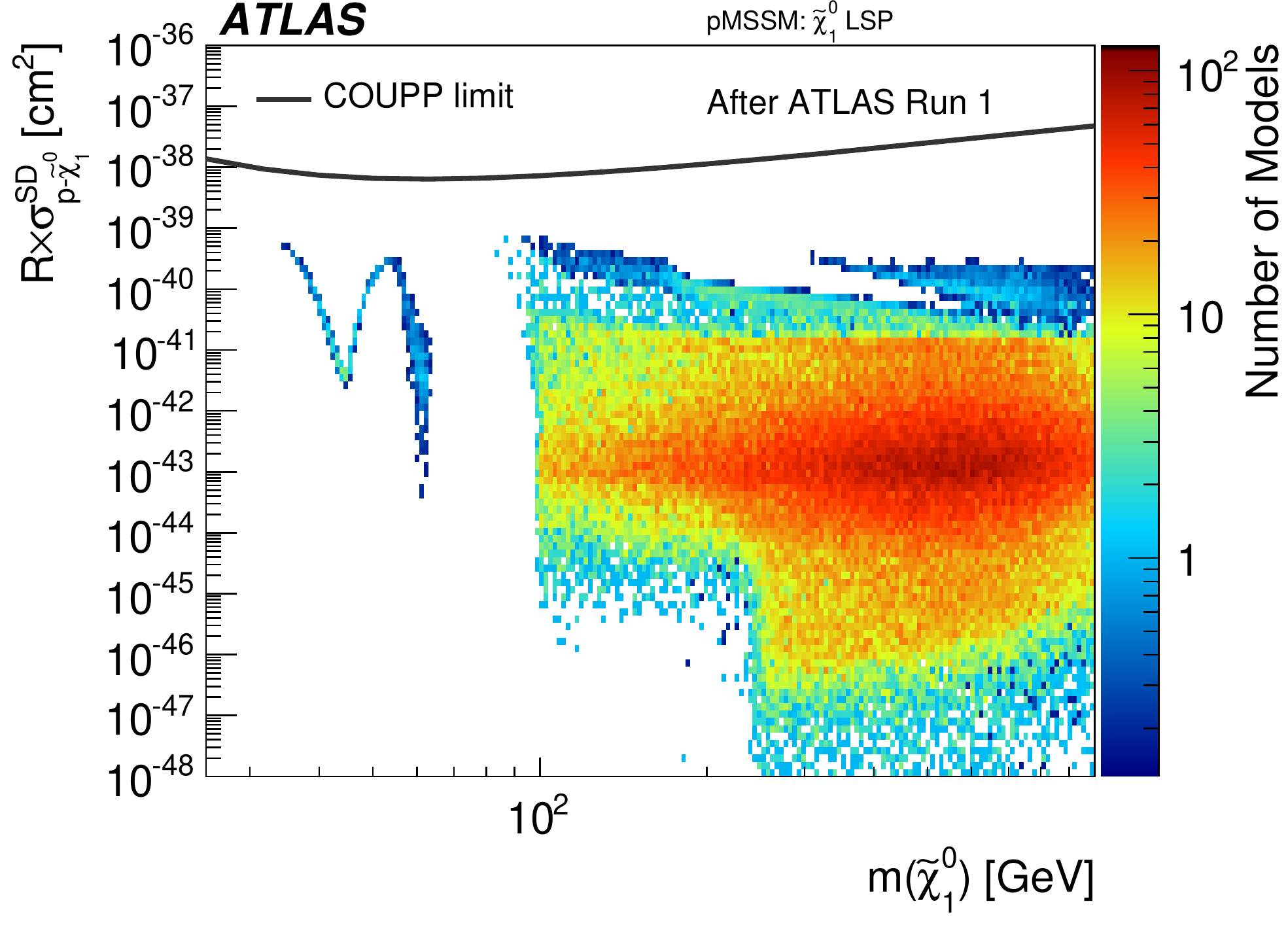}}\\
\end{tabular}
\end{center}
\caption{\label{fig:Xenonxsec_vs_neutralino} 
Left, the distribution of model points generated; right, the distribution of
model points not excluded by ATLAS \runone{} searches, 
as projected onto the scaled
spin-independent (SI) interaction cross-section of nucleons with the neutralino versus
the neutralino mass. The cross-sections are scaled by a factor of
$R_\Omega = \omegahsquared / \Omega_{\rm Planck}h^2$. 
The calculated spin-independent interaction
cross-sections are a weighted average of the contributions from proton and
neutron scattering, corresponding to the Xenon atom (the target nucleus of the LUX experiment)
and normalised to one nucleon. 
The 90\% confidence limit~\cite{Akerib:2013tjd} from the LUX direct detection experiment 
is overlaid, in which it is assumed that the dark matter comprises only 
the LSP, with relic density as measured by the Planck Collaboration~\cite{Ade:2015xua}.
For the spin-dependent cross-sections, the calculated proton cross-section is shown.
It is compared to the direct detection limit from the COUPP experiment~\cite{Behnke:2012ys}.
}
\end{figure}

The limits from experiments searching for direct detection of dark matter
were taken into account when generating the pMSSM points.
Figure~\ref{fig:Xenonxsec_vs_neutralino} shows the
corresponding predictions together with the limits set by the measurements of
the spin-independent (SI) and the proton spin-dependent (SD) cross-sections by the 
LUX~\cite{Akerib:2013tjd} and COUPP~\cite{Behnke:2012ys} experiments respectively.
The direct detection limits include the additional uncertainties described in 
Section~\ref{sec:dmconstraint}.
In each case the interaction cross-section is scaled by a factor of
$R_\Omega = \omegahsquared / \Omega _{\mathrm{Planck}} h^2$,
since in this paper it is assumed that the LSP may be only one of a range of
possible contributors\footnote{These other possible contributors are not
described by the pMSSM points considered in this study.} to the dark matter abundance, 
whereas direct detection experiments interpret their results in a framework
in which the LSP completely saturates the relic density.

Figure~\ref{fig:Xenonxsec_vs_neutralino}(a) shows the model points projected onto the
plane of the spin-independent interaction cross-section versus the 
LSP mass. 
The direct detection limit from the LUX experiment is shown, 
in which, as already mentioned, the relic density of the colliding 
dark matter was assumed to saturate the value measured by the Planck Collaboration.
The only model points with LSP mass less than \SI{100}{\GeV} are those with
bino-like LSPs lying in the Higgs- or $Z$-funnel regions. 
Around 15\% of the bino-like model points that pass the preselection 
(Table~\ref{tab:constraints}) would have been excluded by the nominal LUX limit had 
the additional uncertainty scaling factor of four, as mentioned in Section~\ref{sec:dmconstraint}, not been applied.
Figure~\ref{fig:Xenonxsec_vs_neutralino}(b) shows the same space after 
removing model points excluded (at 95\%~CL) by the ATLAS \runone{} searches. 

Figure~\ref{fig:Xenonxsec_vs_neutralino}(c)
shows the spin-dependent cross-section, again as a function of the neutralino mass.
The plot also shows the corresponding direct detection
exclusion limit, this time from the COUPP experiment.
The same plot is shown after ATLAS \runone{} search constraints in
Figure~\ref{fig:Xenonxsec_vs_neutralino}(d). Again one can see that 
sensitivity from LHC searches can stretch to regions 
with cross-sections several orders of magnitude below those of the current best
direct detection experiments.
Similar conclusions are found for spin-dependent neutron cross-sections,
in this case the best direct detection limit being from Xenon-100~\cite{Aprile:2013doa}.

Figure~\ref{fig:Xenonxsec_vs_neutralino}(b) shows a roughly 
rectangular region excluded by ATLAS searches, 
bounded approximately by 
 $m(\neutralino)\lsim \SI{220}{\GeV}$ and 
$R_\Omega \times \sigma^{\rm SI}_{N-\neutralino} \lsim 10^{-48} {\rm cm}^2$.
A similar region is excluded in the SD case (Figure~\ref{fig:Xenonxsec_vs_neutralino}(d)), 
this time for 
$R_\Omega \times \sigma^{\rm SI}_{N-\neutralino} \lsim 10^{-44} {\rm cm}^2$
for the same LSP mass range.
These regions are dominated by models with an LSP that is almost pure wino.
The winos, having only very small Higgsino admixtures and thus little interaction with the Higgs boson, have small 
direct detection cross-sections. 
Such model points then have small $\chargino$--$\neutralino$ mass splittings,
and are tightly constrained by the \DisappearingTrack{} analysis,
since the lightest chargino (the charged wino) is sufficiently long-lived to be detectable
in that analysis.
As the degree of Higgsino mixing increases, the direct detection cross-sections
increase, and so does the LSP--NLSP mass splitting. This results in a more rapid NLSP decay, and lack of sensitivity by ATLAS.

As one would expect, the direct detection and LHC search techniques 
are found to be very complementary.
The ATLAS searches have sensitivity
for $m(\neutralino)$ up to about \SI{800}{\GeV},
and are particularly effective in constraining low-mass LSPs.
ATLAS shows sensitivity to many SUSY models with direct detection
cross-sections several orders of magnitude below current
direct detection limits.
On the other hand, direct detection experiments rule out 
large $R_\Omega \times \sigma$ up to higher LSP masses
than the LHC can access.

\subsection{Effect of ATLAS Higgs boson coupling measurements}\label{sec:results:higgscoupling}

\begin{figure}
\vspace*{-0.5cm}
\begin{center}
\begin{tabular}{cc}
\subfloat[Light Higgs boson $\kappa_b$]{\includegraphics[width=0.49\textwidth]{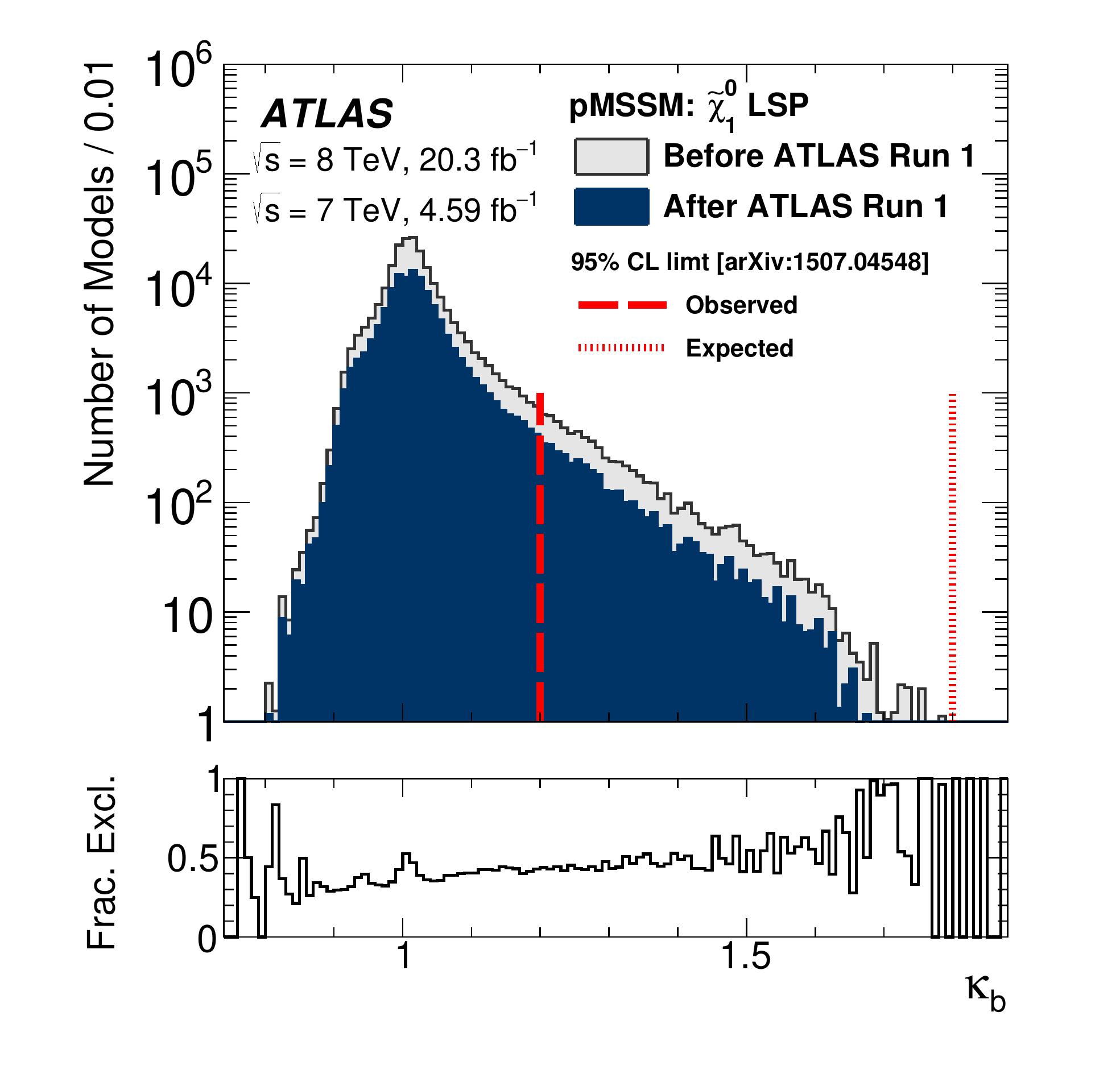}}
\subfloat[Light Higgs boson BR to $\neutralino\neutralino$]{\includegraphics[width=0.49\textwidth]{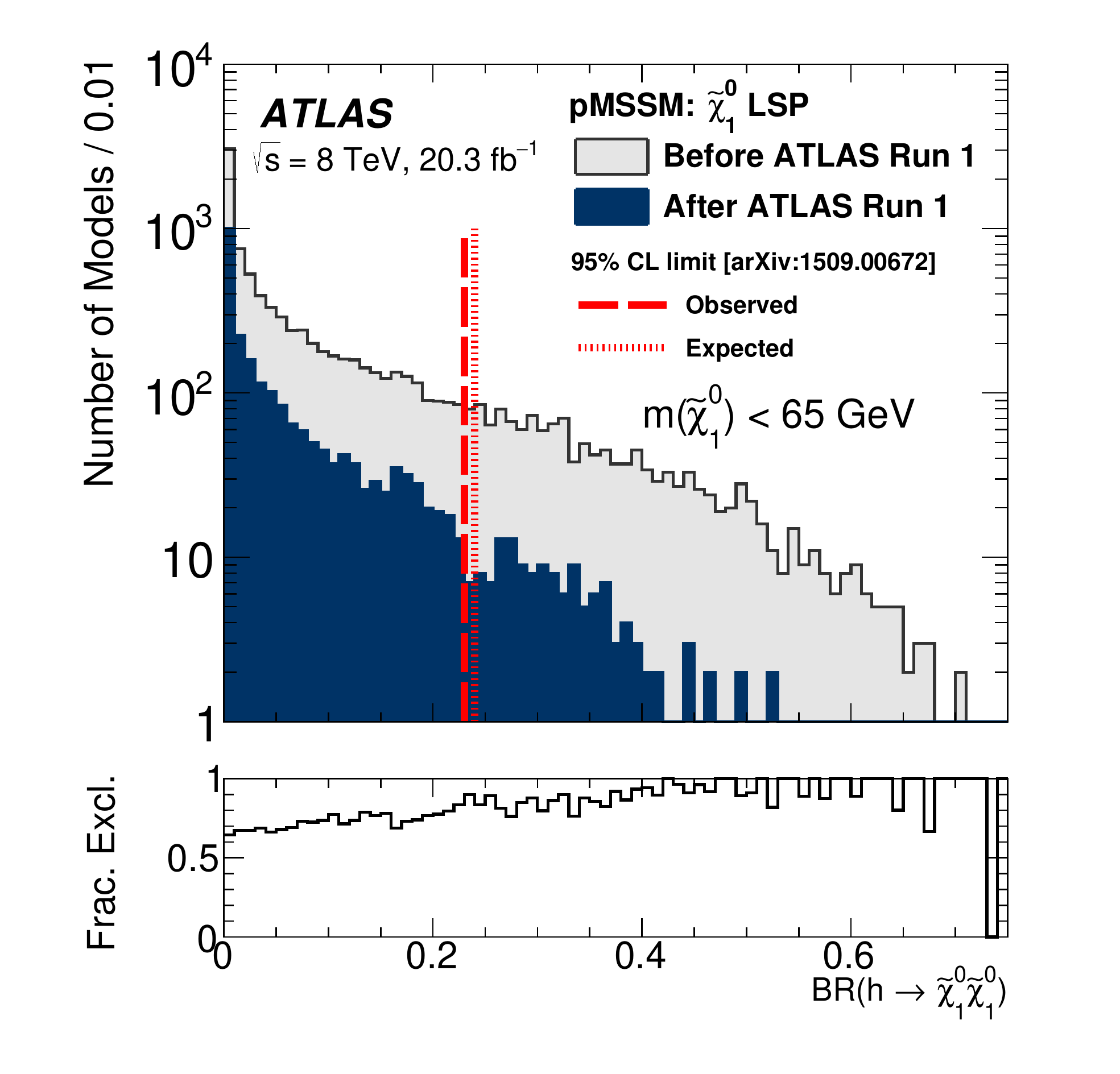}}
\end{tabular}
\caption{\label{fig:kappa_b_beforeAndAfterATLAS}
Distributions of pMSSM points before and after the constraints from the ATLAS direct searches for
observables related to the decays of the observed Higgs boson.
(a) Higgs--$b$-quark coupling ratio $\kappa_b$, for all LSP types. 
(b) Branching ratio of the Higgs boson to the LSP, only for those pMSSM points 
with bino-like LSPs and $m(\neutralino)<\SI{65}{\GeV}$.
The dashed lines show the observed and expected upper bounds of the 95\% confidence limits
on those parameters.
The subplots show the fraction of model points excluded by the 
ATLAS direct searches.
}
\end{center}
\end{figure}

Direct searches for SUSY particles are complemented by indirect searches 
for the effects of new particles and fields.
In SUSY models the presence of the second Higgs doublet generally 
results in light Higgs boson couplings that are modified with respect to their SM values.  
For example, enhancements to the couplings to down-type quarks 
and charged leptons are predicted at large $\tan\beta$.  
In addition, loop contributions
from virtual SUSY particles can produce an observable effect on lower-energy observables,
including Higgs boson decay rates.

Combined fits by ATLAS to the measured Higgs boson production and decay rates
provide a sensitive probe for such additional Higgs fields and 
for SUSY particles contributing either through loop effects or  
through invisible decays of the Higgs boson to unobserved neutralinos.
The relevant ATLAS analyses, statistical methods, and systematic uncertainties
are described in Refs.~\cite{Aad:2013wqa,ATLAS-HIGG-2014-06-002}. 
These measurements are based on up to \SI{4.8}{\ifb} of $pp$ collision data at
$\sqrt{s} = \SI{7}{\TeV}$ and up to \SI{20.3}{\ifb} at
$\sqrt{s} = \SI{8}{\TeV}$.
They are used to constrain coupling scale factors 
\[
\kappa_{i} = \sqrt{ \frac{\Gamma_{h}}{\Gamma_{ h,{\rm SM}}} \times { \frac{{\rm BR}(h\to i+i)}{{\rm BR}(h\to i+i)_{\rm SM}} }},
\]
where $\Gamma_h$ represents the total width of the observed Higgs boson, BR indicates its branching ratio, and a SM subscript indicates the Standard Model prediction.
The measurements of these scale factors were derived using the observed Higgs boson's visible decay channels as reported in Ref.~\cite{ATLAS-HIGG-2014-06-002}.  The upper limit on the invisible branching ratio, taking account of both visible and invisible Higgs boson decay channels is reported in Ref.~\cite{Aad:2015pla}.\footnote{These constraints are based directly on the log-likelihood test statistic, rather than a calculation of $CL_s$, hence these particular limits are not protected against disfavouring models where there is little sensitivity.} 
Most of the coupling parameters are not yet sufficiently well measured to have
an effect on the SUSY parameter space. The exceptions -- those two parameters
which are already sufficiently well measured to constrain the pMSSM -- are
discussed below.

The parameter $\kappa_b$ corresponds to the coupling of the Higgs field to $b$-quarks. 
Measuring a value of $\kappa_b$ differing significantly from unity would be a clear indication of physics beyond the SM, and an important constraint on the allowed pMSSM parameter space.
For the current ATLAS fit, the expected 95\% CL range is $|\kappa_b| < 1.8$. 
This expected region contains almost all the pMSSM points not excluded by the direct searches, 
so the expected constraint from $\kappa_b$ is negligible.
The current ATLAS observed fitted value for $\kappa_b$ is $0.62\pm 0.28$ at 
68\% CL~\cite{Aad:2015pla}.
This value is lower than, but consistent with, the SM value. 
The corresponding observed 95\% CL spans the range $|\kappa_b|<1.2$,
and is more restrictive than the expected range. 
As a result the observed fit disfavours those pMSSM points 
(see Figure~\ref{fig:kappa_b_beforeAndAfterATLAS}(a)) with $\kappa_b>1.2$, 
which typically are those with large $\tan\beta$.
Those models that are disfavoured at the 95\% CL represent a weighted fraction of 3.1\% of all the pMSSM points. 
This fraction is rather similar before and after the ATLAS \runone{} direct searches are considered.
 
The observed Higgs boson can decay to a pair of LSPs if those LSPs are sufficiently light.
This can lead to an enhanced branching ratio of the $h$ to the invisible final state.
The expected branching ratio for decay of a light Higgs boson via $ZZ^*$ to neutrinos is $\mathcal{O} (10^{-3})$, 
much smaller than the present experimental uncertainties, and so is negligible in this context.
No models with a sufficiently light wino- or Higgsino-like LSP 
satisfied the initial constraints of Section~\ref{sec:pMSSMselection},
so only models with a bino-dominated LSP need be considered in this context.
All such LSPs have some wino and Higgsino admixture, allowing them 
to couple to the Higgs boson.
Figure~\ref{fig:kappa_b_beforeAndAfterATLAS}(b) shows the calculated distribution of the 
lightest Higgs boson's invisible branching ratio for those pMSSM points
with bino-like LSPs and having $m(\neutralino) < \SI{65}{\GeV}$.
To account for finite-width effects, this selection allows for models with LSP masses a 
little above $m(h)/2$. 
These light-LSP models represent 4.5\% of the model points with bino-like LSPs
that survive the ATLAS searches in Table~\ref{tab:susySearches}.
The observed ATLAS bound on the $h$ invisible branching ratio is 
${\rm BR}(h\to \neutralino+\neutralino)<0.23$ at the 95\% CL~\cite{Aad:2015pla}.
This observed value is close to the expected upper limit of 0.24.
Considering only those pMSSM points with a bino-like LSP lighter than \SI{65}{\GeV}, 
and which were not excluded by the direct searches,
the fraction disfavoured at the 95\% CL by the invisible branching ratio fit is
6.4\%.
The disfavoured points correspond to a weighted fraction of 0.0056\% of all the pMSSM points
(with any LSP type) not excluded by the ATLAS direct searches.
The corresponding expected fractions are 6.1\% (of those surviving models having bino-like LSPs 
with $m(\neutralino) < \SI{65}{\GeV}$) 
and 0.0053\% (weighted by LSP type) of all surviving models.

There is no overlap between the models disfavoured by the $\kappa_b$ measurement
and those disfavoured by the Higgs boson invisible branching ratio.
This demonstrates that these indirect searches for new physics
complement the direct searches in different ways 
by constraining different parts of the pMSSM space.

\subsection{Impact of ATLAS searches on precision observables}\label{sec:results:precision}

\begin{figure}
\vspace*{-0.5cm}
\begin{center}
\begin{tabular}{cc}
\subfloat[Light Higgs boson mass]{\includegraphics[width=0.49\textwidth]{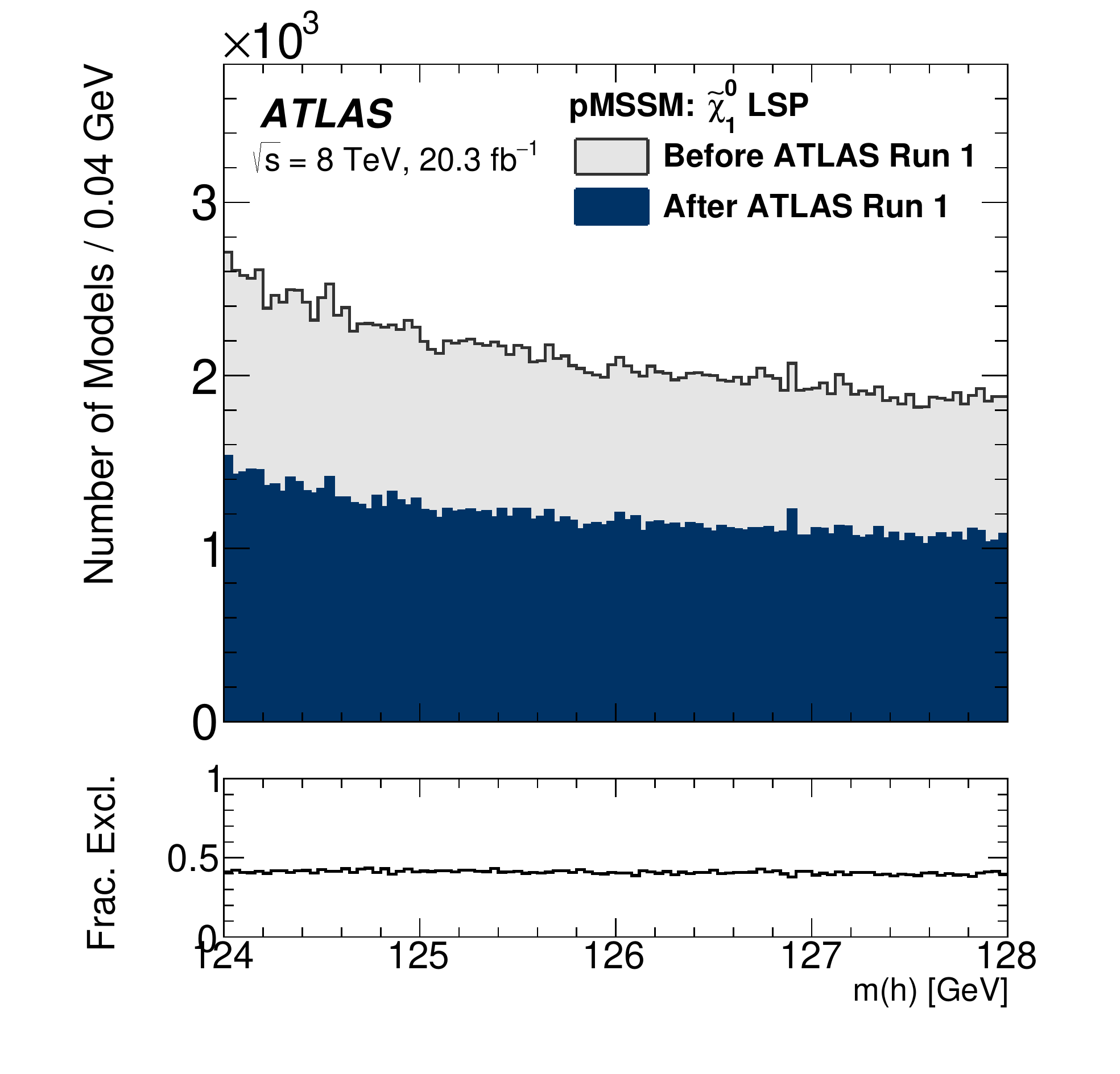}}&
\subfloat[$B_s \to \mu\mu$]{\includegraphics[width=0.49\textwidth]{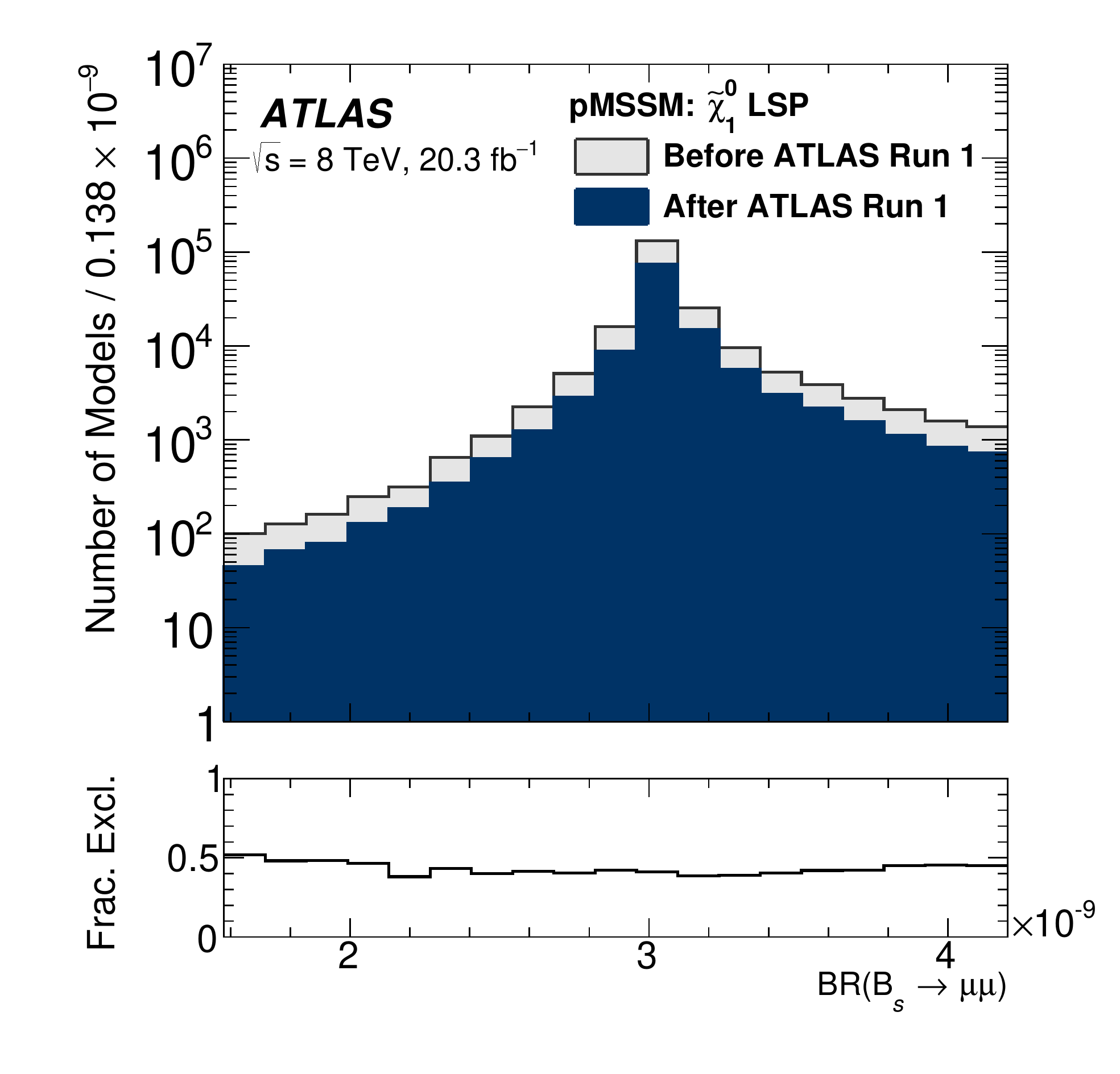}}\\
\subfloat[$b \to s + \gamma$]{\includegraphics[width=0.49\textwidth]{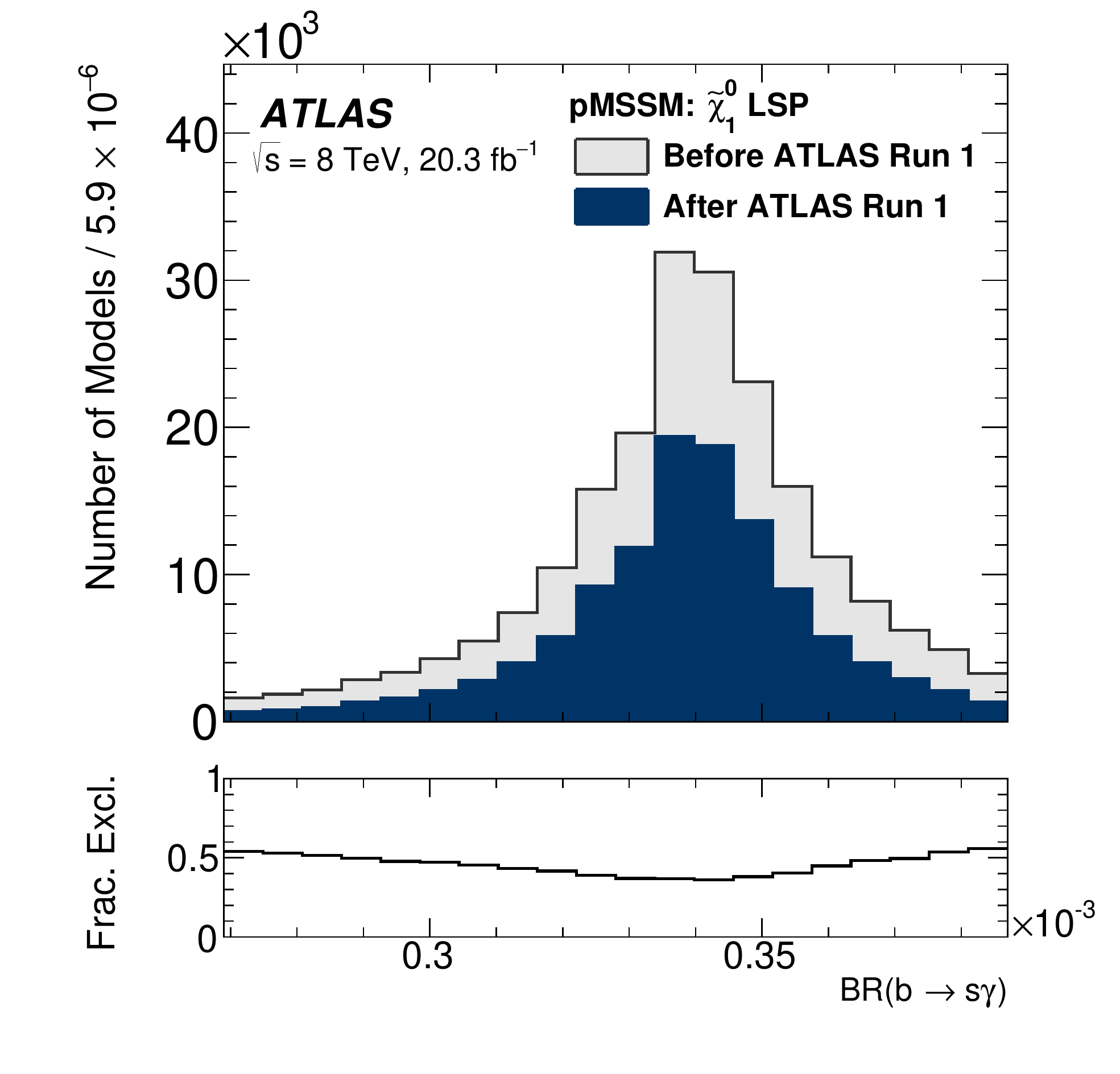}}&
\subfloat[Muon magnetic moment]{\includegraphics[width=0.49\textwidth]{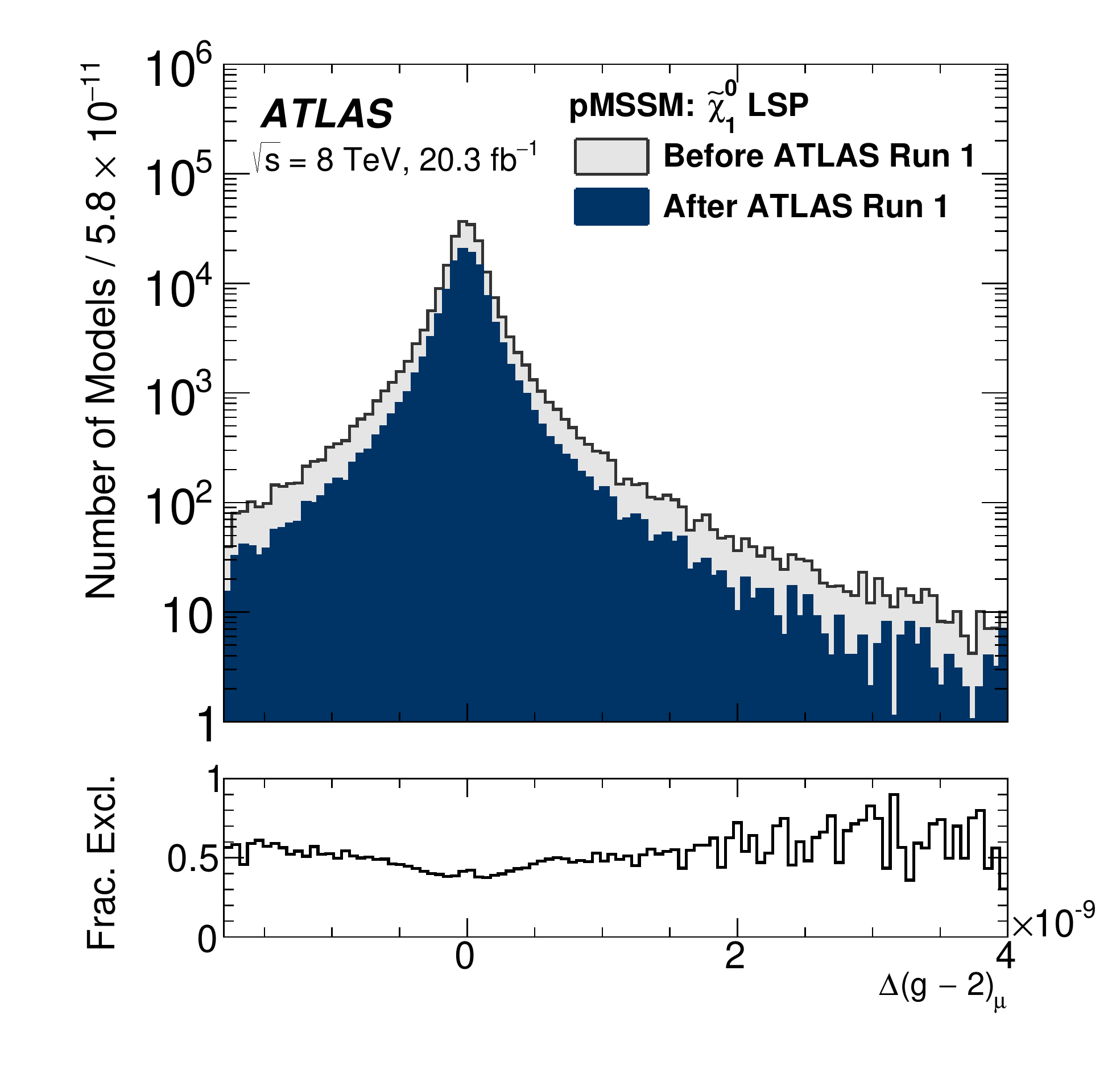}}\\
\end{tabular}
\end{center}
\caption{\label{fig:BrBsmumu_m_h_BRbsg_gminus2_beforeAndAfterATLAS}
Distributions of model points before and after applying ATLAS searches for various
precision observables. Figure  
(a) shows the mass of the light (SM-like) Higgs boson, $m(h)$,
(b) shows the branching ratio of $B_{s} \rightarrow \mu + \mu$,
(c) is the branching ratio of $b \rightarrow s + \gamma$ 
and (d)  is the difference between the predicted value of $g-2$ and the SM value, $\Delta(g-2)_\mu$.
The subplots show the fraction of model points excluded by ATLAS as a function of the observables.
}
\end{figure}

Loop contributions from SUSY particles can also affect lower-energy observables.
The measured values of several precision observables  
were taken as prior constraints on the pMSSM space sampled,
as explained in Section~\ref{par:precision}.

The effect of the ATLAS \runone{} searches on the 
distribution of pMSSM points as projected onto the 
expected values of these precision observables is shown in 
Figure~\ref{fig:BrBsmumu_m_h_BRbsg_gminus2_beforeAndAfterATLAS}.
In each case the most noticeable feature is that the ATLAS
direct searches remove pMSSM points rather uniformly
across the space of each precision variable, demonstrating
the complementarity of the searches.
The direct searches from \runone{} have placed tight constraints
on the strongly interacting sector of the pMSSM, whereas
the precision variables depend on other sparticles,
for example the smuon and electroweakino masses,
which are less tightly constrained by direct searches so far.

\begin{figure}
\centering
\includegraphics[width=0.6\textwidth]{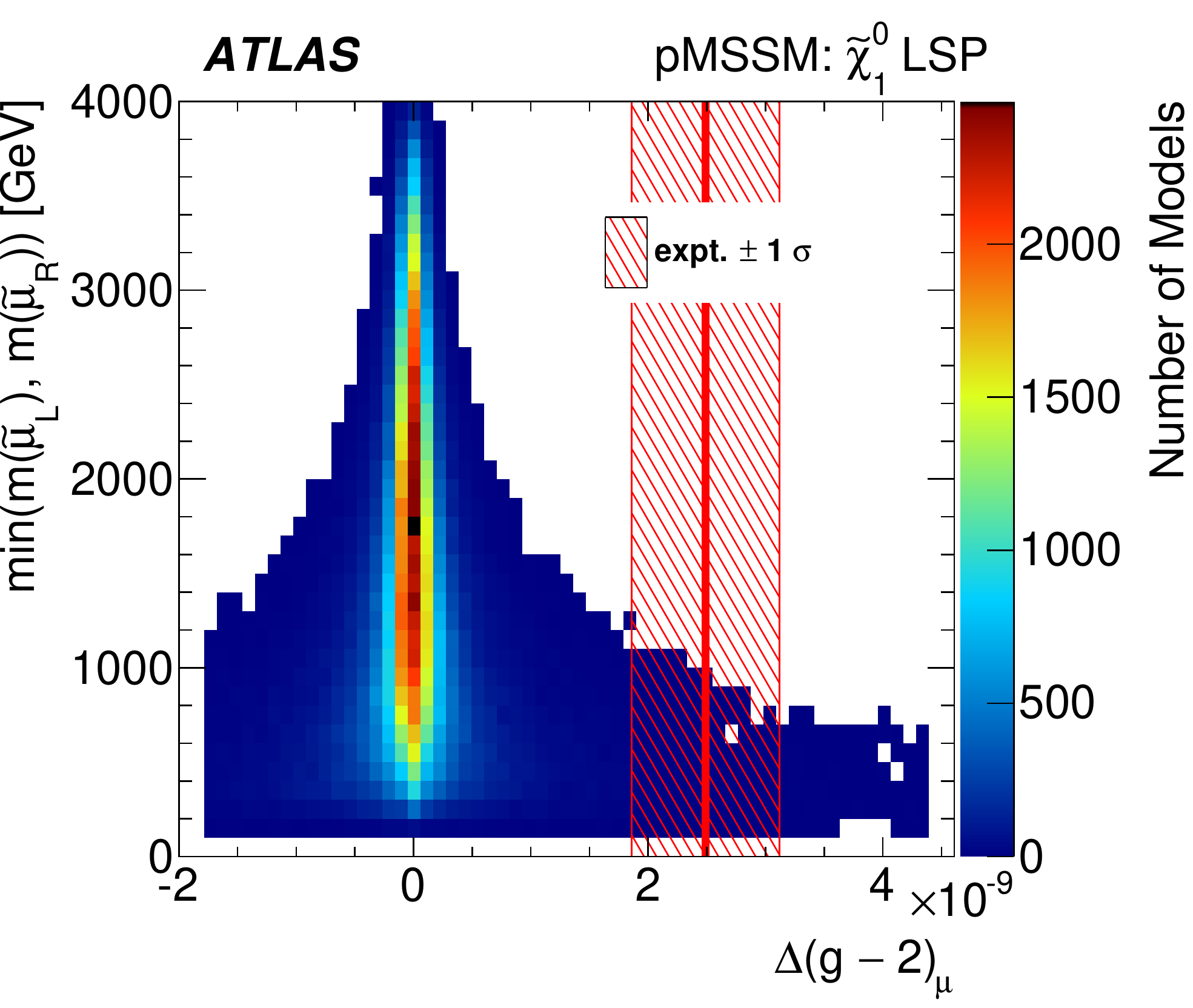}
\caption{
\label{fig:gminus2}
Distribution of model points in the 
plane of the mass of the lightest left-
or right-handed smuon versus $\Delta(g-2)_\mu$. 
The experimental measurement is overlaid as the
hatched band~\cite{Aoyama:2012wk}. 
}
\end{figure}

Figure~\ref{fig:BrBsmumu_m_h_BRbsg_gminus2_beforeAndAfterATLAS}
supports the statement in 
Section~\ref{par:collider} that the exclusion power of the ATLAS searches does
not depend 
significantly on the mass range used for the light (SM-like) Higgs boson, 
the fraction of excluded model points being remarkably flat.
The other three distributions: ${\rm BR}(B_s \to \mu\mu)$, ${\rm BR}(b\rightarrow s \gamma)$, 
and the $\Delta(g-2)_\mu$ also show good exclusion by ATLAS for all values of the observables.
It is noticeable that the majority of the pMSSM points
have only small SUSY contributions to these observables,
so those model points would not be expected to be 
discoverable by using those measurements in the near future. 
Nevertheless there exists, for each precision measurement, 
a small number of model points with 
SUSY contributions large enough to be discovered by each.
One observes that ATLAS also tends to have slightly larger sensitivity
to these `tail' model points.
This increased sensitivity is not surprising
since these model points are generally those with some lighter SUSY particles 
contributing to the precision measurements via loop diagrams.
This correlation becomes more clear in Figure~\ref{fig:gminus2},
which shows the number of generated model points in the 
plane of the $\Delta(g-2)_\mu$ versus the mass of the lighter 
left- or right-handed smuon ($\min(m(\tilde{\mu}_{\rm L}) ,m(\tilde{\mu}_{\rm R}))$).
The experimentally measured value~\cite{Aoyama:2012wk}, 
represented by the hatched band,
only overlaps a region where $m(\tilde{\mu}) < 1$\,\TeV.
As noted previously in Section~\ref{par:precision},  
the allowed range used for $\Delta(g-2)_\mu$ in this paper
is the union of the $3\sigma$ intervals around the SM value
 and the experimental measurement.
The experimentally measured value, if confirmed, 
would be a powerful constraint on the space.

\subsection{Fine-tuning}

\begin{figure}
\begin{center}
\begin{tabular}{cc}
\subfloat[All LSP types]{\includegraphics[width=0.49\textwidth]{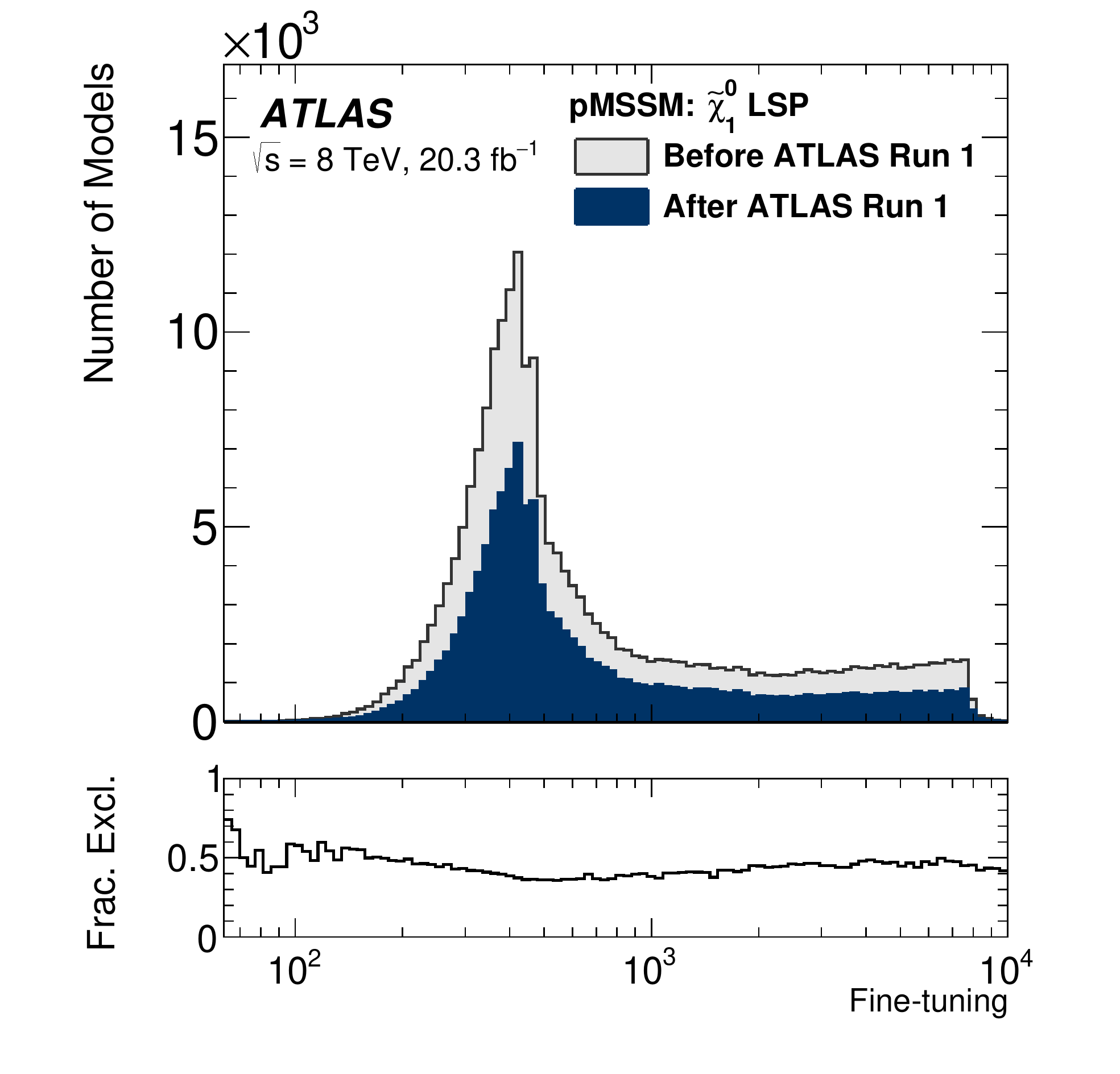}}&
\subfloat[Bino-like LSPs]{\includegraphics[width=0.49\textwidth]{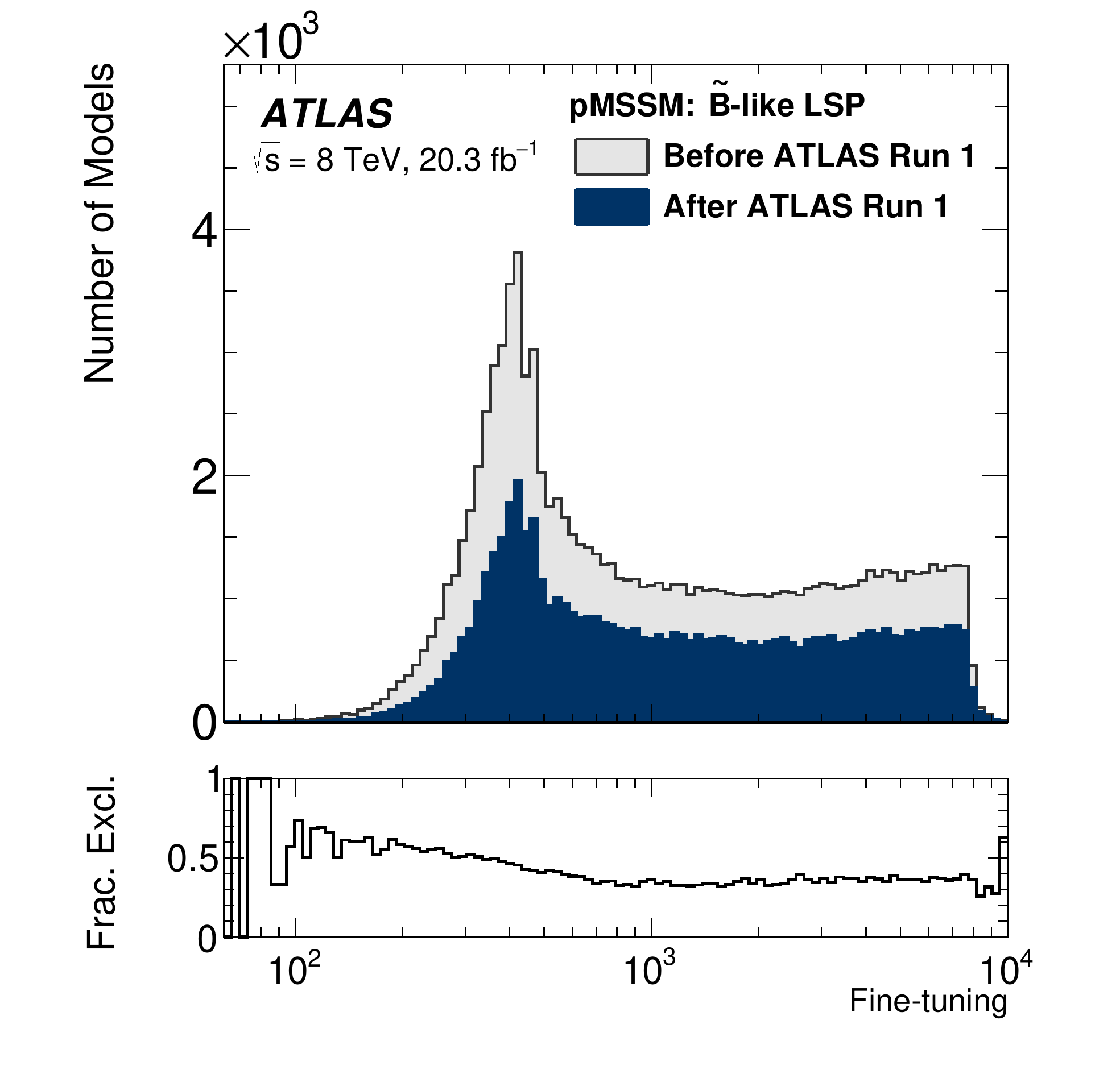}}\\
\subfloat[Wino-like LSPs]{\includegraphics[width=0.49\textwidth]{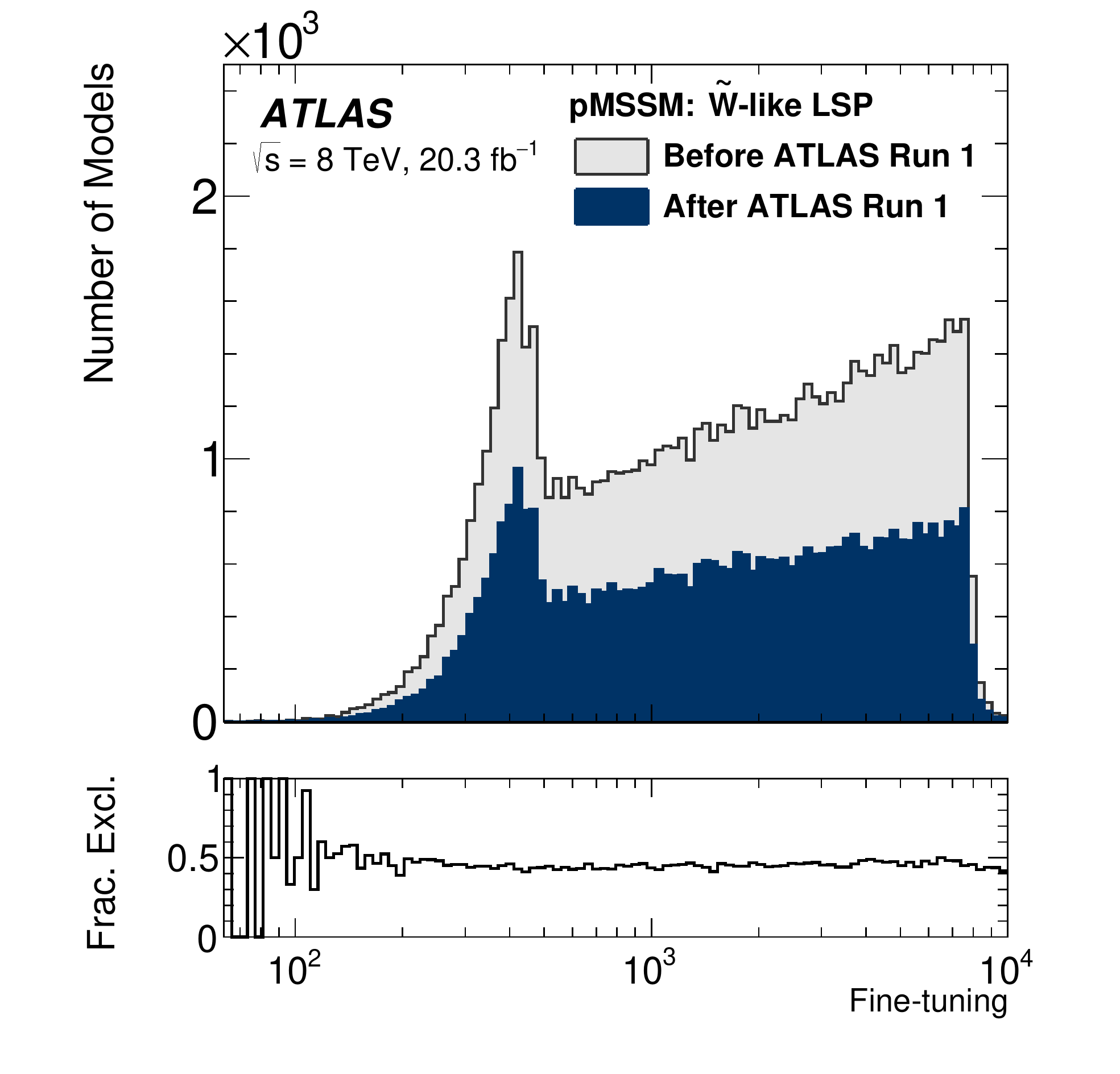}}&
\subfloat[Higgsino-like LSPs]{\includegraphics[width=0.49\textwidth]{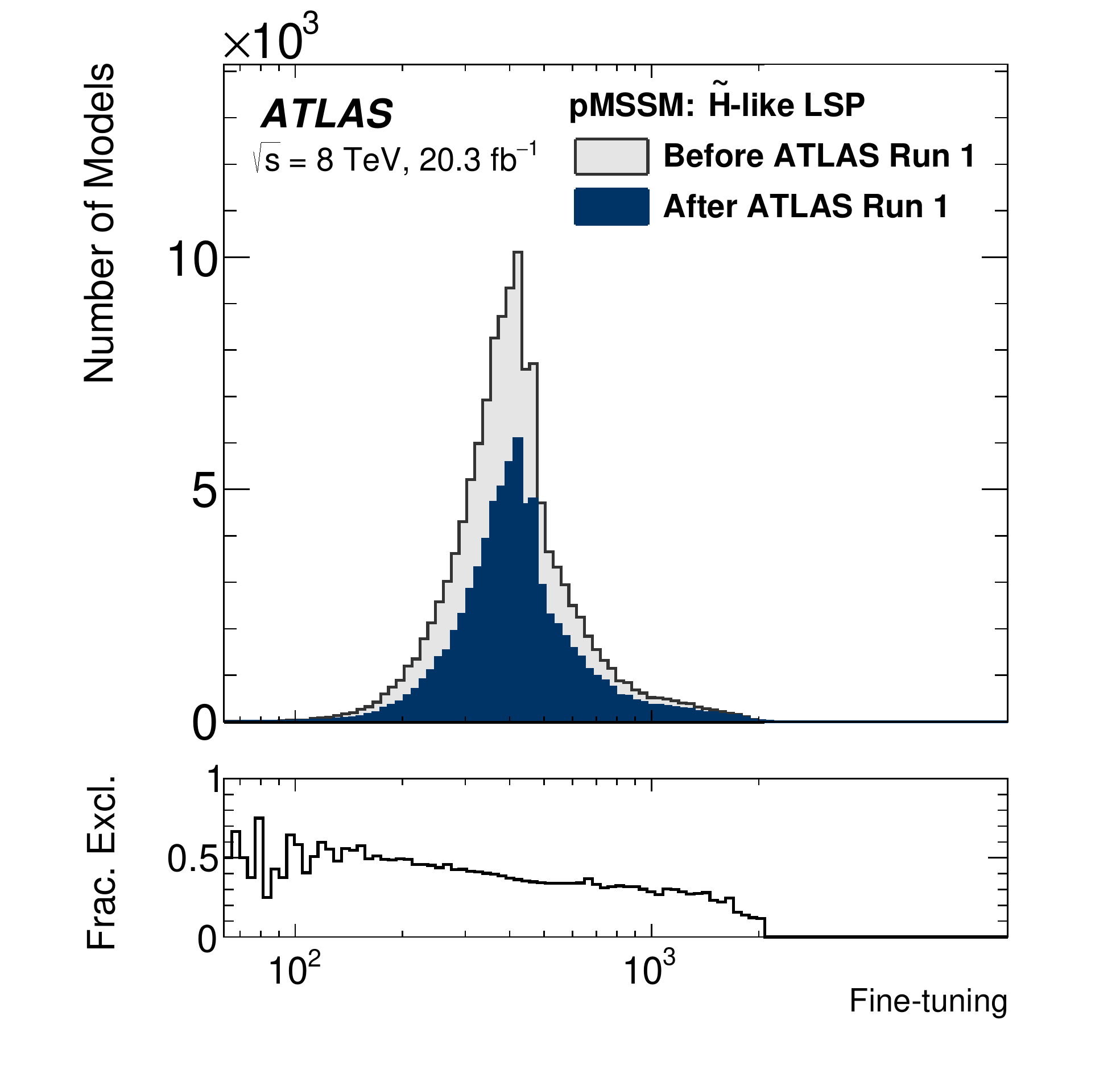}}\\
\end{tabular}
\end{center}
\caption{\label{fig:finetuning_beforeAndAfterATLAS} 
Distribution of fine-tuning (as defined in Ref.~\cite{Barbieri:1987fn}), 
before and after ATLAS exclusion. The subplots show
the fractions of model points excluded by the ATLAS \runone{} searches.
}
\end{figure} 

\begin{figure}
\begin{tabular}{cc}
\subfloat[Point 18898934 (fine-tuning 56)]{\includegraphics[width=0.48\textwidth,height=0.4\textwidth]{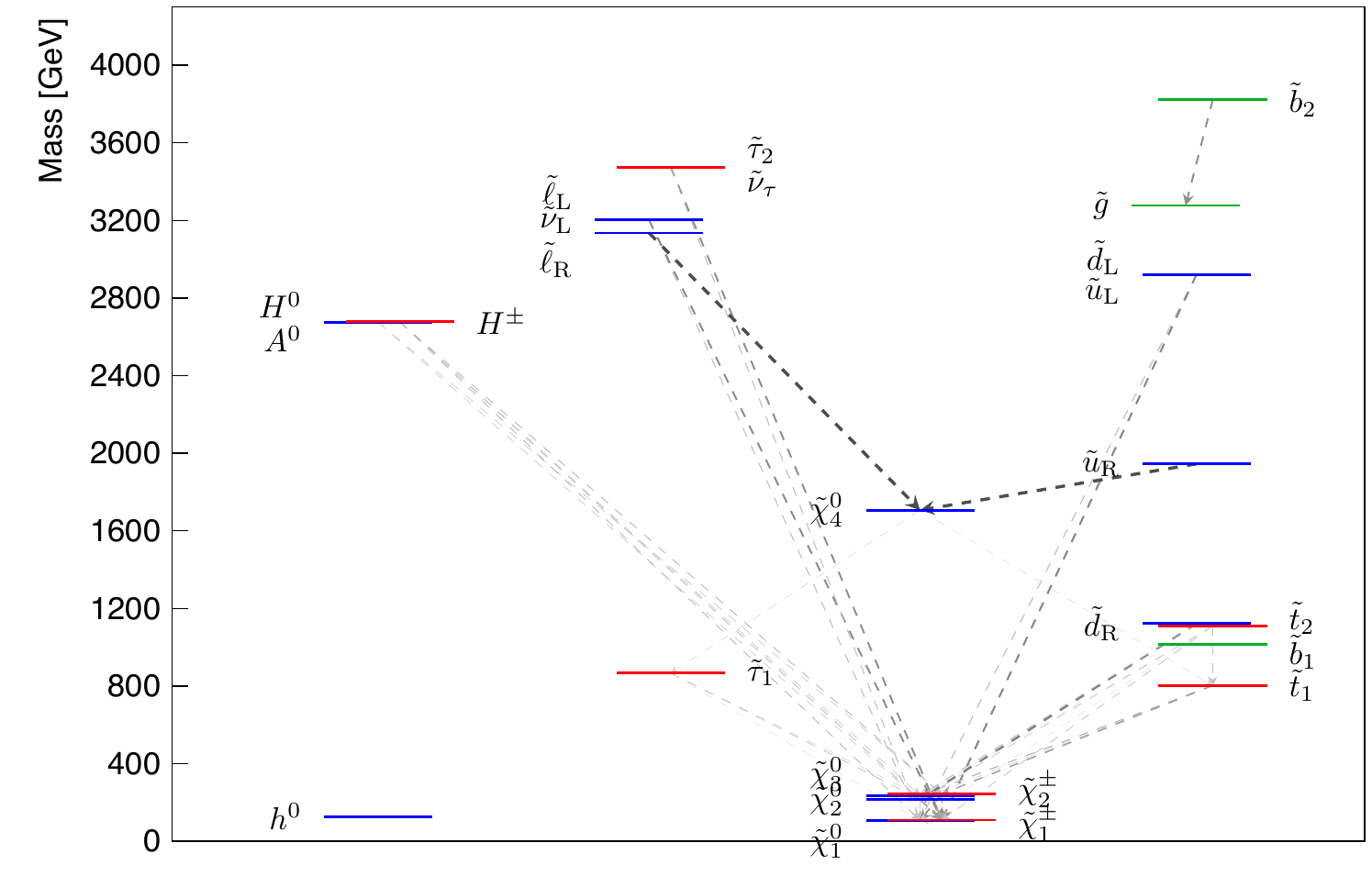}} & 
\subfloat[Point 10407816 (fine-tuning 57)]{\includegraphics[width=0.48\textwidth,height=0.4\textwidth]{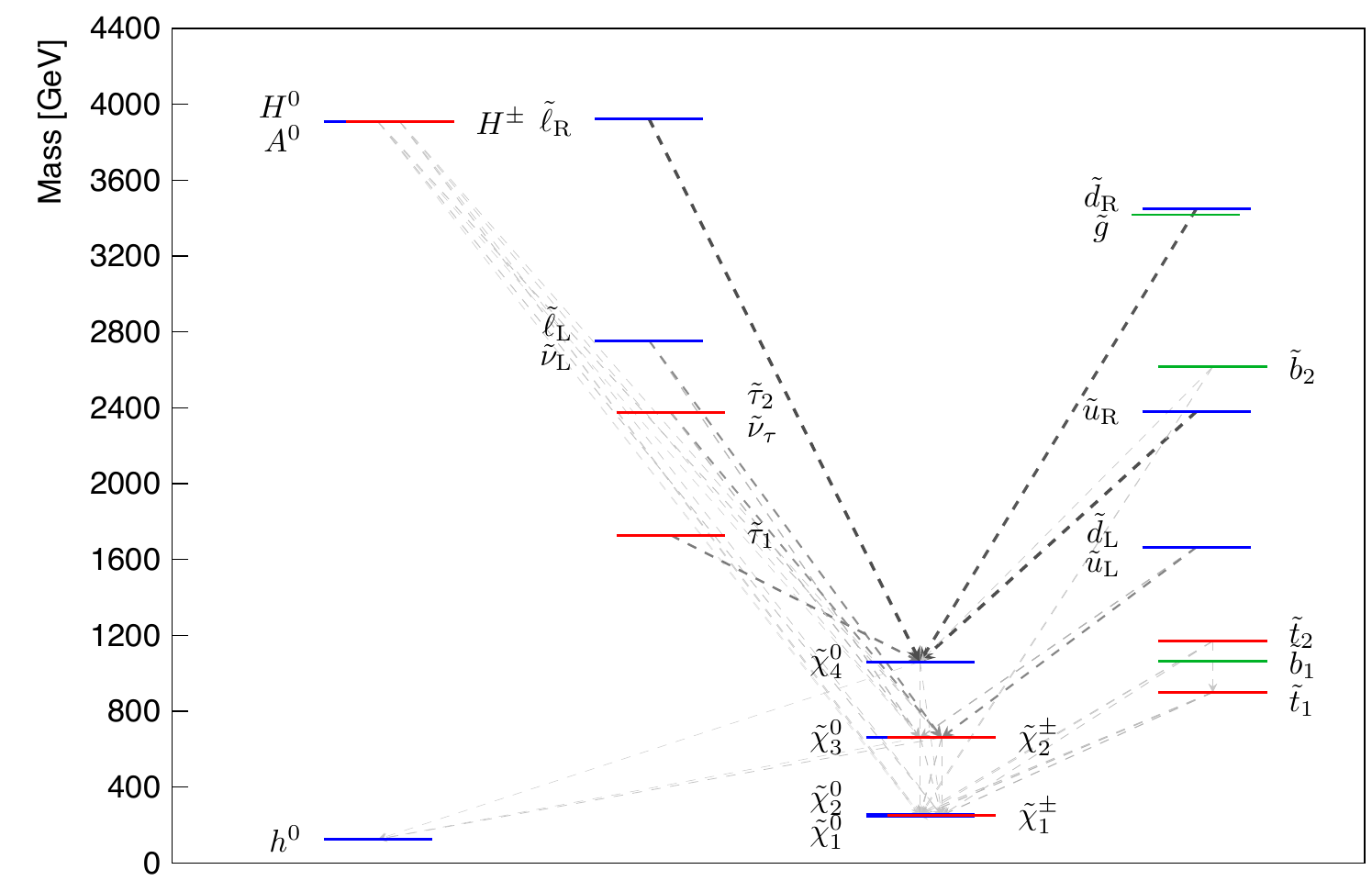}}\\
\subfloat[Point 112647893 (fine-tuning 64)]{\includegraphics[width=0.48\textwidth,height=0.4\textwidth]{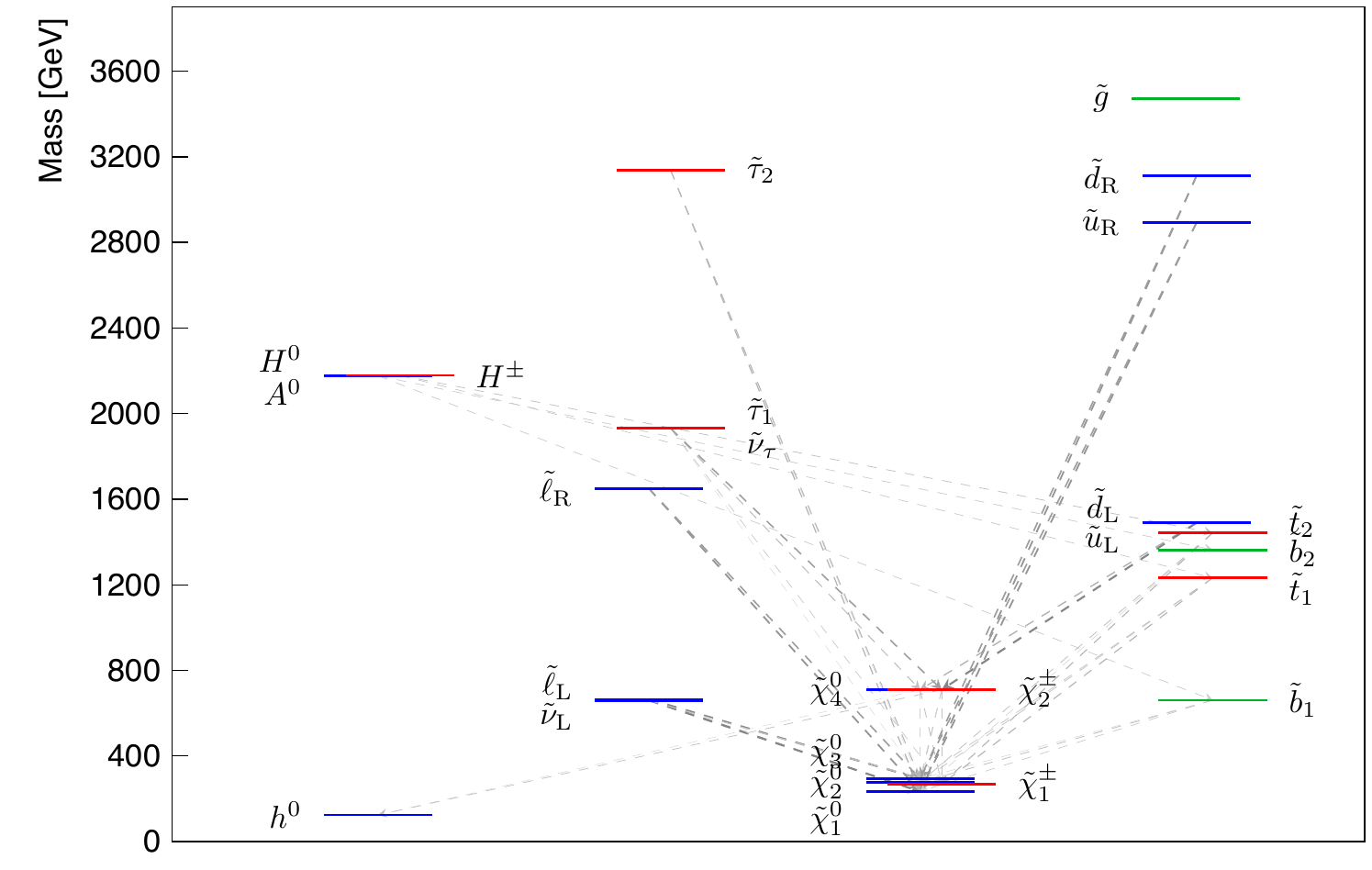}}&
\subfloat[Point 6755879 (fine-tuning 63)]{\includegraphics[width=0.48\textwidth,height=0.4\textwidth]{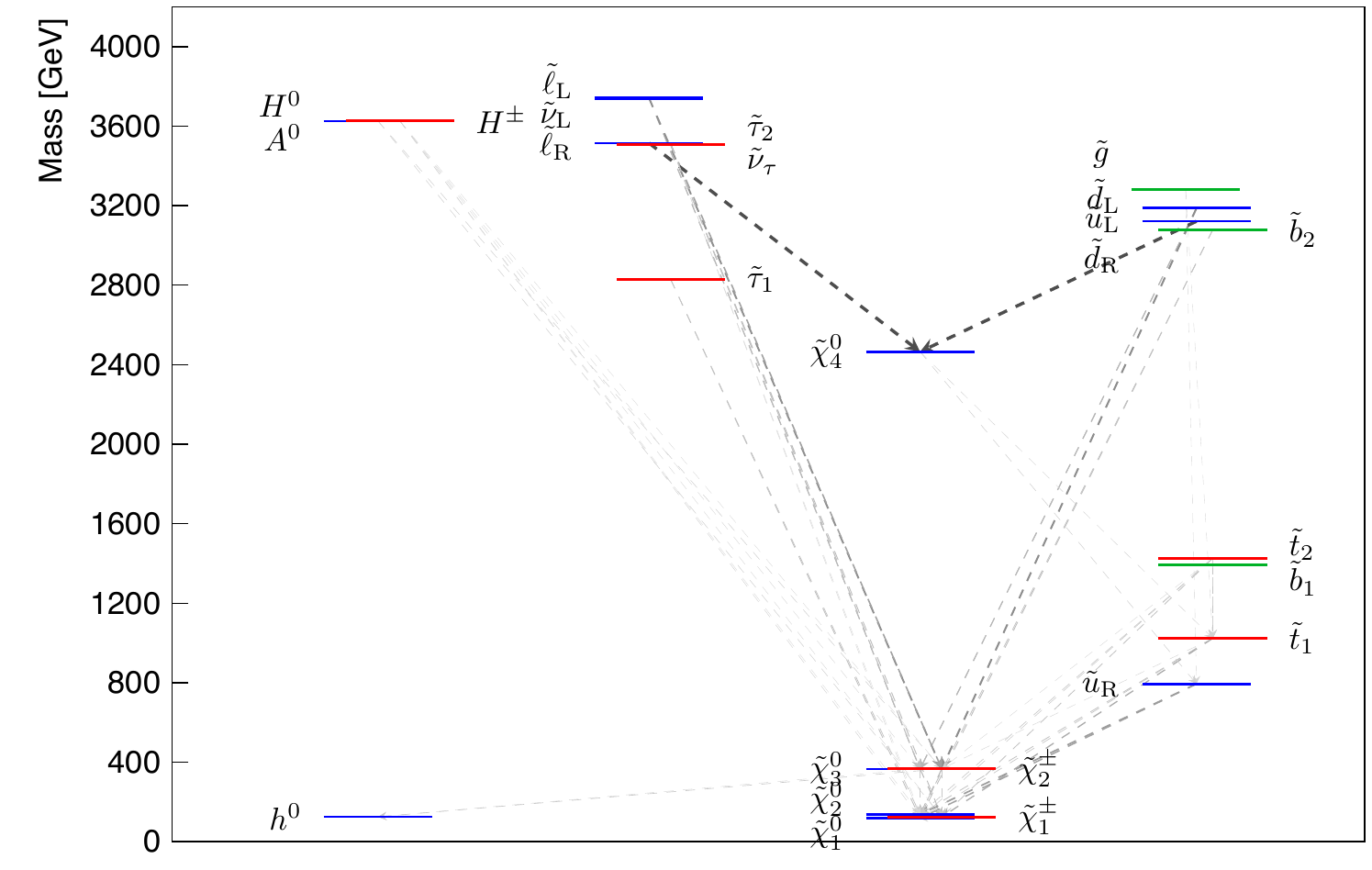}} \\ 
\end{tabular}
\caption{\label{fig:spectra}
The sparticle spectra for some of the 
pMSSM-19 model points with the smallest fine-tuning 
not to have been excluded by the ATLAS \runone{} searches.
The dashed lines indicate the dominant decay modes.
For more details see the text.
}
\end{figure}

The naturalness of pMSSM points is not considered in their generation,
but it is interesting to see just how fine-tuned the model points are after the ATLAS exclusion.
The fine-tuning parameter adopted here is the one defined by Barbieri and
Giudice~\cite{Barbieri:1987fn}.
Figure~\ref{fig:finetuning_beforeAndAfterATLAS} illustrates the prediction of the fine-tuning before and after ATLAS exclusion, for all LSP types (a),
and for each LSP-type separately (b)--(d).
The shapes of the distributions are determined by the ranges of the pMSSM parameters considered.
The dominant contributions to the fine-tuning come from $\mu$ and $A_t$.
It can also be observed that the ATLAS \runone{} searches exclude pMSSM
points with a wide range of different fine-tuning values.
A decrease in sensitivity is observed with increasing fine-tuning for models
with a Higgsino-like LSP.
This is an artifact caused by the fact that the Higgsino is the LSP,
that its mass is driven by the $\mu$ parameter,
and so by requiring large fine-tuning, one is indirectly requiring that there 
be a heavy LSP, reducing ATLAS sensitivity.

The lowest fine-tuning parameter value for a surviving model point is 56.
A representation of the sparticle spectrum of this point can be found in 
Figure~\ref{fig:spectra}(a).
As expected for a generic model point with low fine-tuning, 
the top squarks have sub-\TeV{} masses.
The model point has a wino-dominated LSP with a mass of \SI{107}{\GeV},
which is lighter by \SI{1.6}{\GeV} than its charged partner. 
This mass difference is sufficient to ensure that the chargino 
lifetime is short enough ($<1$\,ps) %(in this case sub picosecond) 
to evade the \DisappearingTrack{} analysis.
The lower part of the mass spectrum includes a Higgsino-dominated  
$\{\tilde{\chi}_2^0,\,\tilde{\chi}_3^0,\,\tilde{\chi}_2^\pm\}$ 
multiplet with masses around \SI{230}{\GeV}
which are mixed with the winos at about the $10\%$ level.
Most strongly interacting sparticles, particularly the gluino
and left-squark doublet, have large masses of around 3 \TeV{}, which is
beyond the \runone{} search reach.
The heavy Higgs bosons are also beyond the \runone{} search reach. 

Similar model points survive with a Higgsino-like LSP,
for example as shown in Figure~\ref{fig:spectra}(b),
which has a fine-tuning of 57.
Model points such as those in both Figure~\ref{fig:spectra}(a) and Figure~\ref{fig:spectra}(b)
would be expected to 
produce observable signals at LHC \runtwo{} through a variety of channels.
In particular in both cases the lighter top squark, with mass of 
about \SI{800}{\GeV} is not far from the current ATLAS \runone{} sensitivity.

Aside from model points with a wino-dominated or a Higgsino-dominated LSP
there also remain model points with low fine-tuning and a mixed LSP.
For example, Figure~\ref{fig:spectra}(c) shows
a model point with a mixed bino-Higgsino LSP
a so-called `well-tempered' neutralino case~\cite{ArkaniHamed:2006mb}.
This model has a bottom squark with mass of about \SI{650}{\GeV},
not far above the \runone{} direct search sensitivity.

Models with a small number of low-mass squarks of the first two generations
also survive. For example, Figure~\ref{fig:spectra}(d) shows
a spectrum of a model point 
with a $\tilde{u}_{\rm R}$ squark with a mass around \SI{800}{\GeV},
which again would be likely to be discoverable at LHC \runtwo.

\FloatBarrier

\section{Conclusion}
\label{sec:conclusion}
The ATLAS Collaboration has performed a wide range of direct searches for 
supersymmetry during the first run of the LHC,
using $pp$ collisions with centre-of-mass energy
up to \SI{8}{\TeV} and an integrated luminosity of up to \maxlumi{}.
The interpretation of those results within the wider framework of the pMSSM gives 
insights into the breadth of sensitivity of those searches. 

From an initial random sampling of 500~million pMSSM points,
generated from the 19-parameter pMSSM,
a total of \numTotal{} model points with \neutralino{} LSP are selected 
each of which satisfies constraints from previous collider searches, precision measurements, 
cold dark matter energy density measurements and direct dark matter searches.
The models are importance-sampled so that there are roughly equal numbers of
models with a bino-like, wino-like or Higgsino-like LSP.
For these model points, more than 30 billion signal events are generated, 
of which more than 600 million events from \numFullSim different model points 
are passed through a {\tt GEANT4}-based fast detector simulation and full reconstruction
to accurately determine which have been excluded (at 95\% CL) by ATLAS searches.

The impact of the ATLAS \runone{} searches on this space is presented, 
showing their overall effect in constraining such supersymmetric models.
The results are particularly clear when considering the
fraction of model points surviving, after projection into 
two-dimensional spaces of sparticle masses.
The constraints on the masses of squarks (including third generation squarks), 
gluinos, electroweakinos, sleptons and heavy neutral Higgs bosons are all
presented.

For the energy and luminosity achieved in LHC \runone{} 
the ATLAS constraints are most 
effective for strongly interacting sparticles,
with weaker constraints on electroweakinos and sleptons.
A general congruence is observed between the pMSSM points 
excluded and the limits determined previously in the
context of simplified models.
Nevertheless, significant differences are observed depending, 
for example, on the number of kinematically accessible squarks,
and on their flavour and couplings,
particularly for direct sleptons production.

The most constraining ATLAS analyses -- for the model points
generated -- were the \ZeroLepton{} analysis,
and the \DisappearingTrack{} analysis, the latter being especially 
powerful in the case of model points with a light wino-like LSP.
Good complementarity is observed between different ATLAS analyses,
with almost all showing regions of unique sensitivity.

When considering dark matter predictions,
the LHC experiments are very complementary
to the direct detection experiments. 
The two techniques have different sensitivities 
across the pMSSM-19 parameter space,
with the ATLAS searches having more sensitivity at lower LSP mass
(with significant sensitivity for $m(\neutralino)\lsim\SI{800}{\GeV}$)
and the direct detection experiments providing bounds 
at larger LSP--nucleon scattering cross-section.
Similarly, the ATLAS experiment provides constraints
which are complementary to those of precision 
measurements of $BR(B_s \to \mu\mu)$, $BR(b\to s \gamma)$,
and $\Delta(g-2)_\mu$. 

Model points with relatively low fine-tuning (of order 50) 
remain viable after LHC \runone{}. Those with the lowest fine-tuning
have relatively light top squarks, 
which indicates that one would expect them to be
accessible by ATLAS searches with the LHC \runtwo{} dataset.

\clearpage
\section*{Acknowledgements}

The ATLAS Collaboration expresses its sincere thanks to Matthew Cahill-Rowley, JoAnne Hewett  
and Ahmed Ismail,
for their sustained efforts in generating the dedicated set of pMSSM points, 
without which this paper would not have been possible. 

% Acknowledgements for papers with collision data
% Version 23-Mar-2015

% Standard acknowledgements start here
%----------------------------------------------
We thank CERN for the very successful operation of the LHC, as well as the
support staff from our institutions without whom ATLAS could not be
operated efficiently.

We acknowledge the support of ANPCyT, Argentina; YerPhI, Armenia; ARC,
Australia; BMWFW and FWF, Austria; ANAS, Azerbaijan; SSTC, Belarus; CNPq and FAPESP,
Brazil; NSERC, NRC and CFI, Canada; CERN; CONICYT, Chile; CAS, MOST and NSFC,
China; COLCIENCIAS, Colombia; MSMT CR, MPO CR and VSC CR, Czech Republic;
DNRF, DNSRC and Lundbeck Foundation, Denmark; EPLANET, ERC and NSRF, European Union;
IN2P3-CNRS, CEA-DSM/IRFU, France; GNSF, Georgia; BMBF, DFG, HGF, MPG and AvH
Foundation, Germany; GSRT and NSRF, Greece; RGC, Hong Kong SAR, China; ISF, MINERVA, GIF, I-CORE and Benoziyo Center, Israel; INFN, Italy; MEXT and JSPS, Japan; CNRST, Morocco; FOM and NWO, Netherlands; BRF and RCN, Norway; MNiSW and NCN, Poland; GRICES and FCT, Portugal; MNE/IFA, Romania; MES of Russia and NRC KI, Russian Federation; JINR; MSTD,
Serbia; MSSR, Slovakia; ARRS and MIZ\v{S}, Slovenia; DST/NRF, South Africa;
MINECO, Spain; SRC and Wallenberg Foundation, Sweden; SER, SNSF and Cantons of
Bern and Geneva, Switzerland; NSC, Taiwan; TAEK, Turkey; STFC, the Royal
Society and Leverhulme Trust, United Kingdom; DOE and NSF, United States of
America.

The crucial computing support from all WLCG partners is acknowledged
gratefully, in particular from CERN and the ATLAS Tier-1 facilities at
TRIUMF (Canada), NDGF (Denmark, Norway, Sweden), CC-IN2P3 (France),
KIT/GridKA (Germany), INFN-CNAF (Italy), NL-T1 (Netherlands), PIC (Spain),
ASGC (Taiwan), RAL (UK) and BNL (USA) and in the Tier-2 facilities
worldwide.
%----------------------------------------------

\clearpage
\appendix
\part*{Appendices}
\addcontentsline{toc}{part}{Appendix}
\label{sec:appendices}
\section{Model point calculation}
\label{sec:modelcalculation}

The calculation of model point properties proceeds as follows.
First,  \program{SoftSUSY 3.4.0}~\cite{Allanach:2001kg}  
is used, embedded in \program{micrOMEGAs 3.5.5}~\cite{Belanger:2006is,Belanger:2010pz}, 
to calculate the sparticle spectrum. This spectrum is used as an input to 
\program{FeynHiggs 2.10.0}~\cite{Heinemeyer:1998yj,Hahn:2013ria}, which recalculates the
light CP-even Higgs boson mass including important 3-loop corrections not
present in \program{SoftSUSY}. For sparticle decay tables,
a modified version of the program \program{SUSY-HIT 1.3}~\cite{Djouadi:2006bz} is used, updated to use \program{HDECAY 5.11}~\cite{Djouadi:1997yw} for the Higgs decays. In some cases,
where relevant or even dominant decay modes are not
included in the \program{SUSY-HIT} output,
\program{MadGraph5\_aMC@NLO 2.1.1}~\cite{Alwall:2014hca} is instead used
to recalculate  those decays as described in detail below. \program{MicrOMEGAs~3.5.5} is used to
calculate the dark matter relic abundance along with the
spin-independent and spin-dependent scattering cross-sections in
direct detection experiments. \program{MicrOMEGAs} is also used to calculate the
following flavour physics and precision electroweak observables:
$\Delta{\rho}$, $\Delta(g-2)_\mu$, BR($b \to s \gamma$) and
BR($B_s \to \mu^+ \mu^-$). Finally, analytic formulas are used to
calculate BR($B^+ \to \tau^+ \nu_{\tau}$) and the $Z$ boson invisible width,
and to check the stability of the vacuum.

\section{Sparticle decay calculation}
\label{sec:decays}

The calculation of sparticle decays is performed as follows.
First, the publicly available version of
\program{SUSY-HIT} is modified by: ($i$) Incorporating light quark and lepton masses
in the calculation of branching ratios and lifetimes for the
various sparticles. Two-body decays implement the full mass
corrections, while three-body decays only include a modified phase-space
cut-off. The mass of the lightest meson of the
appropriate type is included in the relevant phase space calculations to account
for hadronisation effects.  ($ii$) Employing full analytic expressions
from Ref.~\cite{Chen:1999yf} for chargino decays when the
chargino--neutralino mass splitting is $\lsim 1$ \GeV, in which case a
careful treatment of hadronisation is important and can significantly
affect the model phenomenology.  ($iii$) Removing QCD corrections to
decays involving top and bottom squarks due to their tendency to result
in negative decay widths.  ($iv$) Separating the Higgs decays to $\tau$
sneutrinos from those to electron and muon sneutrinos (the public
version of \program{HDECAY} calculates them separately but then averages them).

Several cases have been identified in which \program{SUSY-HIT} does not
calculate important decay channels or predicts a value that differs
significantly from the full matrix element prediction, which are detailed in 
the following.

\subsection{Right-handed sfermion decays}

For right-handed sfermions, \program{SUSY-HIT} only calculates two-body decays
of the form $\tilde{f}_{\rm R} \to f \tilde{\chi}^0$ (and $\tilde{f}_{\rm R} \to
f \tilde{g}$ for right-handed squarks). If the bino component of the
kinematically accessible neutralinos is very small, these decays can
be highly suppressed. As a result, it is common for three-body decays
of the form $\tilde{f}_{\rm R} \to f \tilde{\chi} B$, where $B$ is an
electroweak gauge boson, to dominate over the two-body decays. In cases
where the splitting between $\tilde{f}_{\rm R}$ and the LSP is small, the three-body
decay modes can be forbidden and a four-body decay (with $B$ off-shell)
can potentially be dominant, although numerically this turns out to be
uncommon. Therefore \program{MadGraph5\_aMC@NLO} is employed to calculate the
decays of any right-handed sfermion satisfying the following selection
criteria (designed to select all model points for which the multi-body
decays would be significant while minimising the number of model points
selected):

\begin{description}

\item[1)] The bino content of the LSP is less than 10\%.

\item[2)] Any accessible neutralinos with a bino content above 10\% have a 
mass splitting with $\tilde{f}_{\rm R}$ that is less than 100\,\GeV{} or below 20\% of 
the $\tilde{f}_{\rm R}$--LSP mass splitting.

\item[3)] For squarks, if the gluino is accessible it must have a mass 
splitting with $\tilde{f}_{\rm R}$ that is less than 100 \GeV{} or below 10\% of the 
$\tilde{f}_{\rm R}$--LSP mass splitting.

\end{description}

For model points with a $\tilde{f}_{\rm R}$ -- LSP mass splitting above 100\,\GeV, 
all possible three-body decays are calculated with \program{MadGraph5\_aMC@NLO} and added
to the existing two-body decays. This can lead to double-counting
when three-body decay diagrams contain intermediate particles which can go
on shell. In this case, the on-shell component of the three-body decay is
already included in the two-body decay width (as a tree-level two-body
decay in which the SUSY decay product also undergoes a tree-level two-body
decay). This is resolved by selectively removing redundant
decays. First, the sum of the combined three-body decay modes is compared
with the sum of tree-level two-body decays for which the SUSY decay product
has tree-level two-body decays (the ``on-shell component''). In the case
where the on-shell component comprised over 80\% of the total three-body
width, the three-body decays are discarded and only the two-body decays kept,
since the sequential two-body decays give a more accurate representation
of the decay kinematics. If the on-shell component is less than 80\%
of the total three-body width, the off-shell component of the three-body
decays is deemed significant and the redundant on-shell component
(the selected two-body decays) is discarded.

For model points with a $\tilde{f}_{\rm R}$--LSP mass splitting below \SI{100}{\GeV}, four-body
decays could be important. Here use was made of the MadWidth
package in \program{MadGraph5\_aMC@NLO}~\cite{Alwall:2014hca} by calling the
\texttt{compute\_widths} function; this function automatically determines
whether three-body and four-body decays are required and includes an
automatic mechanism for removing the degeneracy. It is found that
MadWidth reproduces the results obtained using the procedure
described above for several test cases with three-body decays.

In total the three-body decays are recalculated for at least 1 sparticle decay
in 160482 model points. For 86109 model points, $\tilde{u}_{\rm R}$ decays are 
recalculated, while for 86791 model points the $\tilde{d}_{\rm R}$ decays are 
recalculated and for 122127 model points the $\tilde{e}_{\rm R}$ decays are 
recalculated. The four-body decays are calculated for 11633 model points, 
although this ends up making a significant contribution only for a few model points.

\subsection{Wino and Higgsino decays to sfermions}

Although \program{SUSY-HIT} normally calculates three-body decays of neutralinos and
charginos that have mass splittings with the LSP below the $Z$ or $W$ boson
masses, it does not calculate these decays when other two-body decay modes
are available (to avoid the sort of double-counting described
above). This can be a problem when the available two-body decays are
suppressed by a small mixing angle (which can occur if the decaying
gaugino is a neutral wino or Higgsino and the two-body decay is to a
right-handed sfermion) or by kinematics. For this reason
\program{MadGraph5\_aMC@NLO} was employed to calculate the full three-body decays in cases where
the decaying gaugino has a bino content below 10\% and a right-handed
sfermion is accessible, or has an accessible decay to any sfermion
where the sfermion--gaugino splitting is below 20\% of the gaugino--LSP
splitting. This occurs in 716 model points with wino or Higgsino LSP and 2080 
model points with bino LSP.
Once again the issue with double-counting between
the two-body and three-body decays is resolved; in this case any
tree-level two-body decays where the product has a tree-level two-body
decay are removed, since the off-shell three-body decays are expected 
to be important in nearly all of the selected scenarios.

\subsection{Four-body top squark decays}

Although \program{SUSY-HIT} includes a calculation of the four-body top squark decays, 
a significant discrepancy is observed between the \program{SUSY-HIT} results
and the decay width calculated by \program{CalcHEP} in a previous model
set. Therefore \program{MadGraph5\_aMC@NLO} is used to recalculate the top squark
decays for any model point in which \program{SUSY-HIT} predicts a non-zero four-body
decay rate for the light top squark. In this case, there are no concerns
about overlap with other top squark decay modes. This procedure is employed
for 15 model points with wino or Higgsino LSP and 5127 model points with bino LSP.

\section{Calculational pathologies}
\label{sec:pathologies}

To ensure that the spectrum is calculable, any model point with
the following \texttt{SOFTSUSY} errors is discarded: ``No acceptable solution found'',
``Non-perturbative'', ``No convergence'', ``Inaccurate Higgs mass'',
``Numerical problemThrown'', and ``Not achieved desired accuracy''. 
Additionally the accuracy of the spectrum generation is tested by
re-generating the spectrum with \texttt{SuSpect}~\cite{Djouadi:2002ze} and
ensuring that \texttt{SuSpect} does not produce any fatal errors or predict any
sparticle masses to differ by more than 50\% from the \texttt{SoftSUSY}
prediction. Since \texttt{FeynHiggs} is used for the light CP-even Higgs boson  mass,
any model point for which \texttt{FeynHiggs} indicates an error or fails to
write an output file, is discarded. Additionally 
it is required that $m(h)$ be within 5\,\GeV{}
of the \texttt{SOFTSUSY} prediction, since a small tail of model points
is observed for which this deviation is extremely large, suggesting that the
calculation may have been unreliable. \texttt{FeynHiggs} also gives its own estimate
of the uncertainty in the Higgs boson mass calculation. 
When this uncertainty is larger than 5\,\GeV{} the model points are discarded as being
unreliable. Finally, the reliability of the \texttt{SUSY-HIT} decays
is asserted by discarding model points for which any particle has a width larger than 
1\,\TeV{} or a negative branching ratio.

\subsection{Theoretical Constraints}
In addition to the numerical pathologies described above, 
model points that produced the following \texttt{SoftSUSY} errors are discarded as being
theoretically inconsistent: ``tachyon'', ``MuSqWrongsign'',
``m3sq-problem'', and ``Higgs potential ufb''. Additionally it is checked
that the scalar potential does not break colour or charge, using
Equation (6) of Ref.~\cite{Chowdhury:2013dka} for A$_t$ and analogous
formulas for A$_b$ and A$_{\tau}$. It should be noted that Equation (6)
of Ref.~\cite{Chowdhury:2013dka} uses $m(t_{\rm L}$) and $m(t_{\rm R}$), while this \paper{} uses
the Lagrangian parameters $m(Q_3)$ and $m(u_3)$.

\section{Importance sampling by LSP type}
\label{bwh}

As described in Section~\ref{sec:models:properties},
the model points generated are categorised according to LSP type
(see Table~\ref{table:lsptype}), and importance sampled 
so that approximately equal numbers of model points are available for each LSP type.

There are various ways in which over-production of dark matter can be achieved.
A neutralino LSP can annihilate through $s$-channel exchange of a $Z$ or
Higgs boson or through $t$-channel exchange of a chargino or
sfermion. If another sparticle is nearly degenerate with the LSP, that
sparticle can also enhance the effective LSP annihilation rate through
coannihilation.  In general, sparticles with the strongest
self-interactions (particularly coloured sparticles) make the most
effective coannihilators.  Supersymmetric models with a wino-like LSP
or  Higgsino-like LSP (see Table~\ref{table:lsptype})
have a chargino that is nearly degenerate with the LSP, resulting in a
sizeable annihilation rate through chargino exchange; the degenerate
chargino also serves as a coannihilator. The relic density constraint
is therefore satisfied as long as the wino (Higgsino) mass is below
2.7 \TeV{} (1 \TeV{}) \cite{Cirelli:2005uq,Cirelli:2007xd,Hisano:2006nn}. On
the other hand, bino-like LSPs ($N_{11}^2 > $ max$(N_{12}^2, N_{13}^2
+ N_{14}^2)$) have no guaranteed annihilation mechanism. One
possibility is for the LSP to mix with the
Higgsino \cite{Feng:2000gh}, giving couplings to the $Z$ and Higgs
bosons and to the charginos. In order for the annihilation rate to be
large enough, the Higgsino must be somewhat degenerate with the LSP
(resulting in a large mixing angle) or the LSP must be approximately
half the mass of the $Z$ or MSSM Higgs bosons (termed a ``funnel''
region), producing a resonantly enhanced annihilation rate that
compensates for the mixing-angle-suppressed couplings. Alternatively,
a chargino or sfermion can be nearly degenerate with the LSP,
resulting in a sizeable $t$-channel annihilation rate and/or significant
coannihilation~\cite{Edsjo:2003us}. 

Correcting the undersampling of model points with bino-like LSPs is
important for two reasons.  First, they are the only model points in which
the observed relic abundance can be obtained with an LSP light enough
to be seen at the 8 \TeV{} LHC, making them particularly interesting from
the dark matter perspective. Second, the absence of an associated
chargino can significantly alter the available decay modes. As an
example, top squark decays to $W^+ b \tilde{\chi}_1^0$ only occur in models with
bino-like LSPs, since the top squark can simply decay to $b \tilde{\chi}^+$ in wino
or Higgsino LSP models.

\printbibliography
\clearpage
%\newpage
%\newpage 

% ATLAS Collaboration author list
% Data extracted on 13-Aug-2015 for paper reference SUSY-2014-08
% \documentclass[11pt]{article}
% \usepackage{a4wide}\begin{document}
\begin{flushleft}
{\Large The ATLAS Collaboration}

\bigskip

G.~Aad$^{\rm 85}$,
B.~Abbott$^{\rm 113}$,
J.~Abdallah$^{\rm 151}$,
O.~Abdinov$^{\rm 11}$,
R.~Aben$^{\rm 107}$,
M.~Abolins$^{\rm 90}$,
O.S.~AbouZeid$^{\rm 158}$,
H.~Abramowicz$^{\rm 153}$,
H.~Abreu$^{\rm 152}$,
R.~Abreu$^{\rm 116}$
% ...
\bigskip
\\
% ...
$^{11}$ Institute of Physics, Azerbaijan Academy of Sciences, Baku, Azerbaijan\\
% ...
$^{85}$ CPPM, Aix-Marseille Universit{\'e} and CNRS/IN2P3, Marseille, France\\
% ...
$^{90}$ Department of Physics and Astronomy, Michigan State University, East Lansing MI, United States of America\\
% ...
$^{107}$ Nikhef National Institute for Subatomic Physics and University of Amsterdam, Amsterdam, Netherlands\\
% ...
$^{113}$ Homer L. Dodge Department of Physics and Astronomy, University of Oklahoma, Norman OK, United States of America\\
% ...
$^{116}$ Center for High Energy Physics, University of Oregon, Eugene OR, United States of America\\
% ...
$^{151}$ Institute of Physics, Academia Sinica, Taipei, Taiwan\\
$^{152}$ Department of Physics, Technion: Israel Institute of Technology, Haifa, Israel\\
$^{153}$ Raymond and Beverly Sackler School of Physics and Astronomy, Tel Aviv University, Tel Aviv, Israel\\
% ...
$^{158}$ Department of Physics, University of Toronto, Toronto ON, Canada\\
% ...
\end{flushleft}

% \end{document}
% Created with xml2latex.py

%\part*{Auxiliary material}

%\addcontentsline{toc}{part}{Auxiliary material}
%\input{auxiliary}

\end{document}